\documentclass[11pt,A4]{report}

\pdfoutput=1

\usepackage{graphicx}

\usepackage[left=2.5cm,right=2.5cm,top=2.5cm,bottom=2.5cm]{geometry}

\usepackage{amsfonts}
\usepackage{amsmath}
\usepackage{amssymb}

\usepackage{hyperref}

\def\openone{\leavevmode\hbox{\small1 \normalsize \kern-.64em1}}

\begin{document}


\thispagestyle{empty}

\vspace{6cm}

\begin{center}
\textbf{
{\LARGE Quantum Communication} \\
{\large Non-classical correlations and their applications}}
\end{center}

\vspace{3cm}

\begin{center}
{\large Tomasz Paterek} \\
Instytut Fizyki Teoretycznej i Astrofizyki \\
Uniwersytet Gda\'nski
\end{center}

\vspace{8.5cm}

\begin{center}
Doctoral dissertation written under \\
supervision
of\textbf{ Professor Marek \.Zukowski}.
\end{center}

\vspace{2cm}

\begin{center}
Gda\'nsk 2007
\end{center}


\newpage
\textrm{ }

\newpage

\begin{center}
\textbf{{\large Abstract}}
\end{center}

Communication is transfer of information.
Compared to classical physics,
new possibilities arise 
when information is encoded in quantum systems
and processed with quantum operations.

In the first part of this thesis
Bell's theorem is revisited.
It points at a difference between the quantum
and the classical world.
This difference is often behind the
advantages of solutions using quantum mechanics.
New and more general versions of Bell inequalities
are presented.
These inequalities involve multiple settings per observer.
Compared with the two-setting inequalities,
the new ones 
reveal the nonclassical character
of a broader class of states.
Some of them are also proven to be optimal (tight).

Next, we go beyond Bell's theorem.
It is shown,
both in theory and in experiment,
that incompatibility between quantum mechanics
and realistic theories
can be extended into an important class of nonlocal models.
We also show that
the violation of Bell inequalities
disqualifies local realistic models
with a limited lack of the experimenter's freedom.
This, at first glance quite philosophical result,
has its down-to-earth implications
for quantum communication.

In the second part of the thesis
well-known examples
of quantum communication are reviewed.
Next, new results concerning quantum cryptography and 
quantum communication complexity are given.

\newpage

\begin{center}
\textbf{{\large Publications}}
\end{center}

This work is based on the following publications:\footnote{Throughout the thesis these papers are cited as [Pn].}

\begin{itemize}

\item[{[}P1{]}]
S. Gr\"oblacher, T. Paterek, R. Kaltenbaek, {\v C}. Brukner,
M. \.Zukowski, M. Aspelmeyer, and A. Zeilinger \\
\emph{An experimental test of non-local realism} \\
Nature \textbf{446}, 871 (2007).

\item [{[}P2{]}] T. Paterek \\
\emph{Measurements on composite qudits} \\
Phys. Lett. A \textbf{367}, 57 (2007).

\item [{[}P3{]}] K. Nagata, W. Laskowski, and T. Paterek \\
\emph{Bell inequality with an arbitrary number of settings and its applications} \\
Phys. Rev. A \textbf{74}, 62109 (2006).

\item [{[}P4{]}] J. Kofler, T. Paterek, and {\v C}. Brukner \\
\emph{Experimenter's freedom in Bell's theorem and quantum cryptography} \\
Phys. Rev. A. \textbf{73}, 22104 (2006). 

\item [{[}P5{]}] T. Paterek, W. Laskowski, and M. \.Zukowski \\
\emph{On series of multiqubit Bell's inequalities} \\
Mod. Phys. Lett. A. \textbf{21}, 111 (2006).

\item [{[}P6{]}] W. Laskowski, T. Paterek, M. \.Zukowski, and {\v C}. Brukner \\
{\it Tight multipartite Bell's inequalities involving many measurement settings} \\
Phys. Rev. Lett. \textbf{93}, 200401 (2004). 

\item [{[}P7{]}] {\v C}. Brukner, T. Paterek, and M. \.Zukowski \\
{\it Quantum communication complexity protocols based on higher-dimensional entangled systems} \\
Int. J. Quant. Inf. \textbf{1}, 519 (2003).

\end{itemize}

\newpage

\begin{center}
\textbf{{\large Acknowledgements}}
\end{center}

I would like to thank Professor
Marek \.Zukowski for opening me a completely new world.
Many thanks to my friends from the groups of Gda\'nsk and Vienna.\\

This research was supported by several institutions 
(in chronological order):
\begin{itemize}
\item
University of Gda\'nsk\\
Stipend for Ph. D. students\\
Grant No. BW/5400-5-0256-3
\item
Austrian-Polish projects \emph{Quantum Communication and Quantum Information}
\item German-Polish project 
\emph{Novel Entangled States for Quantum Information Processing: Generation and Analysis}
\item
Foundation for Polish Science\\
Stipends under the Professorial Subsidy of Marek \.Zukowski
\item
State Committee for Scientific Research\\
Grant No. PBZ-MIN-008/P03/03\\
Grant No. 1 P03B 04927
\item
The Erwin Schr\"odinger International Institute for Mathematical Physics\\
Junior Research Fellowship
\item
European Union\\
QAP programme Contract No. 015848
\end{itemize}

\tableofcontents

\chapter{Introduction and summary}

\section{Introduction: field of quantum communication}

We are living in the age of information.
One can safely state that the rapid progress
of the last century was connected with 
a growing accessibility of information.
In turn, this accessibility is linked
with the discoveries of the secrets of nature.
Many of them were secrets of quantum physics.
Quantum physics may 
influence our everyday life soon.
Many technological developments
reach the scale of its applicability.
Already today it is estimated that quite a big part
of our tools can be designed only due to
our knowledge of quantum mechanics.
For example, if the progress in the power of personal computers
is going to stay at the current level
(doubles itself every two years -- the so called Moore's law)
quantum effects inevitably start to dominate in the processors soon.

Information and physics cannot be divided.
Information is an intrinsically physical concept.
It relies on physical systems
in which information is stored
and by means of which information is processed or transmitted.
Transmission of information encoded
in quantum systems
and processed by operations
allowed by quantum mechanics
is studied in the field of quantum communication.

Quantum communication is a relatively new sub-branch of physics and
information theory. 
Its main goals include a theory of optimal encoding and decoding
of information into quantum systems,
and their faithful transmission through (possibly noisy)
communications channels.
 It is also aimed at showing communication
 tasks which are either impossible in classical regime
 or their quantum versions outperform
 the best classical solution.
 At least one of such protocols,
quantum cryptography,
is already at the stage of useful real-world applications. 

Quantum cryptography allows secure communication.
Communication which guarantees that transmitted information
is inaccessible to third parties.
The security is due to the laws of quantum physics.
Any disturbance of a quantum system, 
inevitably caused by an eavesdropper, 
changes a state of the system.
This change can in principle be detected by legitimate partners.
Classical crypto-algorithms up to date
make use of problems which are believed to be computationally hard.
Despite of many attempts 
it is not proven that classical cryptography 
cannot be compromised.
Moreover, there exist a quantum algorithm (Shor's algorithm)
which efficiently solves problems at the heart of classical cryptography
(believed to be classically hard).
Thus, quantum cryptography may one day dominate on the market.
There are also attempts to unify cryptography and
computation and there already are proposals
for secure computation.

There is an ongoing debate on which are the properties of quantum physics
that allow to derive benefit from using "quantum" in information processing.
The correlations allowed by quantum mechanics seem to be a good candidate.
The correlations between quantum particles
can be much stronger than the correlations between classical objects.
These non-classical correlations are due to quantum entanglement.
Although there exist superior quantum protocols
which make use of no entanglement,
this purely quantum resource is usually sufficient
for better performance of quantum algorithms.
As soon as one recognizes that a problem requires correlations
similar to those of quantum entanglement,
most probably the problem has efficient quantum solution
utilizing a suitable entangled state.

For example, in the field of communication complexity (introduced in $1979$ by Yao)
one can show problems with quantum solutions
which outperform any classical ones.
In a communication complexity problem,
separated parties performing local computations
exchange information in order to accomplish
a globally defined task,
which is impossible to solve singlehandedly.
These problems find applications in optimization of
data structures or minimization of time required to
perform a computation with large integrated circuits.
An instance of a communication complexity problem
is evaluation of a function dependent on distributed inputs.
Imagine every party receives two bits
in such a way that they do not know
the bits of any other party.
Their common goal is to compute
a value of a function defined on all the bits.
However, each party can communicate only one bit.
Before parties receive their inputs
they can communicate freely.
They can fix the protocol they will use
once the data is obtained,
they can share some correlated strings of numbers
in the classical scenario or entangled states in the quantum case.
The communication starts to be ``expensive'' with the delivery of the bits.
The essence of quantum solutions to communication complexity problems
is to share an entangled state before parties receive their inputs.
The state is such that
when suitably measured
gives correlated results in agreement with the function to be computed.
When the bits are delivered,
parties make local measurements depending on their local inputs,
next they communicate local outcomes (assumed to be bits in our example),
and give as the computed value of the function 
the product of all local results.
There are functions for which this protocol is more efficient in terms of communication complexity
than the best classical protocol.
Evidently, the quantum protocol is linked with the entanglement.\footnote{However, 
it is possible to recast at least some of entanglement-based
problems in terms of a single particle sequentially transmitted
from one party to another.
The quantum feature employed here is superposition.
Information is stored in the relative phases between
elements of superposition.
Thus, entanglement helps to link problems and their quantum solutions.
Nevertheless entanglement is not necessary for quantum advantage.}

Generally, it is a very hard problem to decide 
whether a quantum state is entangled or not.
The problem seems to be even more difficult
when experimental data is taken into account.
It can happen that it is not even clear 
which quantum state
describes the data.
Fortunately, there are operational criteria (entanglement witnesses), 
relying on measurements of correlations,
with a possible outcome from which one can conclude that the
state is entangled.

One of such witnesses is a Bell inequality.
A Bell inequality is satisfied by all 
states which are not entangled.
Thus, if a violation of a Bell inequality is  
observed the state which describes the results is entangled.
Interestingly, Bell inequalities were first introduced in a context
of foundations of quantum mechanics.
They represent constraints on correlations which must be fulfilled
by all models which are based on the classical concepts
and the principle of relativistic causality.
With the emergence of quantum information
Bell inequalities found new applications.
Limits of performance of certain classical protocols
(e.g. already mentioned communication complexity protocols)
can be described in a form of a Bell inequality.
Since entangled states violate Bell inequalities
it is clear that quantum protocols
can beat classical limits.
In this way, human philosophical curiosity
has found applications in applied science.

The link between physics and information
also brings new insight into physics itself.
It is always good to view problems
from different perspectives.
Quantum information puts forward such a new perspective.

\section{Summary of the results}

In the first part of this dissertation
Bell's theorem is revisited.
The evolution of Bell inequalities is described.
We starting with the problem of the possibility of 
local realistic models of quantum predictions,
which was posed by Einstein, Podolsky and Rosen \cite{EPR}.
Next, we scan through some known versions of Bell's impossibility theorem 
\cite{BELL,CHSH,CH,GHZ},
and finish with the necessary and sufficient condition
for the local realistic description
of correlation experiments performed on many qubits \cite{WW,WZ,ZB}.
Next, basing on the assumptions of Bell,
we present new multisetting inequalities for many qubits [P3,P5,P6].
The inequalities are derived using two different techniques.
Inequalities [P5,P6] are proven to be optimal
but they cannot involve arbitrary number of settings per party.
The other inequalities [P3] incorporate in a compact form
any number of settings per party, but they are not always optimal.
In both cases, we derive conditions for violation
of the inequalities and present examples of states which (do not)
violate them.
It is shown that the 
multisetting inequalities
reveal non-classical character of certain states
which satisfy all two-setting Bell inequalities
with correlation functions \cite{WW,WZ,ZB}.

Violation of Bell inequalities
was observed in many experiments.
The results
agree with quantum predictions 
(the milestones are those of \cite{FC,ASPECT,STRICT_LOCALITY,DETECTION_LOOPHOLE}).
Certain loopholes are known to exist
which still allow to explain experimental data in a local realistic way.
However, these were all separately closed in the cited experiments.
Thus, the violation of local realism
is usually considered as a well established fact.
We go beyond Bell's assumptions and
describe a class of nonlocal realistic theories
still incompatible with quantum mechanics.
This class was introduced by Leggett \cite{LEGGETT_NONLOCAL}.
We have also performed an experiment
with entangled photons
which disqualifies this class of theories.
This was the first experimental demonstration
which invalidates some nontrivial nonlocal realistic models.
The considered theories (i) model all experiments
in which a violation of two-setting Bell inequalities 
(e.g. the widely studied CHSH inequality \cite{CHSH}) is observed;
(ii) model perfect correlations in the complementary bases,
which is \emph{the} feature of the Bell singlet state;
(iii) although nonlocal do not allow to transmit information
faster than speed of light [P1].

In an independent line of research
we have relaxed, 
often tacit in the derivation of Bell inequalities,
the ``free-will'' assumption \cite{GILL1,GILL2}.
This assumption is essential for Bell experiments,
as lack of freedom to choose between different experimental arrangements
allows one to explain a violation
of Bell inequalities within local realism.
We argue that within a local realistic model
this freedom can experimentally be checked.
If one wants to keep such a picture,
the experimental evidence of a violation of Bell inequalities
sets the minimal amount to which the freedom has to be abandoned.

In the second part of the thesis some examples of the superiority of quantum communication are presented.
We review quantum teleportation \cite{TELEPORTATION},
quantum dense coding \cite{DENSE_CODING},
quantum cryptography \cite{BB84,EKERT,BBM92},
and quantum communication complexity \cite{BRASSARD_REVIEW,BZPZ}.
In the last two fields, both linked with Bell's theorem,
new results are presented.

In the case of quantum cryptography we show,
following the freedom considerations,
that one can relax to some extend the assumption
that laboratories of authenticated parties are not vulnerable
and still secure quantum key distribution is possible [P4].
It is also shown that quantum cryptography
with higher-dimensional quantum systems \cite{MOHAMED},
proven to be more secure than qubit-based protocol,
is relatively easy to realize
using system composed of two subsystems [P2].

In the field of quantum communication complexity
we present the general link between Bell inequalities
for qubits and communication complexity problems \cite{BZPZ},
associated with one of the multisetting inequalities [P3].
Next, we construct a communication
complexity task for higher-dimensional quantum systems [P7].
The quantum solution outperforms a broad class
of classical protocols and it may be conjectured,
based on the recent results for qubits \cite{EXP_CCP}, 
that the class of classical solutions includes the optimal one.

\chapter{Non-classical correlations}

This part of the thesis is devoted to Bell's theorem \cite{BELL}.
It states that no local realistic (classical)
model exists which explains all quantum predictions.
Thus, Bell's discovery points at a difference
between the quantum and the classical world.
Some conditions necessary for the local realistic models 
are given in form
of Bell inequalities.
Quantum states which violate these inequalities
are a valuable resource in 
quantum communication
and 
quantum
information processing in general \cite{ACIN_IJQI}.

First we present a brief history of Bell's discovery 
and a few well-known versions of his theorem.
Further on, a new family of tight\footnote{The concept of tight inequality is described in section 2.1.7}
(optimal)
Bell inequalities is discussed,
which enlarges the class
of quantum states that do not admit a local
hidden-variable (realistic) description.
Next, we give a compact formula for Bell
inequalities involving an arbitrary number of measurement
settings and an arbitrary number of observers.
Many previously known inequalities
are special cases of this
general one.
We also present the violation conditions for these inequalities
and examples of states which (do not) violate them.

In the following section
it is proven, and experimentally confirmed,
that quantum mechanical predictions
are incompatible with certain plausible
classes of \emph{nonlocal} hidden-variable theories.
This program was initiated by Leggett \cite{LEGGETT_NONLOCAL}.

We also relax the assumption
of experimenter's freedom to choose
between different measurement settings.
A measure of the lack of this freedom is developed,
and the minimal extend of this lack,
which allows to explain the violation
of Bell inequalities within a local realistic picture,
is derived.

\section{Overview of earlier works on Bell's theorem}

\subsection{Bell's theorem}

Quantum mechanics gives predictions in form of probabilities.
Already some of the fathers of the theory were puzzled with the question
whether there can exist
a deterministic structure beyond quantum mechanics which recovers quantum statistics
as averages over ``hidden variables'' (see a beautiful review by Clauser and Shimony \cite{CS}).
In this way, it was hoped, one could get a classical-like description
which would solve the problems
with the interpretations of quantum mechanics.
In his famous impossibility proof Bell made precise assumptions about the form of 
a possible underlying hidden variable structure. 
Spatially separated systems and laboratories were assumed to be independent of one another \cite{BELL}.
He derived an inequality which must be satisfied by all such (local realistic) structures.
Next, he presented example of quantum predictions which violate it.
In this way the famous Einstein-Podolsky-Rosen (EPR) paradox \cite{EPR} was solved.
Bell proved that EPR elements of reality 
cannot be used to describe quantum mechanical systems.

The noncommutativity of quantum theory precludes simultaneous deterministic predictions 
of measurement outcomes of complementary observables.
For EPR this indicated that  
``the wave function does not provide a complete description
of physical reality''.
They expected the complete theory 
to predict outcomes of all possible measurements,
prior to and independent of the measurement (realism),
and not to allow
``spooky action at a distance'' (locality).
Such a completion was disqualified by Bell.

A more general version of Bell's theorem for two qubits (two-level systems) was given 
by Clauser, Horne, Shimony, and Holt (CHSH), 
and extended by Clauser and Horne (CH) \cite{CHSH,CH}.
The important feature of the CHSH and CH inequalities,
which hold for \emph{all} local realistic theories,
is that they can not only be compared with ideal quantum predictions,
but also with experimental results.
Thus, a debate that seemed quasi-philosophical could be moved into the lab!

The three or more qubit versions of Bell's
theorem were presented by Greenberger, Horne, and Zeilinger (GHZ),
surprisingly $25$ years after the original paper of Bell \cite{GHZ, GHSZ}.
In contradistinction with the two particle case,
now the contradiction between local realism and quantum mechanics
could be shown for perfect correlations.
Immediately after that, Mermin produced a series of inequalities for arbitrary many particles,
which cover the GHZ case, and made the GHZ paradox directly testable
in the laboratory \cite{MERMIN}.
A complementary series of inequalities was introduced by Ardehali \cite{ARDEHALI}.
In the next step Belinskii and Klyshko gave series of two-settings inequalities, 
which contained the \emph{tight} inequalities of Mermin and Ardehali \cite{BELINSKII}.
Finally, the full set of tight two-setting
Bell inequalities for dichotomic observables, involving correlations between $N$ partners, 
was described independently by Werner and Wolf \cite{WW},
and in the papers by Weinfurter and \.Zukowski \cite{WZ}, and \.Zukowski and Brukner \cite{ZB}. 

Intuitively, one would link the violation of
Bell inequalities with entanglement.
Indeed, only entangled states can violate them.
Surprisingly, there are \emph{pure} entangled states
whose multiparticle correlations, 
obtained in a two-setting Bell experiment,
can be modelled in a local realistic way \cite{ZBLW}.
To reveal non-classical behaviour of such states
one needs to perform Bell experiments with many settings per party.
Various methods were
proposed to obtain multisetting inequalities
\cite{PITOWSKYSVOZIL, SLIWA, COLLINS, MAREK3X3, ZUK93,GISIN_PLA_99,MASSAR, MASSAR2, MASSAR3}.
Here, we present the simple and efficient method of [P6].
It allows a derivation of tight Bell inequalities
involving various
combinations of the number of settings per party,
and an arbitrary number of parties.
Finally, an inequality
incorporating an arbitrary number of settings
and arbitrary number of observers is given [P3].
However, this inequality is not always tight.

\subsection{Einstein-Podolsky-Rosen}

In their original paper Einstein, Podolsky, and Rosen (EPR)
considered quantum predictions for measurements of position and momentum \cite{EPR}.
We explain their reasoning with a simpler example of two
maximally entangled qubits.
This approach was first presented by Bohm \cite{BOHM}.

EPR assume there exists an objective reality
independent of any physical theory.
Theoretical concepts help us to understand this reality,
``by means of these concepts we picture this reality to ourselves''.
By EPR the necessary condition for completeness of a physical theory is that 
\emph{every element of physical reality must have a counterpart in the physical theory}.
Next, they define elements of physical reality
in the following way: 
\emph{If, without in any way disturbing the system, we can predict with certainty
(i.e., with probability equal to one)
the value of a physical quantity, 
then there exists an element of physical reality
corresponding to this physical quantity}.
Within these definitions quantum theory seems to be incomplete
because according to EPR one can show the existence of elements of physical reality,
whereas quantum mechanics does not use this concept.

Consider two observers, Alice and Bob, in two distant laboratories [Fig. \ref{EPR_SETUP}].
They perform measurements on spin-$\frac{1}{2}$ particles which used to interact in the past.
The quantum mechanical description of their joint state of spins reads:
\begin{equation}
|\psi^- \rangle = \frac{1}{\sqrt{2}}\Big[ |z+ \rangle_A | z- \rangle_B - |z- \rangle_A |z+ \rangle_B \Big],
\label{SINGLET}
\end{equation}
where $|z+ \rangle$ and $|z- \rangle$ denote the eigenstates of the $(\sigma_3 \equiv) \sigma_z$ operator.
The remarkable property of the state (\ref{SINGLET}) is its invariance
under the same rotations of observables in the two labs.
In particular, if Alice and Bob measure the same observable, 
whatever
outcome of Alice, the outcome of Bob is always opposite.
If Alice measures $\sigma_z$ 
then she can predict with certainty the outcome 
of Bob's $\sigma_z$ measurement.
Thus, according to EPR there exists an element of physical reality
connected with the $\sigma_z$ measurement.
Just as well Alice could measure $(\sigma_1 \equiv) \sigma_x$ and predict
with certainty, without in any way disturbing the system,
the outcome of a possible $\sigma_x$ measurement by Bob.
Again, seemingly there exists an element of reality connected with the $\sigma_x$ measurement.
Locality is assumed here:
the physical reality at Bob's site 
is independent of everything that happens at Alice's site.
Since quantum mechanics does not allow simultaneous
knowledge of both $\sigma_x$ and $\sigma_z$, it misses
some concepts which are necessary for the theory to be complete.

\subsection{Bell and Clauser-Horne-Shimony-Holt}

Twenty nine years after the EPR paper,
Bell proved that the completion of quantum mechanics 
expected by EPR is impossible \cite{BELL}.
In his original proof Bell utilized the perfect anticorrelations, which arise
whenever Alice and Bob measure local spins (with respect to the same direction)
on the two-qubit system in the state (\ref{SINGLET}).
However, unavoidable experimental imperfections imply that 
correlations are never perfect.
To illustrate the essence of Bell's theorem we re-derive the CHSH inequality \cite{CHSH}.
The validity of this inequality does not require perfect correlations 
and thus it can be directly experimentally checked.

\begin{figure}
\begin{center}
\includegraphics{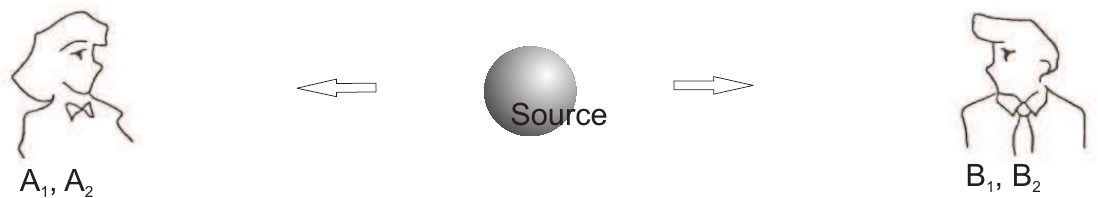}
\end{center}
\caption{EPR-Bell gedanken experiment.
Two distant observers (Alice and Bob) measure particles
which used to interact in the past.
Alice and Bob choose between two alternative settings
of the local measurement apparatuses.}
\label{EPR_SETUP}
\end{figure}
Consider the experiment proposed by EPR [Fig. \ref{EPR_SETUP}] and studied by Bell.
The pair emission begins an experimental run.
In each run Alice and Bob can choose between two alternative settings
of the local measurement apparatuses.
Their choices what to measure are absolutely free,
uncorrelated with (statistically independent of) the operation of the source.
According to realism the outcomes of all possible measurements
exist prior to and independent of the acts of measurement.\footnote{This can be relaxed to the assumption of
the existence of a joint probability distribution of results of incompatible measurements.}
Locality assumes that the outcomes of Alice depend on her setting only,
and the same for Bob.
For a given run, denote the predetermined local realistic results as $A_1$, $A_2$ for Alice,
and $B_1$, $B_2$ for Bob.\footnote{Note that the assumptions are already present in this notation.}
For example, if Alice chooses to measure setting ``1''
she obtains outcome $A_1$, if she chooses to measure ``2''
she obtains $A_2$.
Under the assumption of realism the outcomes of
all possible measurements are defined,
even if only some of them are actually measured.
Experiments on qubits can give one of two results, to which
we ascribe numbers, $+1$ and $-1$, i.e. $A_k,B_l = \pm 1$,
with indices $k,l=1,2$ denoting the settings.
The following identity holds in every experimental run:
\begin{equation}
A_1 (B_1 + B_2) + A_2 (B_1 - B_2) = \pm 2.
\label{CHSH_IDENTITY}
\end{equation}
All variables in this expression are dichotomic
(of values $\pm 1$), thus either $B_1 + B_2 = \pm 2$ and $B_1 - B_2 = 0$,
or the other way around.

After averaging over many experimental runs
expression (\ref{CHSH_IDENTITY})
reads:
\begin{equation}
-2 \le \langle A_1 B_1 + A_1 B_2 + 
A_2 B_1 -  A_2 B_2 \rangle \le + 2.
\end{equation}
The bounds follow from the fact that with averaging one cannot exceed
the extremal values of the averaged expression.
Since the average of a sum is a sum of averages,
the last inequality transforms to:
\begin{equation}
\Big| \langle A_1 B_1 \rangle + \langle A_1 B_2 \rangle + 
\langle A_2 B_1 \rangle - \langle A_2 B_2 \rangle \Big| \le 2.
\end{equation}
Note that within realistic theories
a single experimental run contributes
to \emph{all} averages in this expression.
After $R$ runs, 
the average of the product of predetermined results,
the local realistic correlation function,
reads:
\begin{equation}
E_{kl}^{LR} \equiv \langle A_k B_l \rangle = \frac{1}{R} \sum_{n = 1}^{R} A_k^{(r)} B_l^{(r)},
\end{equation}
where $A_k^{(r)}$, $B_l^{(r)}$ denote the predetermined results
in the $r$th run.
Finally, one arrives at the famous
Clauser-Horne-Shimony-Holt inequality \cite{CHSH}:
\begin{equation}
S_{CHSH} \equiv \Big| E_{11}^{LR} + E_{12}^{LR} + 
E_{21}^{LR} - E_{22}^{LR} \Big| \le 2,
\label{CHSH}
\end{equation}
which is satisfied by the correlations
of all local realistic models.

To complete the proof of Bell's theorem,
let us give an example of quantum predictions
which violate the CHSH inequality.
One replaces the local realistic correlation functions
 in (\ref{CHSH}) with their quantum counterparts, $E_{kl}^{QM}$,
for the singlet state (\ref{SINGLET}).
The quantum correlation function reads (Appendix A):
\begin{equation}
E_{kl}^{QM} = - \vec a_k \cdot \vec b_l,
\end{equation}
where dot stands for a scalar product
between vectors $\vec a_k$ and $\vec b_l$,
which
parameterize the measurement settings of Alice and Bob, respectively.
Thus, quantum mechanics predicts for the left-hand side of (\ref{CHSH}):
\begin{equation}
S_{CHSH}^{QM} = \Big|- \vec a_1 \cdot \vec b_1 - \vec a_1 \cdot \vec b_2 - \vec a_2 \cdot \vec b_1 + \vec a_2 \cdot \vec b_2 \Big|,
\end{equation}
which can be directly transformed to:
\begin{equation}
S_{CHSH}^{QM} = \Big|\vec a_1 \cdot (\vec b_1 + \vec b_2) + \vec a_2 \cdot (\vec b_1 - \vec b_2) \Big|.
\label{DERIV_CHSH_VIOL}
\end{equation}
We are looking for a maximum of this expression.
Since the $\vec b_k$ vectors are normalized, the vectors in the brackets are orthogonal:
\begin{equation}
(\vec b_1 + \vec b_2) \cdot (\vec b_1 - \vec b_2) = |\vec b_1|^2 - |\vec b_2|^2 = 0.
\end{equation}
Further, note that:
\begin{equation}
|\vec b_1 + \vec b_2|^2 + |\vec b_1 - \vec b_2|^2 = 2 (|\vec b_1|^2 + |\vec b_2|^2) = 4.
\end{equation}
Thus, one can parameterize the length of these vectors with a single angle, $\alpha$.
Finally, one can introduce normalized orthogonal vectors $\vec b_+$ and $\vec b_-$ such that:
\begin{eqnarray}
\vec b_1 + \vec b_2 &=& 2 \cos \alpha \textrm{ } \vec b_+, \\
\vec b_1 - \vec b_2 &=& 2 \sin \alpha  \textrm{ }  \vec b_-.
\end{eqnarray}
Using this decomposition, expression (\ref{DERIV_CHSH_VIOL})
transforms to:
\begin{equation}
S_{CHSH}^{QM} = \Big|2 \cos \alpha  \textrm{ }  \vec a_1 \cdot \vec b_+ + 2 \sin \alpha  \textrm{ }  \vec a_2 \cdot \vec b_- \Big|.
\end{equation}
The scalar products are maximal (and equal to one) if one chooses $\vec a_1 = \vec b_+$
and $\vec a_2 = \vec b_-$. After this choice 
one needs to find a maximum of $2|\cos\alpha + \sin\alpha|$.
The maximum is attained for $\alpha = \pi/4$, and gives a corresponding
maximal quantum value for the CHSH expression
\begin{equation}
S_{CHSH}^{QM}(max) = 2 \sqrt{2},
\end{equation}
clearly  above  the local realistic bound of $2$.
This value
was confirmed in numerous experiments, 
e.g. \cite{FC,ASPECT,STRICT_LOCALITY,DETECTION_LOOPHOLE}.

To reach the maximal violation one constraints
the measurement vectors for Alice and for Bob to lie in the same plane.
In this case the quantum correlation function
can be written as
\begin{equation}
E_{kl}^{QM} = - \cos(\varphi_k^A-\varphi_l^B),
\end{equation}
where $\varphi_k^A$ and $\varphi_l^B$
parameterize the position of the measurement vectors
within the plane,
relative to some fixed axis.
The maximum is achieved, for example, if Alice sets her angles to
\begin{equation}
\varphi_1^A = 0, \quad {\rm and} \quad
\varphi_2^A = \pi/2,
\end{equation} 
and Bob sets his angles to
\begin{equation}
\varphi_1^B = \pi/4, \quad {\rm and} \quad
\varphi_2^B = -\pi/4.
\end{equation}

\subsection{Assumptions}

Let us gather together the assumptions behind
the derivation of Bell inequalities,
and their experimental tests.

To derive the CHSH inequality (\ref{CHSH})
one assumes:
\begin{itemize}
\item \emph{realism} \\
Unperformed measurements have well-defined, yet unknown, results.

A picture behind realism is that
there exist objective properties 
of particles, which predetermine measurement outcomes. 
These properties,
as well as the properties of the measurement apparatus,
 are described by hidden variables.

\item \emph{locality} \\
Measurement outcomes at one location
depend on the measurement setting in this location only.

\end{itemize}

In the experimental tests of Bell inequalities
an additional assumption is unavoidable:

\begin{itemize}

\item \emph{freedom}\\
Statistical independence between the choice of measurement settings
and the workings of the source.

It is assumed that the correlations \emph{measured},
given the settings $k$ and $l$,
are the same 
up to insignificant statistical fluctuations
as the \emph{hypothetical}
local realistic correlations $E_{kl}^{LR}$ 
(cf. section on experimenter's freedom).
Otherwise one could not derive the inequality:
\begin{equation}
|E_{11} + E_{12} + E_{21} - E_{22}| \le 2,
\end{equation}
in which the experimental correlations appear.\footnote{To see 
how lack of freedom can lead to a violation of
CHSH inequality
consider the following simple model.
Let Bob decide what setting he chooses \emph{after} he knows
his potential outcomes $B_1,B_2$.
Take a local realistic model in which
with probability $\frac{1}{2}$ the predetermined results are
$(A_1,A_2,B_1,B_2) = (1,1,1,1)$ and 
otherwise they read $(A_1,A_2,B_1,B_2) = (1,-1,-1,1)$.
If $B_1=B_2$ Bob chooses setting $l=1$,
in the other case he chooses setting $l=2$.
In this way both terms $(A_1+A_2)B_1$ and $(A_1-A_2)B_2$
are equal to $2$, and one
reaches the algebraic limit of four
for the CHSH expression $S_{CHSH}$.}

Note that this assumption is fundamental,
and cannot be removed in any experimental setup.

\end{itemize}

\subsection{Loopholes}

Additionally, there exist certain experimental imperfections
which still allow to describe measured correlations
in a local realistic way.
Although there is no experiment
up to date which closes all these loopholes
simultaneously,
every loophole was closed in separate experiments.
Therefore, it is unlikely
that a ``final'' test would fail.
However, as usual in physics,
final verdict belongs to experiment.

\begin{itemize}

\item \emph{locality loophole} \\
A natural locality requirement
comes from the relativity theory.
If the detection event of, say, Bob
lies within a light-cone initiated
by the choice 
of the measurement setting
of Alice,
the outcome of Bob can be a function
of the setting of Alice.
Arbitrary violation of inequality (\ref{CHSH})
can be explained in this case
since the identity (\ref{CHSH_IDENTITY}) no longer
holds. Instead one has:
\begin{equation}
A_{11} B_{11}+A_{12} B_{21} + A_{21} B_{12} - A_{22} B_{22},
\end{equation}
where $A_{kl}$ and $B_{lk}$ are functions
of both, the settings of Alice and Bob.
This identity can achieve even the value of four.
One simply sets $B_{11} = B_{21} = B_{12} = - B_{22} = 1$
and $A_{11} = A_{12} = A_{21} = A_{22} = 1$.

There is a subtle issue connected to the locality loophole.
In the famous experiment of Aspect \cite{ASPECT}
the settings were chosen during the flight of the photons
such that detection event and the choice of the settings
were separated by a spacelike interval.
However, in this experiment the measurement choices were \emph{predictable}.\footnote{An
acousto-optical method was used to direct photons to differently oriented polarization
analyzers. The acoustic wave is not a random process.}
In principle, one can build a local hidden variable model taking advantage of this predictability
and again the outcomes of Bob could effectively depend on the setting
of Alice.
This possibility was disproved
in the Innsbruck experiment, 
in which it is impossible to predict the settings
in advance, due to their inherent randomness \cite{STRICT_LOCALITY}.

\item \emph{detection loophole} \\
Most of the experiments testing Bell inequalities
are performed with photons.
Unfortunately, we still lack efficient photo-detectors,
and only a tiny fraction of all
particles emitted is finally detected.
For detector efficiencies below a certain threshold,
the violation observed for a detected fraction
of particles
does not imply that 
the violation would still be observed
if all particles were detected.
There exist local realistic models
in which one violates Bell inequalities
in the subensemble of all runs \cite{LO,BELL_BOOK,GARG,LARSSON}.
A possible way out of this loophole (not the only one) are experiments with
atoms, in which the detection efficiency is nearly perfect
\cite{DETECTION_LOOPHOLE}.

\end{itemize}

For several other more specialized
and setup dependent loopholes see e.g. \cite{BARRETT,KENT}.
They will not be discussed here.
Finally, there are several proposals for loophole-free
Bell experiments \cite{LOOPHOLE_FREE1,LOOPHOLE_FREE2,LOOPHOLE_FREE3}.

\subsection{Greenberger-Horne-Zeilinger}

The CHSH inequality (\ref{CHSH})
represents a bound on possible local realistic
correlation functions.
 Greenberger, Horne, and Zeilinger (GHZ) 
 show that the joint assumption of locality and realism 
is inconsistent with specific \emph{perfect} correlations 
for systems with at least three qubits \cite{GHZ,GHSZ}.

Consider three separated two-level systems,
and accordingly three observers: Alice, Bob, and Carol.
This time we first give a quantum mechanical correlations of a certain state,
and then show that there can be no local realistic model for these correlations.
Consider a so-called three-particle GHZ state in the form given by Mermin \cite{MERMIN}:
\begin{equation}
|GHZ \rangle = \frac{1}{\sqrt{2}} \Big[ |z+\rangle_A |z+\rangle_B |z+\rangle_C 
+ i |z- \rangle_A |z- \rangle_B |z- \rangle_C \Big],
\label{GHZ_STATE}
\end{equation}
This state is an eigenstate of the following operators:
\begin{eqnarray}
\sigma_2^A  \sigma_1^B  \sigma_1^C  |GHZ \rangle &=& |GHZ \rangle, \nonumber \\
\sigma_1^A  \sigma_2^B  \sigma_1^C  |GHZ \rangle &=& |GHZ \rangle, \nonumber \\
\sigma_1^A  \sigma_1^B  \sigma_2^C  |GHZ \rangle &=& |GHZ \rangle, \nonumber \\
\sigma_2^A  \sigma_2^B  \sigma_2^C  |GHZ \rangle &=& - |GHZ \rangle.
\label{GHZ_PARADOX}
\end{eqnarray}
In the ideal case, without any experimental imperfections,
quantum mechanics predicts that any of the above joint measurements 
always gives perfect correlations 
(in the three cases correlations are equal to $+1$, in the last one they are given by $-1$).

Can there exist a local realistic explanation for these correlations?
According to local realism the outcomes of all possible measurements are predetermined.
In particular, the system carries definite answers to both: 
measurement of $\sigma_1$ and $\sigma_2$.
Let us denote these predetermined results as
$A_1, A_2$, for Alice, $B_1, B_2$ for Bob, and $C_1, C_2$ for Carol.
The first three equations of (\ref{GHZ_PARADOX}) 
define the following relations between the predetermined results:
\begin{eqnarray*}
A_2  B_1  C_1 &=& 1, \\
A_1  B_2  C_1 &=& 1, \\
A_1  B_1  C_2 &=& 1.
\label{GHZ_LR123}
\end{eqnarray*}
Since the square of $\pm1$ is always equal to $+1$,
multiplication of these gives the local realistic prediction for the last product:
\begin{equation}
A_2  B_2 C_2 = 1.
\label{GHZ_LR4}
\end{equation}
This strongly contradicts $-1$, the product of the outcomes predicted by quantum mechanics.
This apparent paradox was given the name
of "Bell's theorem without inequalities".

To verify experimentally these predictions
one needs to take care of unavoidable imperfections.
Inequalities appear to be a handy way of
dealing with experimental data.
The Bell inequality equivalent to the GHZ paradox
was first derived by Mermin \cite{MERMIN}.
Simply note that:
\begin{equation}
A_2 B_1 C_1 + A_1 B_2 C_1 + A_1 B_1 C_2 - A_2 B_2 C_2 = \pm 2,
\end{equation}
holds for all possible combinations of local realistic results $A_k,B_l,C_m = \pm 1$.
An average over many experimental runs results in the inequality:
\begin{equation}
\Big| E_{211}^{LR} + E_{121}^{LR} + E_{112}^{LR} - E_{222}^{LR} \Big| \le 2.
\label{MERMIN}
\end{equation}
According to (\ref{GHZ_PARADOX}) the maximum quantum value
of the left-hand side, after replacing local realistic
correlations with their quantum counterparts,
reaches four.
A violation of this inequality
was experimentally observed using
three-photon polarization entanglement \cite{GHZ_EXP_NATURE}.

\subsection{Polytope of local realistic theories}

The experimental violation of the CHSH inequality (\ref{CHSH}) 
or the Mermin inequality 
(\ref{MERMIN}) implies that no local realistic
explanation for the observed correlations is possible.
But what if the inequality is satisfied?
Can one then build a local realistic model for the observations?
The answer is negative.
A necessary and sufficient condition for a local realistic model
involves a set of inequalities, not a single one.

Consider the following 
geometrical picture
of a Bell scenario with two observers
choosing between two alternative measurement settings each.
The predetermined results are denoted by $A_k$ and $B_l$.
One can form a ``vector'' out of the predetermined results of each observer:
$\vec A = (A_1,A_2)$ and $\vec B = (B_1,B_2)$ in this case.  
One can also define a ``vector'' 
(or a ``tensor'')
of the local realistic correlation functions,
$\hat E_{LR}$, with components $E^{LR}_{kl} = \langle A_k B_l \rangle$.
All such local realistic models, $\hat E_{LR}$, can be written as:
\begin{equation}
\hat E_{LR} = \sum_{\vec A, \vec B = (\pm1,\pm1)} P(\vec A,\vec B) \vec A \otimes \vec B,
\label{CORRELATION_POLYTOPE}
\end{equation}
where $P(\vec A,\vec B)$ is the local realistic probability with which a certain quadruple of predetermined results
$\{A_1,A_2,B_1,B_2\}$ appears.
That is, every local realistic model of the correlation functions
is a convex combination of the extreme points $\vec A \otimes \vec B$,
and thus lies within a convex polytope, spanned by the vertices $\vec A \otimes \vec B$.
The necessary and sufficient condition for a local realistic description
is a set of inequalities which define the interior of the polytope
and are saturated at the border hyperplanes of it.
Such inequalities are called \emph{tight} Bell inequalities.

One can consider 
deterministic and stochastic local hidden variable theories.
In the stochastic theory,
in contrast to the deterministic theory,
one lacks the knowledge of some hidden variables.
As a result, measurement outcomes
are not exactly predetermined.
Instead, each particle \emph{separately carries probabilities} of certain outcomes.
All such theories give
predictions which lie inside the polytope.
To disprove stochastic local hidden variable models
it is sufficient to disprove deterministic models.

\subsection{All Bell inequalities for two qubits}

Let us present a construction of the necessary and sufficient condition for 
the possibility of a local realistic description of correlation functions 
obtained in standard Bell experiments with two qubits.
This approach was first given by \.Zukowski and Brukner \cite{ZB}.
The word ``standard'' refers to experiments 
in which observers choose between two settings.
First, one derives a necessary condition for a local realistic model,
then proves that the condition is also sufficient.
For future use we introduce a more elaborated notation.
The two local dichotomic observables are 
parameterized by vectors $\vec{n}_1^j$ and $\vec{n}_2^j$
(Appendix A), for party $j$.
In the case of two observers 
$j=1,2$ ($1$ for Alice, $2$ for Bob).
The predetermined results for the $j$th party are denoted by
$A_j(\vec{n}_1^j) = \pm 1 $ and $A_j(\vec{n}_2^j) = \pm 1$.
Since $A_j(\vec{n}_{k}^j)$ are dichotomic, for each observer
$j$ one has either $|A_j(\vec{n}_1^j)+A_j(\vec{n}_2^j)|=0$ and
$|A_j(\vec{n}_1^j)-A_j(\vec{n}_2^j)|=2$, or vice versa.
Therefore, for all sign choices of $s_1,s_2=\pm 1$ the product
$[A_1(\vec{n}_{1}^1) + s_1 A_1(\vec{n}_{2}^1)][A_2(\vec{n}_{1}^2) + s_2 A_2(\vec{n}_{2}^2)]$ vanishes
except for \emph{one} sign choice, for which it is equal to $\pm 4$.
If one sums up all such four products, with an arbitrary sign in
front of each of them, the sum is always  equal to the value of
the only non-vanishing term, i.e., it is $\pm 4$.
Thus the following algebraic identity holds for the predetermined results:
\begin{equation}
A_{12,12;S} \equiv \sum_{s_1,s_2 =\pm1} 
S(s_1,s_2)[ A_1(\vec{n}_{1}^1) + s_1 A_1(\vec{n}_{2}^1)]
[A_2(\vec{n}_{1}^2) + s_2 A_2(\vec{n}_{2}^2)] =\pm 4, \label{INEQ2}
\end{equation}
where $S(s_1,s_2)$ stands for an arbitrary ``sign'' function of the
summation indices $s_1,s_2$ [$ S(s_1,s_2) = \pm1$]. 
The notation $A_{12,12;S}$ describes the situation in which two parties
choose between two settings ``1'' or ``2''.

After averaging expression (\ref{INEQ2}) over the ensemble of
the runs one
obtains the following set of Bell inequalities:
\begin{equation}
\Big|
 \sum_{s_1,s_2 = \pm 1} S(s_1,s_2) 
 \sum_{k_1,k_2 = 1,2}
 s^{k_1-1}_1 s^{k_2-1}_2 E_{k_1 k_2}^{LR} \Big| \leq
4. \label{allbellineq}
\end{equation}
Since there are 16 different functions $S(s_1,s_2)$,
inequality (\ref{allbellineq}) represents a set of $16$
Bell inequalities for the correlation functions. 
A specific choice of the sign function,
$S(s_1,s_2) = \frac{1}{2}(1+s_1+s_2-s_1s_2)$,
leads to the well-known CHSH inequality (\ref{CHSH}).
Note that this function is non-factorable,
i.e. it cannot be written as $S(s_1,s_2) = S_1(s_1)S_2(s_2)$.
Putting factorable sign functions
into (\ref{allbellineq})
results in trivial inequalities
--- inequalities which cannot be violated.
To illustrate this consider e.g. $S(s_1,s_2) = s_1$.
Performing the sums of (\ref{allbellineq})
results in $|E_{21}| \le 1$.
Other factorable sign functions lead to 
trivial inequalities $|E_{kl}| \le 1$.

There is only one type of nonfactorable
sign functions of two bit-valued arguments:
\begin{equation}
S(s_1,s_2) = \pm \frac{1 \pm s_1}{2} \pm s_2 \frac{1 \mp s_1}{2},
\end{equation}
where the signs in front of the two fractions are free,
and those in the numerators have to be different.
Thus,
all Bell inequalities in this case are of the CHSH form --
different inequalities have a minus sign in front of different correlation functions.
In general, the set of all $16$ inequalities represented by (\ref{allbellineq}) is
equivalent to a {\it single} Bell inequality:
\begin{equation}
 \sum_{s_1,s_2  = \pm 1}  \Big|
 \sum_{k_1,k_2 = 1,2} s^{k_1-1}_1 s^{k_2-1}_2  E_{k_1k_2}^{LR} \Big| \leq 4.
\label{thebellineq}
\end{equation}
The equivalence of (\ref{thebellineq}) and (\ref{allbellineq}) is
evident once one  recalls that for real numbers, $|a+
b|\leq c$ and $|a-b|\leq c$  if and only if $|a|+|b|\leq c$, and
writes down a generalization of this property to sequences of an
arbitrary length.

Inequality (\ref{thebellineq}) is satisfied by all local realistic models.
It forms a necessary condition for the possibility of a local
realistic description.
To prove the sufficiency of this condition
one can construct a local realistic model
for any set of \emph{experimental} correlation functions, $E_{k_1k_2}$, which satisfy it.
In other words one is interested in the local realistic models
$E^{LR}_{k_1k_2}$ such that they fully agree with the measured correlations $E_{k_1k_2}$
for all possible observables $k_1,k_2=1,2$.
Recall that the set of local realistic correlation functions
can be put as (\ref{CORRELATION_POLYTOPE}).
Put
\begin{equation}
\vec A = A_1(\vec{n}_{1}^1)
\left( \!\!
\begin{array}{c}
1   \\
s_1
\end{array}
\!\!
\right),
\quad {\rm and} \quad
\vec B = A_2(\vec{n}_{1}^2)
\left( \!\!
\begin{array}{c}
1   \\
s_2
\end{array}
\!\! \right).
\end{equation}
Let us ascribe for fixed $s_1,s_2$, a hidden probability that $A_j(\vec{n}_{1}^j) = s_j A_j(\vec{n}_{2}^j)$ 
 in the form familiar from Eq. (\ref{thebellineq}):
\begin{equation}
P(s_1,s_2)=\frac{1}{4} \Big|\sum_{k_1,k_2=1,2} s_1^{k_1-1} s_2^{k_2-1} E_{k_1k_2} \Big|.
\label{PROB}
\end{equation}
Obviously these probabilities are positive.
However they sum up to identity only if inequality (\ref{thebellineq}) is saturated.
Otherwise there is a ``probability deficit'', $\Delta P$.
First, let us prove that the local realistic model,
$\hat E_{LR}$, is a valid model for the correlations measured
$\hat E = (E_{11},E_{12},E_{21},E_{22})$,
i.e. $\hat E_{LR} = \hat E$.
Next, it will be shown how one can compensate the probability
deficit without affecting the correlation functions.

In the four dimensional real space where both $\hat E_{LR}$ and $\hat E$
are defined one can find an orthonormal basis set $\hat S_{s_1s_2} = \frac{1}{2}(1,s_1)\otimes(1,s_2)$.
Using this basis the hidden probabilities acquire a simple form:
\begin{equation}
P(s_1,s_2) = \frac{1}{2} |\hat S_{s_1s_2} \cdot \hat E|,
\end{equation}
where the dot denotes the scalar product in $\mathcal{R}^4$.
The local realistic model, $\hat E_{LR}$, expressed as (\ref{CORRELATION_POLYTOPE}), reads:
\begin{equation}
\hat E_{LR} = \sum_{s_1,s_2=\pm1} 
|\hat S_{s_1s_2} \cdot \hat E| A_1(\vec{n}_{1}^1) A_2(\vec{n}_{1}^2) \hat S_{s_1s_2}.
\end{equation}
The modulus of any real number $|x|$ can be split into $|x| = x \textrm{ sign}(x)$.
Further, one can always demand the product $A_1(\vec{n}_{1}^1) A_2(\vec{n}_{1}^2)$
to have the same sign as the expression inside the modulus.\footnote{This choice is a part of the local realistic model.}
Thus one has:
\begin{equation}
\hat E_{LR} = \sum_{s_1,s_2=\pm1} 
(\hat S_{s_1s_2} \cdot \vec E) \hat S_{s_1s_2}.
\end{equation}
The expression in the bracket is the coefficient of the tensor $\hat E$
in the basis $\hat S_{s_1s_2}$. These coefficients are then summed
over the same (complete) basis vectors.
Therefore, the equivalence is proven:
\begin{equation}
\hat E_{LR} = \hat E.
\end{equation}

If inequality (\ref{thebellineq}) is not saturated,
that is
$\Delta P > 0$, one adds a ``tail'' to the local realistic model (\ref{CORRELATION_POLYTOPE})
\begin{equation}
\frac{\Delta P}{16} \sum_{\vec A, \vec B= (\pm 1,\pm 1)} \vec A \otimes \vec B
\end{equation}
which represents fully random noise.
Since each vertex $\vec A \otimes \vec B= (\pm 1,\pm 1,\pm 1,\pm 1)$
comes in the ``tail'' with the same probability
the ``tail'' does not contribute to the correlation functions.
However, each probability $P(s_1,s_2)$
is increased by $\frac{\Delta P}{4}$
such that now they sum up to identity,
as it should be.

In this way the set of inequalities (\ref{allbellineq}),
or its equivalent --- the single inequality (\ref{thebellineq}) ---
is proven to be sufficient and necessary for the possibility of local realistic
description of correlation experiments on two qubits,
in which both Alice and Bob measure one of two local settings.
This kind of reasoning can also be applied to an arbitrary number of qubits.

\subsection{All Bell inequalities for many qubits}

A generalization of the approach presented for two qubits to many qubits is straightforward 
and was presented in the same paper by \.Zukowski and Brukner \cite{ZB}.
For $N$ particles the generalization of identity (\ref{INEQ2})
consists of the sum of $N$ products of local identities $A_j(\vec{n}_{1}^j) + s_j A_j(\vec{n}_{2}^j) = \pm 2$.
The summation is now taken with a 
more general sign function, $S(s_1,...,s_N)$, of $N$ parameters:
\begin{equation}
A_{12,...,12;S} \equiv \sum_{s_1,...,s_N =\pm1} 
S(s_1,...,s_N) \prod_{j=1}^N [ A_j(\vec{n}_{1}^j) + s_j A_j(\vec{n}_{2}^j)] =\pm 2^N, \label{INEQN}
\end{equation}
Since there are $2^{2^N}$ different sign functions of $N$ two-valued arguments, 
the above formula leads to a set of $2^{2^N}$ Bell inequalities. 
Using the trick described above, one can write a \emph{single} inequality
equivalent to the whole set \cite{WW,WZ,ZB}:
\begin{equation}
 \sum_{s_1,...,s_N  = \pm 1} \Big|
 \sum_{k_1,...,k_N = 1,2} s^{k_1-1}_1 ... s^{k_N-1}_N  E_{k_1...k_N}^{LR} \Big| \leq 2^N.
\label{THEBELLINEQ_N}
\end{equation}
Many of these inequalities are trivial. 
For example, if $S(s_1,...,s_N)=1$ for all arguments, we get the condition
$|E_{1...1}|\leq1$. 
Specific nonfactorable choices of $S(s_1,...,s_N)$ give non-trivial inequalities. 
For example, for $S(s_1,...,s_N) = \sqrt{2}\cos[(s_1+...+s_N-N+1)\frac{\pi}{4}]$, 
one recovers the tight inequalities of \cite{MERMIN,ARDEHALI,BELINSKII}.

Up to now we have shown that if 
a local realistic model exists,
the general Bell inequality (\ref{THEBELLINEQ_N}) follows. The
converse is also true: whenever inequality (\ref{THEBELLINEQ_N})
holds, one can construct a local realistic model for the
correlation functions, in the case of a standard Bell experiment.
For $N$ particles the hidden probability that the predetermined outcomes 
of the $j$th observer are $A_j(\vec{n}_{1}^j) = s_j A_j(\vec{n}_{2}^j)$
is given by the form familiar from Eq.\ (\ref{THEBELLINEQ_N}):
\begin{equation}
P(s_1,...,s_N) = \frac{1}{2^N} \Big|\sum_{k_1,...,k_N=1,2} s_1^{k_1-1}...s_N^{k_N-1}E_{k_1...k_N}\Big|. 
\end{equation}
The same steps as for two qubits above (now in the $\mathcal{R}^{2^N}$ space)
lead to the result that any correlation experiment satisfying (\ref{THEBELLINEQ_N})
can be explained within a local realistic picture.
That is, one can claim that the set of Bell
inequalities represented by (\ref{THEBELLINEQ_N}) is complete.
This completeness implies that all series of Mermin $N$-qubit inequalities,
which give tight inequalities, are a subset of the inequalities generated
by (\ref{THEBELLINEQ_N}).
This also applies to the tight Ardehali inequalities
and the full set of Belinskii-Klyshko inequalities \cite{ARDEHALI,BELINSKII}.

\subsection{Violation condition of Horodeckis}

In this section one finds a derivation of a necessary and sufficient
condition for the violation of a general \emph{bipartite} Bell inequality 
(\ref{thebellineq})
with an arbitrary (mixed) quantum state.
This is a reformulation of a condition
first given by the Horodecki family \cite{CHSH_NS}.
This reformulation
allowed \.Zukowski and Brukner to generalize the violation condition
to the multiparticle case, which will be described later \cite{ZB}.

A reader not familiar with the
correlation tensor formalism
is strongly encouraged to read Appendix A first.
The full set of inequalities for the $2 \times 2$ 
problem (two observers choose between two settings each) 
is
derivable from the CHSH inequality (see discussion below (\ref{allbellineq})):
\begin{equation}
\Big| \Big\langle 
(A_1+ A_2)B_1 + (A_1 - A_2)B_2 \Big\rangle \Big| \le 2. \label{full22}
\end{equation}
The quantum correlation function  $E^{QM}(\vec a_k,\vec b_l)$ is given by
the scalar product of the correlation tensor $\hat T$ with the
tensor product of the local measurement settings represented by
unit vectors $\vec a_k \otimes \vec b_l$, 
i.e. $E^{QM}(\vec a_k,\vec b_l) = \hat T \circ \vec a_k \otimes \vec b_l$. 
Thus, the condition for a
quantum state endowed with the correlation tensor $\hat T$ to
satisfy the inequality (\ref{full22}), is that for all directions
$ \vec a_1, \vec a_2, \vec b_1, \vec b_2$ one has
\begin{equation}
\Big| \Big[ \Big( \frac{\vec a_1 + \vec a_2}{2} \Big) \otimes \vec b_1
+ \Big( \frac{\vec a_1 - \vec a_2}{2} \Big) \otimes \vec b_2 \Big] \circ \hat T \Big| \le 1,
\label{QUANTUM_INEQ}
\end{equation}
where both sides of (\ref{full22}) were divided by $2$.

Note that the pairs of local vectors define the
``local measurement planes''.
Here we shall find the conditions for (\ref{QUANTUM_INEQ}) to hold for two,
arbitrary but fixed, measurement planes, one for each observer.
Therefore, only those components of $\hat T$ are relevant which describe
measurements in these two planes.
Thus $\hat T$ is effectively described by a $2 \times 2$ matrix, or tensor $\hat T'$.

Let us denote the vectors in the round brackets of (\ref{QUANTUM_INEQ}) as:
\begin{equation}
\vec A_{\pm} = \frac{1}{2}(\vec a_1 \pm \vec a_2)
\end{equation}
These vectors 
satisfy the following relations: $\vec A_+ \cdot \vec A_- = 0$ 
(orthogonality) and 
$||\vec A_+ ||^2 + ||\vec A_- ||^2 = 1$ (normalization).
Thus $\vec A_+ + \vec A_-$ is a unit vector,
and $\vec A_\pm$ represent its decomposition into two orthogonal vectors.
If one introduces the unit vectors $\vec a_\pm$ such that
$\vec A_\pm = a_\pm \vec a_\pm$, one has $a_+^2 + a_-^2 = 1$.
Thus one can put (\ref{QUANTUM_INEQ}) into the following form:
\begin{equation}
|\hat S \circ \hat T'| \le 1,
\label{SCAL_PROD_S_AND_T}
\end{equation}
where $\hat S = a_+ \vec a_+ \otimes \vec b_1 + a_- \vec a_- \otimes \vec b_2$.
Since $\vec a_+ \cdot \vec a_- = 0$,
one has $\hat S \circ \hat S = 1$, i.e. $\hat S$ is a tensor of unit norm.
Any $2 \times 2$ tensor of unit norm, $\hat U$, has the following Schmidt decomposition:\footnote{A simple
and intuitive proof of Schmidt decomposition can be found in the book of Peres \cite{PERES_BOOK}}
\begin{equation}
\hat U = \lambda_1 \vec v_1 \otimes \vec w_1 + \lambda_2 \vec v_2 \otimes \vec w_2,
\quad {\rm where} \quad 
\vec v_i \cdot \vec v_j = \delta_{ij}, \quad \vec w_i \cdot \vec w_j = \delta_{ij} \quad
\textrm{and} \quad  \lambda_1^2 + \lambda_2^2 = 1.
\end{equation}
The freedom of  the choice of the measurement directions
$\vec b_1$ and $\vec b_2$, allows one, 
by choosing $\vec b_2$ orthogonal to $\vec b_1$,
to find $\hat S$ of a form isomorphic with $\hat U$.
The freedom of choice of $\vec a_1$ and $\vec a_2$ allows
$\vec a_+$ and $\vec a_-$ to be arbitrary orthogonal unit vectors,
and $a_+$ and $a_-$ to be also arbitrary.
Thus $\hat S$ can be equal to any unit tensor.
Therefore, to get the maximum of the left hand side of (\ref{SCAL_PROD_S_AND_T})
we put $\hat S$ parallel to $\hat T'$, $\hat S= \frac{1}{||\hat T'||} \hat T'$.
The maximum reads $||\hat T'|| = \sqrt{\hat T' \cdot \hat T'}$.
Thus,
\begin{equation}
\max\big[\sum_{k,l=1,2} T_{kl}^2 \big] \le 1,
\label{t2-22}
\end{equation}
where 
the maximization
is taken over all local coordinate systems of two observers,
is the necessary and sufficient condition for the inequality (\ref{THEBELLINEQ_N})
to hold for quantum predictions.
Since the inequality (\ref{THEBELLINEQ_N}) itself
is a necessary and sufficient condition
for the possibility of a local realistic model,
the inequality (\ref{t2-22}) also forms such a condition.

\subsection{Gisin's theorem}

The theorem of Gisin states that {\it any} pure non-product state violates local realism.
There are sets of measurements that can be performed on the state
which cannot be described within a local realistic picture \cite{GISIN,GISINPERES}.
This theorem formalizes the intuition that entanglement is a purely quantum phenomenon.
Using the approach presented here, one can write down 
a simple proof of Gisin's theorem for \emph{two qubits}. 
Any state of two qubits is given
in its Schmidt basis $| z \pm \rangle$ by: 
\begin{equation}
|\psi \rangle = 
\cos{\alpha} |z+\rangle_A |z+\rangle_B + \sin{\alpha} |z-\rangle_A |z-\rangle_B, 
\quad {\rm with} \quad \alpha \in [0,\pi/4],
\end{equation}
The following correlations do not vanish for this state:
\begin{equation}
T_{xx} = \sin{2\alpha}, \quad T_{yy} =  - \sin{2\alpha}, \quad T_{zz} = 1.
\end{equation}
Therefore the necessary and sufficient condition for local realism is violated
for all entangled ($\alpha \neq 0$) states:
\begin{equation}
\sum_{k,l=\{x,z\}} T_{kl}^2 = 1 + \sin^2{2\alpha} >1,
\quad {\rm for} \quad \alpha \in (0,\pi/4].
\end{equation}

\subsection{Violation of standard Bell inequalities}

Surprisingly, the intuitive result
that all pure entangled bipartite states violate standard Bell inequalities
does not hold in the multiparticle case.
There exist pure entangled states
the correlations of which,
obtained in a standard Bell experiment 
can be explained in a local realistic way \cite{ZBLW}.
To see this we follow \.Zukowski and Brukner and derive
conditions for a violation
of the general inequality (\ref{THEBELLINEQ_N}).

If one replaces the local realistic correlations of (\ref{THEBELLINEQ_N}) 
by the quantum predictions, one gets:
\begin{equation}
\sum_{s_1,...,s_N=\pm 1} 
\Big| \sum_{k_1=1}^2 ... \sum_{k_N=1}^2 s_1^{k_1-1}...s_N^{k_N-1} 
\vec a_{k_1}^1 \otimes ... \otimes \vec a_{k_N}^N \circ \hat T \Big| \le 2^N,
\label{INEQ_N_TO_DEVIDE}
\end{equation}
where $\vec a_{k_j}^j$ is a vector describing setting $k_j$ of party $j$ (Appendix A).
Writing the sums over $k_j$ explicitly
and dividing both sides by $2^N$ brings this inequality to the form:
\begin{equation}
\sum_{s_1,...,s_N=\pm 1} 
\Big| \frac{\vec a_{1}^1 + s_1 \vec a_{2}^1}{2} \otimes 
... \otimes \frac{\vec a_{1}^N + s_N \vec a_{2}^N}{2} \circ \hat T \Big| \le 1,
\label{PRE_CHANGE_OF_COORD}
\end{equation}
Similarly to the two-qubit case,
one can introduce for each party new (orthogonal) local  
coordinate systems built out of vectors $\vec \alpha_{1}^j$ and $\vec \alpha_{2}^j$,
such that:
\begin{equation}
\begin{array}{c}
\frac{1}{2}(\vec a_{1}^j + \vec a_{2}^j) = \alpha_{1}^j \vec \alpha_{1}^j, \\
\frac{1}{2}(\vec a_{1}^j - \vec a_{2}^j) = \alpha_{2}^j \vec \alpha_{2}^j,
\end{array}
\quad {\rm with} \quad
(\alpha_1^j)^2 + (\alpha_2^j)^2 = 1.
\label{NEW_COORDINATE_SYSTEMS}
\end{equation}
Thus,
\begin{equation}
\sum_{x_1,...,x_N= 1}^2
\Big| \alpha_{x_1}^1... \alpha_{x_N}^N  \vec \alpha_{x_1}^1 \otimes 
... \otimes \vec \alpha_{x_N}^N \circ \hat T \Big| \le 1.
\end{equation}
Note that $\vec \alpha_{x_1}^1 \otimes 
... \otimes \vec \alpha_{x_N}^N \circ \hat T = T_{x_1...x_N}$
is a component of a tensor $\hat T$
in the new local bases.
Thus, the condition:
\begin{equation}
\max \Big[ \sum_{x_1,...,x_N=1}^2 \alpha_{x_1}^1...\alpha_{x_N}^N |T_{x_1...x_N}| \Big] \le 1,
\label{NS_N_QUBITS}
\end{equation}
where the maximization is taken over all possible
parameters $\alpha_{x_n}^n$ and bases of the correlation tensor,
is the necessary and sufficient condition
for a violation of inequality (\ref{THEBELLINEQ_N}).
Note that putting the numbers $\alpha_{x_j}^j$
in the condition (\ref{NS_N_QUBITS})
outside the moduli  does not change the maximum.

The left-hand side of condition (\ref{NS_N_QUBITS})
can be estimated using the Cauchy inequality.
The sum
can be thought of as a scalar product $\vec \alpha \cdot \vec \tau$.
The 
vector $\vec \alpha = (\alpha_{1}^1...\alpha_{1}^N,\alpha_{1}^1...\alpha_{2}^N,...,\alpha_{2}^1...\alpha_{2}^N)$
is built out of all possible products $\alpha_{x_1}^1...\alpha_{x_N}^N$, with $x_j = 1,2$.
The corresponding components of vector $\vec \tau = (|T_{1...1}|,|T_{1...2}|,...,|T_{2...2}|)$,
are given by the moduli of the correlation tensor elements.
The scalar product is bounded by:
\begin{equation}
\vec \alpha \cdot \vec \tau \le || \vec \alpha|| \textrm{ } ||\vec \tau||.
\end{equation}
Due to properties (\ref{NEW_COORDINATE_SYSTEMS})
vector $\vec \alpha$ is normalized.
The norm of $\vec \tau$ reads:
\begin{equation}
||\vec \tau|| = \sqrt{ \sum_{x_1,...,x_N=1}^2 T_{x_1...x_N}^2}.
\end{equation}
Thus, one obtains the following simple and useful
\emph{sufficient} condition for a local realistic description:
\begin{equation}
\max \Big[ \sum_{x_1,...,x_N=1}^2 T_{x_1...x_N}^2 \Big] \le 1,
\label{S_N_QUBITS}
\end{equation}
in which maximization is taken over all local coordinate systems.
If this condition is satisfied,
then also (\ref{NS_N_QUBITS}) is satisfied
and one can build local realistic model.

\subsubsection{Example}

Surprisingly, one can build local realistic model
for correlation experiments in which
pure entangled state was measured.
The state (so-called generalized GHZ state) is given by
\begin{equation}
| \psi_{GHZ} \rangle = 
\cos \alpha |z+ \rangle_1 ... |z+ \rangle_N
+ \sin\alpha |z- \rangle_1 ... |z- \rangle_N,
\quad {\rm with} \quad 0 \le \alpha \le \pi/4.
\label{GEN_GHZ}
\end{equation}
For parameter $\alpha$ satisfying
\begin{equation}
\sin 2\alpha \le 1/\sqrt{2^{N-1}}
\quad \textrm{ and} \quad N \textrm{ odd},
\end{equation}
the state $| \psi_{GHZ} \rangle$ satisfies condition (\ref{S_N_QUBITS}).
For the details of the proof consult \cite{ZBLW}.
Here we give an intuitive argument
for the different
behaviour of odd and even particle systems (no violation/violation).
The non-vanishing correlations between all the parties
measuring the
generalized GHZ state (\ref{GEN_GHZ})
read:
\begin{equation}
T_{z...z} = 
\Bigg\{
\begin{array}{ccc}
1 & {\rm for} & N \textrm{ even}, \\
\cos 2\alpha & {\rm for} & N \textrm{ odd},
\end{array}
\qquad
T_{x...x} = \sin 2\alpha,
\label{T_GEN_GHZ}
\end{equation}
and all components with $2 \xi$ indices
equal to $y$ and the rest equal to $x$
take the value $(-1)^{\xi} \sin 2\alpha$.\footnote{There are 
$\sum_{\xi =1}^{\lfloor N/2 \rfloor} {N \choose 2\xi} = 2^{N-1}-1$ such components,
$\lfloor N/2 \rfloor$ denotes the integer part of $N/2$.}
For example, for $N=3$, one has:
\begin{equation}
T_{yyx} = T_{yxy} = T_{xyy} = - \sin 2 \alpha.
\end{equation}
The expression $\sum_{k_1,...,k_N=x,z} T_{k_1...k_N}^2$,
which appears in the condition (\ref{S_N_QUBITS})
can be understood as a ``total measure of the strength
of correlations'' in mutually complementary sets of local
measurements (as defined by the summation over $1$ and $2$) \cite{ESSENCE}.
The unity on the right-hand side of
the condition is the classical limit for the amount of correlations.
Specifically, pure product states cannot exceed the limit of $1$, as
they can show perfect correlations in one set of local
measurement directions only.
In contrast, entangled states can
show perfect correlations for more than one such set.
Only if $N$ is even the state (\ref{GEN_GHZ})
 shows perfect correlations already between measurements along
$z$-directions. Therefore, reaching the classical limit.
Yet, they also show additional
correlations in other, complementary, directions. 
However,
in the case of
$N$ odd, there are no perfect correlations 
along $z$-direction and the correlations in the
complementary directions do not suffice to violate the bound of $1$.

\section{Multisetting Bell inequalities [P3,P5,P6]}

\subsection{Multisetting Bell inequalities [P5,P6]}

The non-classicality of the generalized GHZ states can be shown using
multiple settings per party.
We present an efficient method for generation
of tight multisetting Bell inequalities, which however do not form a complete set.
This method was invented by Wu and Zong \cite{WUZONG1,WUZONG2}, 
and generalized in [P5,P6].

We start with the case of $N=3$ observers. Suppose that the first
two observers choose between four settings, and the third one chooses
between two settings. Such a problem is denoted here as $4 \times 4 \times 2$. 
As described before, the local realistic values for the first two observers 
satisfy the following algebraic
identity:
\begin{equation}
A_{12,12,S'} \equiv \sum_{s_1,s_2=\pm 1}
S'(s_1,s_2) [A_1(\vec n_1^1)+ s_1 A_1(\vec n_2^1)] 
[A_2(\vec n_1^2)+ s_2 A_2(\vec n_2^2)]=\pm 4, \label{kraj}
\end{equation}
where $S'(s_1,s_2)$ is any sign function,
In an analogous way one defines $A_{34,34,S''}$, by replacing
$A_1(\vec n_1^1),A_1(\vec n_2^1),A_2(\vec n_1^2),A_2(\vec n_2^2)$ by 
$A_1(\vec n_3^1),A_1(\vec n_4^1),A_2(\vec n_3^2),A_2(\vec n_4^2)$, respectively, and $S'$ by
$S''$. Depending on the value of $s= \pm 1$ one has $(A_{12,12,S'} +
s A_{34,34,S''})=\pm 8,$ or $0$. By analogy to (\ref{kraj}) one has:
\begin{equation}
A_{1234,12} \equiv
\sum_{s_1,s_2=\pm 1} \hspace{-2mm}  
S(s_1,s_2) [A_{12,12,S'}+s_1 A_{34,34,S''}][A_3(\vec n_1^3)+ s_2 A_3(\vec n_2^3)]=\pm 16.
\label{GEN}
\end{equation}
After averaging over many runs of the experiment, and introducing
the correlation functions 
$E_{ijk}^{LR} \equiv \langle A_1(\vec n_i^1) A_2(\vec n_j^2) A_3(\vec n_k^3) \rangle$ 
one obtains multisetting Bell inequalities.
Because of the freedom to choose the sign functions $S, S', S''$,
there are $(2^4)^3 = 2^{12}$ Bell inequalities in this case.

All of these inequalities which are nontrivial
can be reduced to a single ``generating'' inequality,
in which all the sign functions $S, S', S''$
are non-factorable.
It will be shown that the choice of a factorable sign function
is equivalent to having a non-factorable one, and some of the
local measurement settings equal.
In general, a 
sign function $S(s_1,s_2)$,
which is a two-valued function of two bit-valued arguments,
has the following useful discrete Fourier transform:
\begin{equation}
S(s_1,s_2) = a(s_1) + b(s_1)s_2,
\quad {\rm with} \quad
a(s_1)b(s_1) = 0,
\textrm{ and }
|a(s_1)| + |b(s_1)| = 1.
\end{equation}
The factorable $S(s_1,s_2)$ is defined by the condition $|a(s_1)| \equiv 1$
or $|b(s_1)| \equiv 1$, which implies that $|b(s_1)| \equiv 0$ or
$|a(s_1)| \equiv 0$, respectively. 
For example, take the last factor
of (\ref{GEN}). Since
\begin{equation}
 \sum_{s_2 = \pm 1} S(s_1,s_2) [A_3(\vec n_1^3)+ s_2 A_3(\vec n_2^3)] 
= 2 a(s_1) A_3(\vec n_1^3) + 2 b(s_1) A_3(\vec n_2^3), \label{GEN_SIGN_F}
\end{equation}
one has for the factorable case, say when $b(s_1) \equiv 0$,
\begin{equation}
\sum_{s_2 = \pm 1} S(s_1,s_2) [A_3(\vec n_1^3)+ s_2 A_3(\vec n_2^3)] = 2 a(s_1) A_3(\vec n_1^3).
\label{FACTORABLES}
\end{equation}
The setting ``2" for the third observer drops out. Note that
a similar result can also be obtained for a non-factorable
$S$ in (\ref{GEN_SIGN_F}) by putting $\vec n_1^3 = \vec n_2^3$. 
For non-factorable $S(s_1,s_2)$ and for given $s_1$
either $a(s_1)$ or $b(s_1)$ does not vanish.
Further, if one inserts this result
into (\ref{GEN}), and, say, $a(s_1) \equiv const = 1$, then after
the summation over $s_1$ the whole term with the settings $3,4$ for the
two observers vanishes. What we get is a trivial extension of the
CHSH inequalities. 

The whole family can be reduced to one ``generating" inequality
which is obtained for non-factorable $S, S', S''$. In such cases
\begin{equation}
a(s_1) = \pm \frac{1 \pm s_1}{2},
\quad {\rm and} \quad
b(s_1) = \pm \frac{1 \mp s_1 }{2}, 
\end{equation}
where the front signs are free, and those in the numerators
are different for the two functions. 
Any other cases are
obtainable by the sign changes $X_i \to - X_i$ ($X = A,B,C$).
Thus, the ``generating" inequality of the whole family can be
chosen as
[here all the sign functions are equal to $S(s_1,s_2) = \frac{1}{2}(1+s_1+s_2-s_1s_2)$]:
\begin{eqnarray}
&& \Big| \Big \langle 
\Big[ A_3(\vec n_1^3) + A_3(\vec n_2^3)\Big] 
\Big[ A_1(\vec n_1^1)[(A_2(\vec n_1^2) + A_2(\vec n_2^2)] + A_1(\vec n_2^1)[A_2(\vec n_1^2) - A_2(\vec n_2^2)] \Big] 
 \nonumber \\
&& + \Big[ A_3(\vec n_1^3) - A_3(\vec n_2^3) \Big] 
\Big[ A_1(\vec n_3^1)[A_2(\vec n_3^2) + A_2(\vec n_4^2)] + A_1(\vec n_4^1) [A_2(\vec n_3^2) - A_2(\vec n_4^2)] \Big] 
\Big\rangle \Big| \nonumber \\
&& \le 4.
\label{442_GEN_INEQ}
\end{eqnarray}
Other inequalities can be obtained by making some settings equal.
For example, the inequalities involving \emph{three} settings for the first two
observers and two settings for the last one can be obtained by
choosing settings 1 and 2 identical for the two observers 
(and renaming $3 \to 2$ and $4 \to 3$):
\begin{eqnarray}
\Big|
2 E_{111} + 2 E_{112} + E_{221} - E_{222}
+E_{231} - E_{232} + E_{331} - E_{322} - E_{331} + E_{332}
\Big| \le 4,
\end{eqnarray}
where $E_{klm} = \langle A_1(\vec n_k^1) A_2(\vec n_l^2) A_3(\vec n_m^3) \rangle$
denotes a three-particle correlation function.

The method can be generalized to various choices of the number of
parties and the measurement settings. We shall present the
$2^{N-1} \times 2^{N-1} \times 2^{N-2}\times ... \times 2$ case. 
Consider $N=4$ observers. 
One starts with the identity (\ref{GEN}). 
Next, one introduces a similar
formula for the settings $\{5,6,7,8\}$, for the first two
observers, and $\{3,4\}$, for the third one. The fourth observer
chooses between two settings with local realistic values $A_4(\vec n_1^4)$ and $A_4(\vec n_2^4)$.
Applying the same method as before, one obtains an identity which
generates Bell inequalities of the $8 \times 8 \times 4 \times
2$ type:
\begin{eqnarray}
&&\sum_{s_1,s_2 = \pm 1} S(s_1,s_2) 
[A_{1234,12}+s_1 A_{5678,34}] [A_4 (\vec n_1^4) + s_2 A_4 (\vec n_2^4)] = \pm 64, \nonumber 
\end{eqnarray}
where $A_{1234,12}$ and  $A_{5678,34}$ depend on some three sign
functions. One may apply this method iteratively,
increasing the number of observers by one, to obtain inequalities
involving an exponential (in $N$) number of measurement settings.

As another example we construct the inequalities involving $N$ partners,
where the first $N-1$ observers choose one of $4$ settings and the last one
chooses between $2$ settings.
We use the local realistic quantity $A_{12,...,12}$ defined in Eq. (\ref{INEQN})
for $N-1$ parties choosing between $2$ settings each:
\begin{equation}
A_{12,...,12} \equiv 
\sum_{s_1,...,s_{N-1} =\pm1} 
S_{12,...,12}(s_1,...,s_{N-1}) 
\prod_{j=1}^{N-1} [ A_j(\vec{n}_{1}^j) + s_j A_j(\vec{n}_{2}^j)] =\pm 2^{N-1},
\end{equation}
and analogically introduce $A_{34,...,34}$ for another pair of observables
available to each party. 
By including the $N$th observer, who can choose between $2$ measurement settings,
one obtains:
\begin{equation}
\sum_{s_1,s_1 =\pm1} 
S_{4,...,42}(s_1,s_2)
(A_{12,...,12} + s_1 A_{34,...,34})(A_N(\vec{n}_{1}^N) + s_2 A_N(\vec{n}_{2}^N)) = \pm 2^{N+1}.
\end{equation}
One can use this expression for generating Bell inequalities
for $N$ observers in the same way as it was previously done.

In order to show the full strength  of the method the 
next example gives a family of  Bell inequalities 
for $N=5$ qubits, which involves eight settings for the first two
observers and four settings for the other three.
We take the identity $A_{1234,12}$ defined in (\ref{GEN}), 
valid for the $4 \times 4 \times 2$ case of three observers,
and define a similar quantity for another set of $4 \times 4 \times 2$ observables,
namely $A_{5678,34}$.
Note that the sign functions entering $A_{5678,34}$
can be different from those entering $A_{1234,12}$.
For the other two observers we introduce:
\begin{equation}
A_{12,12} \equiv \sum_{s_1,s_2 = \pm 1}
S_{12,12}(s_1,s_2)[A_4(\vec{n}_{1}^4) + s_1 A_4(\vec{n}_{2}^4)]
[A_5(\vec{n}_{1}^5) + s_2 A_5(\vec{n}_{2}^5)]=\pm 4
\end{equation}
and a similar expression, $A_{34,34}$, for another pair of observables $A_4(\vec{n}_{3}^4),A_4(\vec{n}_{4}^4)$ 
and $A_5(\vec{n}_{3}^5),A_5(\vec{n}_{4}^5)$.
In the next step we get the following algebraic identity which
can be used, via averaging, to generate a family of Bell
inequalities:
\begin{equation}
\sum_{s_1, s_2 = \pm 1}S_{88444}(s_1,s_2)
(A_{1234,12}+s_1 A_{5678,34})
(A_{12,12} + s_2 A_{34,34}) = \pm 256.
\end{equation}

It is clear that there is no bound in extending this type of derivations. 
Finally, let us recall that all the inequalities with a lower number of settings 
can be obtained from our construction by making some
of the local settings identical.

The multisetting inequalities constructed by the above
procedure are tight. Consider the case of $4 \times 4 \times 2$ inequalities.
The left hand side of the identity
(\ref{GEN}) is equal to $\pm 16$ for any combination of
predetermined local realistic results. In a 32 dimensional real space, one can
build a convex polytope, containing all possible local realistic models of the
correlation functions for the specified settings, with vertices
 given by the tensor products of 
$\hat v=(A_1(\vec n_1^1),A_1(\vec n_2^1),A_1(\vec n_3^1),A_1(\vec n_4^1))
\otimes(A_2(\vec n_1^2),A_2(\vec n_2^2),A_2(\vec n_3^2),A_2(\vec n_4^2))
\otimes(A_3(\vec n_1^3), A_3(\vec n_2^3))$. 
Since the factor $\xi_1 = A_1(\vec n_1^1) A_2(\vec n_1^2) A_3(\vec n_1^3)$
can be put  in front of the tensor product:
$\hat v=\xi_1 
(1,\xi_2,\xi_3,\xi_4)
\otimes(1,\xi_5,\xi_6,\xi_7)
\otimes(1, \xi_	8)$,
with all $\xi_i = \pm 1$,
the politope has $256 = 2^8$ different vertices. 
Tight Bell inequalities define the
half-spaces in which is the polytope,  which contain a face of it
in their border hyperplane. If 32 linearly independent vertices
belong to a hyperplane, this hyperplane defines a tight
inequality. Half of the vertices in (\ref{GEN}) give the value
$16$ and  the other half gives $-16$.
Every vertex $\hat v$ from the first set has a
partner $- \hat v$ in the second one. 
Next notice that 
any set of $128$
vertices $\hat v$, which does not contain pairs $\hat v$ and $- \hat v$ contains
a set of 32 linearly independent points. Thus, each
inequality is tight. This reasoning can be adapted to all
inequalities discussed here.

The multisetting inequalities reveal a violation
of local realism of classes of states, for which standard inequalities,
with two measurement settings per side, are satisfied.

\subsection{Violation of multisetting Bell inequalities [P6]}

Let us derive 
\emph{necessary and sufficient}
conditions for the violation of 
$2^{N-1} \times 2^{N-1} \times ... \times 2$ inequalities.
First, consider the case of three qubits.
All $4 \times 4 \times 2$ inequalities
are generated by the
following inequality [compare (\ref{442_GEN_INEQ})]:
\begin{equation}
\Big| \Big\langle
A_{12,12;S'} [A_3(\vec n_1^3)+ A_3(\vec n_2^3)]
+ A_{34,34;S''} [A_3(\vec n_1^3)- A_3(\vec n_2^3)] 
\Big\rangle \Big| \le 16,
\label{442_INEQ}
\end{equation}
where $A_{12,12;S'}$ and $A_{34,34;S''}$
are known from the $2 \times 2$ case, (\ref{INEQ2}).
The
condition for the $4 \times 4 \times 2$ inequalities to hold, in
the quantum case, transforms to:
\begin{equation}
|[\hat A_{12,12;S'} \otimes (\vec a_1^3 + \vec a_2^3) + \hat A_{34,34;S''}
\otimes (\vec a_1^3 - \vec a_2^3)] \circ \hat T |\le 8, \label{KO}
\end{equation}
where e.g. 
\begin{equation}
\hat A_{12,12;S'} =\sum_{s_1,s_2= \pm 1} S'(s_1,s_2) (\vec a_1^1 + s_1 \vec a_2^1) \otimes
(\vec a_1^2 + s_2 \vec a_2^2),
\end{equation}
with $S'(s_1,s_2)$ being some 
non-factorable sign function.
The aim is to find the maximum, 
over choices of local measurement settings,
of the left-hand side
of (\ref{KO}), given an arbitrary quantum state
(correlation tensor).

By defining $\frac{1}{2}(\vec a_1^3 + \vec a_2^3) = \alpha_1^3 \vec \alpha_1^3$
and $\frac{1}{2}(\vec a_1^3 - \vec a_2^3) = \alpha_2^3 \vec \alpha_2^3$
as before in Eq. (\ref{NEW_COORDINATE_SYSTEMS}), inequality (\ref{KO})
transforms to:
\begin{equation}
|[\alpha_1^3 \hat A_{12,S'} \otimes \vec \alpha_1^3 
+ \alpha_2^3 \hat A_{34,S''} \otimes \vec \alpha_2^3]
 \circ \hat T |\le 4.
 \label{KO1}
\end{equation}
The three qubit correlation tensor can be Schmidt decomposed into:
\begin{equation}
\hat T = \hat P_1 \otimes \vec \gamma_1 
+ \hat P_2 \otimes \vec \gamma_2
+ \hat P_3 \otimes \vec \gamma_3,  
\end{equation}
where the three unit vectors $\vec \gamma_i$
form a basis in $\mathcal{R}^3$
and the unnormalized rank two tensors
are also orthogonal:
\begin{equation}
\hat P_i \circ \hat P_j = 0
\quad {\rm for } \quad
i \ne j.
\end{equation}
Further, one can assume that the rank two tensors
are ordered by their indices in accordance with decreasing norms.
Thus, if one specifies
\begin{equation}
\vec \alpha_1^3 = \vec \gamma_1
\quad {\rm and } \quad
\vec \alpha_2^3 = \vec \gamma_2,
\end{equation}
the value of the left hand side of (\ref{KO1}) is maximized
and the whole inequality depends on rank two tensors only:
\begin{equation}
|[\alpha_1^3 \hat A_{12,12;S'} \circ \hat P_1 
+ \alpha_2^3 \hat A_{34,34;S''} \circ \hat P_2 |\le 4.
\label{RANK_TWO_CONDITION}
\end{equation}
One can interpret the expression within the moduli
as the scalar product between two two-dimensional vectors.
Namely, between vector $\vec \alpha^3 \equiv (\alpha_1^3,\alpha_2^3)$
and vector $\vec P \equiv (\hat A_{12,12;S'} \circ \hat P_1,\hat A_{34,34;S''} \circ \hat P_2)$.
Since vector $\vec \alpha^3$
is an arbitrary normalized vector,
to maximize the left-hand side of this expression
one chooses it to be equal to:
\begin{equation}
\vec \alpha^3 = \frac{\vec P}{||\vec P||}.
\end{equation}
Thus, maximum of the left-hand side is given
by the norm $\frac{\vec P \cdot \vec P}{||\vec P||} = ||\vec P||$.
The condition (\ref{RANK_TWO_CONDITION}) can be written as:
\begin{equation}
[\hat A_{12,12;S'} \circ \hat P_1]^{2} 
+ [\hat A_{34,34;S''} \circ \hat P_2]^{2} \leq 4^2,
\end{equation}
where we have squared both sides.
Since
$\hat A_{12,12;S'}$ depends on different vectors than 
$\hat A_{34,34;S''}$, one can maximize the two terms \emph{independently}. 
Furthermore, the problem of maximization of
each of them is equivalent to the $2 \times
2$ case studied earlier. The overall maximization process gives
the following \emph{necessary and sufficient} condition for
quantum correlations to satisfy the inequality (\ref{442_INEQ}):
\begin{equation}
\max  \Big[ \sum_{x=1,2} \sum_{k_x,l_x=1,2} T_{k_x l_x x}^2 \Big] \le 1.
\label{NS_CONDITION_MULTI}
\end{equation}
When compared with the \emph{sufficient} condition for
$2 \times 2 \times 2$ inequalities to hold, namely \cite{ZB}: 
\begin{equation}
\max \Big[ \sum_{k,l,m=1,2} T_{klm}^2 \Big] \le 1,
\label{S_COND_THREE}
\end{equation}
the new 
condition is {\em more demanding} because the
Cartesian coordinate systems denoted by the indices $k_1,l_1$ and
$k_2,l_2$ do not have to be the same.

In a similar way one can reach analogous conditions for
violation of $2^{N-1} \times 2^{N-1} \times 2^{N-2} \times ...
\times 2$ inequalities by quantum predictions.
The problem of maximization of the Bell expression
with a rank $N$ correlation tensor
can be split into problems
considering lower rank tensors.
In the Table \ref{CONDITIONS_TABLE} we present these conditions for small $N$.
\begin{table}
\caption{Examples of necessary and sufficient conditions for violation of
multisetting inequalities.}
\label{CONDITIONS_TABLE}
\begin{center}
\begin{tabular}{c|c|c}
\hline  \hline
$N$ & case & $C_N$  (the condition) \\ \hline \hline
$2$ & $2 \times 2$ & $\sum_{k,l=1,2} T_{kl}^2 \le 1$ \\ \hline
$3$ & $4 \times 4 \times 2$ & $\sum_{k,l=1,2} T_{kl2}^2 + \sum_{k',l'=1,2} T_{k'l'1}^2 \le 1$  \\ \hline
$4$ & $8 \times 8 \times 4 \times 2$ & 
$\sum_{k_1,l_1=1,2} T_{k_1l_122}^2 + \sum_{k_2,l_2=1,2} T_{k_2l_212}^2 +$ \\
&&
$\sum_{k_3,l_3=1,2} T_{k_3l_321}^2 + \sum_{k_4,l_4=1,2} T_{k_4l_411}^2 \le 1 $
\\ \hline 
\end{tabular}
\end{center}
\end{table}
One can see a useful recurrence that can be used to write down the condition for arbitrary $N$. 
Let us define:
\begin{equation}
\mathcal{C}_2 \equiv \sum_{k,l=1,2} T_{kl}^2.
\end{equation}
Then the condition for two qubits reads: $\max (\mathcal{C}_2) \le 1$.
Next, let us put a recursive definition:
\begin{equation}
\mathcal{C}_N = [\mathcal{C}_{N-1}]_{\oplus 2} + [\mathcal{C}_{N-1}]_{\oplus 1}',
\end{equation}
where $[\mathcal{C}_{N-1}]_{\oplus k}$ is the expression in the condition for $N-1$ qubits
in which the correlation tensor elements $T_{i_1...i_{N-1}}$
are replaced by $T_{i_1...i_{N-1}k}$, i.e.\ elements of the $N$-qubit correlation tensor.
The ``prime'' denotes the fact that the second term can involve components
of $\hat T$ in a different set of coordinate systems 
(for the first $N-1$ observers) as the unprimed term.

The sufficient and necessary condition for $N$ qubits to satisfy
all $2^{N-1} \times 2^{N-1} \times 2^{N-2} \times ... \times 2$
inequalities, within this convention reads:
\begin{equation}
\max(\mathcal{C}_N) \le 1.
\end{equation}

\subsubsection{Examples}

Let us give examples of states
for which multisetting inequalities
form a more stringent constraint
on local realism than standard inequalities.

First, consider the generalized GHZ state, as given in Eq.\ (\ref{GEN_GHZ}):
\begin{equation}
| \psi_{GHZ} \rangle = 
\cos \alpha |z+ \rangle_1 ... |z+ \rangle_N
+ \sin\alpha |z- \rangle_1 ... |z- \rangle_N,
\quad {\rm with} \quad 0 \le \alpha \le \pi/4.
\end{equation}
Such states satisfy all standard correlation Bell inequalities
for small values of angle $\alpha$ and odd $N$ \cite{ZBLW}.
The condition to satisfy multisetting Bell inequalities
for $N$ partners,
in which the last party chooses between
settings $x$ and $z$ can be put as ($\mathcal{C}_N \le 1$):
\begin{equation}
\sum_{k_1,...,k_{N-1}=x,y} T^2_{k_1...k_{N-1}x}
+ \sum_{k_1, ..., k_{N-1} = x,z} T^2_{k_1...k_{N-1}z}
\le 1.
\end{equation}
Inserting the values of the correlation tensor elements
of the generalized GHZ state (for odd $N$)\footnote{For even $N$ the state obviously violates the inequality as in this case $T_{z...z} = 1$
and one has additional correlations in the $xy$ plane.}, 
given in (\ref{T_GEN_GHZ})
and below that formula,
results in the left-hand side equal to:
\begin{equation}
2^{N-2} \sin^2{2\alpha} + \cos^2{2\alpha} > 1,
\quad \textrm{ for}  \quad 0 < \alpha \le \pi/4.
\end{equation}
Out of $2^{N-1} - 1$ non-zero elements of the correlation tensor in the $xy$ plane
there are $2^{N-2}$ components with $x$ as the last index.
Thus,  the multisetting Bell inequalities are violated 
for the whole range of $\alpha$ and for
arbitrary $N$, in contrast to the case of standard Bell
inequalities.

Consider the so-called $|W \rangle$ state
of $N$ qubits:
\begin{equation}
|W \rangle = \frac{1}{\sqrt{N}}
\Big[ |z+ \rangle_1 |z- \rangle_2 ... |z- \rangle_N
+ |z- \rangle_1 |z+ \rangle_2 ... |z- \rangle_N
+ |z- \rangle_1 |z- \rangle_2 ... |z+ \rangle_N
\Big].
\end{equation}
It has the following nonvanishing correlation tensor elements,
which involve correlations between all the subsystems:
\begin{eqnarray}
T_{z...z} &=& (-1)^{N-1}, \\
T_{xxz...z} &=& ... = T_{z...zxx} = \frac{2}{N} (-1)^N, \nonumber \\
T_{yyz...z} &=& ... = T_{z...zyy} = \frac{2}{N} (-1)^N. \nonumber
\end{eqnarray}
The terms with only two indices
equal to $x$ or $y$ and all other indices
equal to $z$ are given by $\frac{2}{N} (-1)^N$.
To get better
results than in the standard case it is enough to allow observers
to choose between observables in the $yz$ plane. The
condition in such a case reduces to:
\begin{equation}
\sum_{k_1,k_2,...,k_N=y,z}
T^2_{k_1...k_N} \leq 1.
\end{equation}
Using the correlation tensor elements given above,
the quantum value of this expression
is, at least (no optimization):
\begin{equation}
1 + {N \choose 2} \frac{4}{N^2} = 3 - \frac{2}{N} > 1.
\end{equation}
Thus, if one considers a noise admixture to the
$|W\rangle$ states, in such a form that one arrives at a mixed
state 
$\rho_{|W\rangle} = (1-V) \rho_{noise} + V |W \rangle \langle W|$, 
with $\rho_{noise} = \openone / 2^N$, then the new
inequalities show that for $V \geq 1/\sqrt{3- 2/N}$ there is no
local realistic description for the correlations. The identical threshold for the
standard inequalities \cite{ADITI}, is, however, \emph{only
necessary} for them to be violated.
The range of $V$ for which there is no local realistic
description for the observed correlations grows.

Finally consider, recently produced \cite{EIBL}, the four-qubit state
first introduced by Weinfurter and \.Zukowski \cite{WZ}:
\begin{eqnarray*}
|\Psi \rangle &=& \sqrt{1/3} \Big( 
| z+ \rangle_1 | z+ \rangle_2 | z+ \rangle_3 | z+ \rangle_4
+ | z- \rangle_1 | z- \rangle_2 | z- \rangle_3 | z- \rangle_4
 \nonumber \\
&+& \frac{1}{2}(| z+ \rangle_1 | z- \rangle_2 | z+ \rangle_3 | z- \rangle_4 
+ | z- \rangle_1 | z+ \rangle_2 | z- \rangle_3 | z+ \rangle_4 \\
&+& | z+ \rangle_1 | z- \rangle_2 | z- \rangle_3 | z+ \rangle_4 +
| z- \rangle_1 | z+ \rangle_2 | z+ \rangle_3 | z- \rangle_4) \Big) \\
&=&\sqrt{2/3}|\mbox{GHZ}\rangle_{1234} +
\sqrt{1/3}|\mbox{EPR}\rangle_{12}|\mbox{EPR}\rangle_{34} \nonumber
\end{eqnarray*}
where $|\mbox{EPR}\rangle=1/\sqrt{2} \Big( | z+ \rangle_1 | z- \rangle_2 + | z- \rangle_1 | z+ \rangle_2 \Big)$.
The non vanishing correlation tensor
components of $| \Psi \rangle$ read: 
\begin{eqnarray}
&T_{xxxx}\!=\!T_{yyyy}\!=\!T_{zzzz}\!=\!1,& \nonumber \\
&T_{xxyy} \! = \! T_{xxzz} \! = \! T_{yyxx} \! = \! T_{yyzz}
 = T_{zzxx} \! = \! T_{zzyy} \! = \! -1/3,& \nonumber \\
&T_{xzxz}=T_{xzzx}=T_{zxxz}=T_{zxzx}=2/3,& \nonumber \\
&T_{xyxy}=T_{xyyx}=T_{yxxy}=T_{yxyx} 
=T_{yzyz}=T_{yzzy}=T_{zyyz}=T_{zyzy}=-2/3.& \nonumber
\end{eqnarray}
The left-hand side of 
the condition $\mathcal{C}_4$
given in the Table \ref{CONDITIONS_TABLE}
is equal to 4,  e.g.\ for all 
local summations over $x$ and $y$.
Thus the $8 \times 8 \times 4 \times 2$ inequality is violated by
the factor $2$ (recall that the quantum value is given by the square root of the left-hand side). 
Therefore a state $(1-V) \rho_{noise} + V |\Psi
\rangle \langle \Psi|$ gives non-classical correlations for $V > \frac{1}{2}$. 
In contrast, standard Bell inequalities cannot be
violated for $V \leq 0.5303$ 
(this value was obtained using numerical method described in \cite{KASZLIKOWSKI_QUDITS}).

\subsection{Arbitrary number of settings [P3]}

Multisetting inequalities described in previous sections
cannot involve an arbitrary number of settings.
For example, the $3 \times 3$ case
is not included in this formalism.
In this section, basing on a geometrical argument by \.Zukowski \cite{ZUK93},
a Bell inequality for many observers,
each choosing between an arbitrary number of dichotomic 
observables, is derived.
Many previously known inequalities are special cases of the new inequality,
e.g.\ the Clauser-Horne-Shimony-Holt inequality \cite{CHSH}
or two-setting multiparty inequalities \cite{MERMIN,ARDEHALI,BELINSKII}.
The new inequalities
are maximally violated by the Greenberger-Horne-Zeilinger (GHZ) states \cite{GHZ}.
Many other states violate them,
including the states which satisfy two-settings inequalities \cite{ZBLW}
and bound entangled states \cite{DUR}.
This is shown using the necessary and sufficient condition
for the violation of the inequalities.
Finally, it is proven that the Bell operator
has only two non-vanishing eigenvalues
which correspond to the GHZ states,
and thus has a very simple form.

Consider $N$ separated parties making measurements on two-level systems.
Each party can choose one of $M$ dichotomic observables.
In this scenario the parties can measure $M^N$ correlations $E_{m_1...m_N}$,
where the index $m_n=0,...,M-1$ denotes the setting of the $n$th observer.
A general Bell expression, which involves these correlations
with some coefficients $c_{m_1...m_N}$,
can be written as:
\begin{equation}
\sum_{m_1, ..., m_N =0}^{M-1} c_{m_1...m_N} E_{m_1...m_N} = \vec C \cdot \vec E.
\label{GENERAL_BELL}
\end{equation}
In what follows we assume a certain form of the coefficients $c_{m_1...m_N}$,
defining our Bell inequality,
and compute the local realistic bound as the maximum of the scalar product
$|\vec C \cdot \vec E^{LR}|$. 
The components of the vector
$\vec E^{LR}$ have the usual form:
\begin{equation}
E_{m_1...m_N}^{LR} = \int d \lambda \rho(\lambda) I_{m_1}^1(\lambda)...I_{m_N}^N(\lambda),
\label{E_LR}
\end{equation}
where $\lambda$ denotes a set of hidden variables, $\rho(\lambda)$ their distribution,
and $I_{m_n}^{n}(\lambda) = \pm 1$ the predetermined result of the $n$th observer under setting $m_n$.

The quantum prediction for the Bell expression (\ref{GENERAL_BELL})
is given by a scalar product of $\vec C \cdot \vec E^{QM}$.
The components of $\vec E^{QM}$, 
according to quantum theory,
are given by (Appendix A):
\begin{equation}
E_{m_1...m_N}^{QM} = {\rm Tr}\left( \rho \textrm{ }\vec m_1 \cdot \vec \sigma^1 \otimes ... \otimes \vec m_N \cdot \vec \sigma^N \right),
\end{equation}
where $\rho$ is a density operator (general quantum state),
$\vec \sigma^n = (\sigma_x^n,\sigma_y^n,\sigma_z^n)$ is a vector of local Pauli operators
for the $n$th observer,
and $\vec m_n$ denotes a normalized vector 
which parameterizes the observable $m_n$ for the $n$th party.

Assume that the local settings are parameterized by a single angle:
$\phi_{m_n}^n$.
In the quantum picture we restrict the observable vectors $\vec m_n$
to lie in the equatorial plane of the Bloch sphere:
\begin{equation}
\vec m_n \cdot \vec \sigma^n = \cos\phi_{m_n}^n \sigma_x^n + \sin\phi_{m_n}^n \sigma_y^n.
\end{equation}
Take the coefficients $c_{m_1...m_N}$
of the form
\begin{equation}
c_{m_1...m_N} = \cos(\phi_{m_1}^1 + ... + \phi_{m_N}^N),
\end{equation}
with the angles given by
\begin{equation}
\phi_{m_n}^n = \frac{\pi}{M} m_n + \frac{\pi}{2MN} \eta.
\label{ANGLES}
\end{equation}
The number $\eta=1,2$ is fixed for a given experimental
situation, i.e.\ $M$ and $N$, and equals:
\begin{equation}
\eta = [M+1]_2 [N]_2 + 1,
\label{ETA}
\end{equation}
where $[x]_2$ stands for $x$ modulo $2$.
The maximum is attained for deterministic local realistic models,
as they correspond to the extremal points of the correlation polytope.
Thus, the following inequality appears:
\begin{eqnarray}
&&|\vec C \cdot \vec E^{LR}| \le 
\max_{I_{0}^1,...,I_{M-1}^N = \pm 1} \left[
\sum_{m_1,...,m_N=0}^{M-1} \! \! \! \! \! \!
\cos(\phi_{m_1}^1 + ... + \phi_{m_N}^N)
I_{m_1}^1...I_{m_N}^N \right]
\label{INEQ_DETER}
\end{eqnarray}
where we have shortened the notation $I_{m_n}^n \equiv I_{m_n}^n(\lambda)$.
Since 
$\cos(\phi_{m_1}^1 + ... + \phi_{m_N}^N) = {\rm Re} \left( \prod_{n=1}^N \exp{(i \phi_{m_n}^n)} \right)$
and the predetermined results, $I_{m_n}^n = \pm 1$, are real,
the expression to be maximized can be written as:
\begin{equation}
\sum_{m_1,...,m_N=0}^{M-1} 
{\rm Re} \left( \prod_{n=1}^N \exp{(i \phi_{m_n}^n)} I_{m_n}^n \right).
\end{equation}
Moreover, since inequality (\ref{INEQ_DETER}) involves
the sum of all possible products of local results respectively
multiplied by the cosines of all possible sums of local angles,
the right-hand side can be further reduced to
involve the product of sums:
\begin{equation}
{\rm Re} \left( \prod_{n=1}^N \sum_{m_n=0}^{M-1}\exp{(i \phi_{m_n}^n)} I_{m_n}^n \right).
\end{equation}
Inserting the angles (\ref{ANGLES}) into this expression results in:
\begin{equation}
{\rm Re} \left( \exp{(i \frac{\pi}{2M} \eta)} \prod_{n=1}^N \sum_{m_n=0}^{M-1}\exp{(i \frac{\pi}{M}m_n)} I_{m_n}^n \right),
\label{RE_ANGLES}
\end{equation}
where the factor $\exp{(i \frac{\pi}{2M} \eta)}$
comes from the term $\frac{\pi}{2MN} \eta$ in (\ref{ANGLES}),
which is the same for all parties.  

One can decompose a complex number given by the sum in (\ref{RE_ANGLES})
into its modulus $R_n$, and phase $\Phi_n$:
\begin{equation}
\sum_{m_n=0}^{M-1}\exp{(i \frac{\pi}{M}m_n)} I_{m_n}^n = R_n e^{i\Phi_n}.
\label{VECTOR}
\end{equation}
We maximize the length of this vector on 
the complex plane.
The modulus of the sum of any two complex numbers
$|z_1 + z_2|^2$ is given by the cosine law
as $|z_1|^2+|z_2|^2 + 2 |z_1| |z_2| \cos \varphi$,
where $\varphi$ is the angle between the corresponding vectors.
To maximize the length of the sum
one should choose the summands as close as possible
to each other.
Since in our case all vectors being summed
are rotated by multiples of $\frac{\pi}{M}$
from each other, the simplest optimal choice
is to put all $I_{m_n}^n = 1$.
In this case one has:
\begin{equation}
R_n^{\max} = \left| \sum_{m_n=0}^{M-1}\exp{(i \frac{\pi}{M}m_n)} \right|
= \left| \frac{2}{1-\exp{(i\frac{\pi}{M})}} \right|,
\end{equation}
where the last equality follows from the finite sum of numbers in the
geometric progression (any term in the sum is given by the preceding term multiplied by $e^{i
\pi/M}$).
The denominator inside the modulus can be transformed to
$\exp{(i\frac{\pi}{2M})} \left[ \exp{(-i\frac{\pi}{2M})} - \exp{(i\frac{\pi}{2M})} \right]$,
which reduces to $- 2i \exp{(i\frac{\pi}{2M})} \sin\left(\frac{\pi}{2M}\right)$.
Finally, the maximal length reads:
\begin{equation}
R_n^{\max} = \frac{1}{\sin\left(\frac{\pi}{2M}\right)},
\end{equation}
where there is no longer need for the modulus 
since the argument of the sine is small.
Moreover, since the local results for each party can be chosen independently, 
the maximal length $R_n^{\max}$ does not depend on the particular $n$, i.e.\ $R_n^{\max} = R^{\max}$.

Since $R^{\max}$ is a positive real number 
its $N$th power can be put
to multiply the real part in (\ref{RE_ANGLES}),
and one finds $|\vec C \cdot \vec E^{LR}|$ to be bounded by:
\begin{equation}
|\vec C \cdot \vec E^{LR}| \le
\left[ \sin \left(\frac{\pi}{2M} \right) \right]^{-N} \! \! \! \! 
\cos \left( \frac{\pi}{2M} \eta + \Phi_1 + ... + \Phi_N \right),
\end{equation}
where the cosine comes from the phases of the sums in (\ref{RE_ANGLES}).
These phases can be found from the definition (\ref{VECTOR}).
As only vectors rotated by a multiple of $\frac{\pi}{M}$ are summed 
(or subtracted) in (\ref{VECTOR}), each phase $\Phi_n$ can acquire
a restricted set of values.
Namely:
\begin{equation}
\Phi_n = 
\Bigg\{
\begin{array}{lc}
\frac{\pi}{2M} + \frac{\pi}{M}k & \textrm{ for } M \textrm{ even}, \\
 & \\
\frac{\pi}{M}k & \textrm{ for } M \textrm{ odd},
\end{array}
\end{equation}
with $k=0,...,2 M-1$,
i.e.\ for $M$ even, $\Phi_n$ is an odd multiple of $\frac{\pi}{2 M}$;
and for $M$ odd, $\Phi_n$ is an even multiple of $\frac{\pi}{2 M}$.
Thus, the sum $\Phi_1 + ... + \Phi_N$
is an even multiple of $\frac{\pi}{2 M}$,
except for $M$ even and $N$ odd.
Keeping in mind the definition of $\eta$, given in (\ref{ETA}),
one finds the argument of
$\cos \left( \frac{\pi}{2M}\eta + \Phi_1 + ... + \Phi_N \right)$
is always an odd multiple of $\frac{\pi}{2 M}$,
which implies the maximum value of the cosine is equal to
$\cos \left(\frac{\pi}{2M} \right)$.
Finally the multisetting Bell inequality reads:
\begin{equation}
|\vec C \cdot \vec E^{LR}| \le \left[ \sin \left(\frac{\pi}{2M}\right) \right]^{-N} \cos \left(\frac{\pi}{2M} \right).
\label{MS_INEQUALITY}
\end{equation}
This inequality,
when reduced to two parties choosing between two settings each,
recovers the Clauser-Horne-Shimony-Holt inequality (\ref{CHSH}).
For a higher number of parties, still choosing between two observables,
it reduces to tight two-setting inequalities \cite{MERMIN,ARDEHALI,BELINSKII}.
When $N$ observers choose between three observables 
the inequalities of \.Zukowski and Kaszlikowski are obtained \cite{ZUK_KASZ},
and for a continuous range of settings ($M \to \infty$) it recovers the inequality of \.Zukowski \cite{ZUK93}.

One can derive a simple and useful form of a Bell operator
associated with the Bell expression (\ref{MS_INEQUALITY}).
It will be used to 
derive the necessary and sufficient condition
for the violation of the inequality.

The form of the coefficients $c_{m_1...m_N} = \cos(\phi_{m_1}^1 + ... + \phi_{m_N}^N)$
we have chosen is exactly the same
as the quantum correlation function
$E_{m_1...m_N}^{GHZ} = \cos(\phi_{m_1}^1 + ... + \phi_{m_N}^N)$
for the Greenberger-Horne-Zeilinger state:
\begin{equation}
|\psi^+ \rangle = \frac{1}{\sqrt{2}}
\Big[ |z+\rangle_1 ... |z+\rangle_N + |z-\rangle_1 ... |z-\rangle_N \Big],
\end{equation}
For this state the two vectors $\vec C$ and $\vec E^{GHZ}$ are equal (thus parallel),
which means that 
the state $|\psi^+ \rangle$ maximally violates inequality (\ref{MS_INEQUALITY}).
The value of the left hand side of (\ref{MS_INEQUALITY}) is given
by the scalar product of $\vec E^{GHZ}$ with itself:
\begin{equation}
\vec E^{GHZ} \cdot \vec E^{GHZ}
= \sum_{m_1,...,m_N=0}^{M-1} \cos^2(\phi_{m_1}^1 + ... + \phi_{m_N}^N).
\label{COS_SQUARE}
\end{equation}
Using the trigonometric identity $\cos^2 \alpha = \frac{1}{2}(1+\cos2\alpha)$
one can rewrite this expression into the form:
\begin{equation}
\vec E^{GHZ} \cdot \vec E^{GHZ}
= \frac{1}{2}M^N + \frac{1}{2}\sum_{m_1,...,m_N=0}^{M-1} 
\! \! \! \! \! \! \! \! \cos[2(\phi_{m_1}^1 + ... + \phi_{m_N}^N)].
\end{equation}
As before, the second term can be written as a real part of the complex number.
Putting the values of angles (\ref{ANGLES})
one arrives at:
\begin{equation}
\frac{1}{2} {\rm Re}
\left( \exp{(i \frac{\pi}{M} \eta)} \prod_{n=1}^N \sum_{m_n=0}^{M-1}\exp{(i \frac{2\pi}{M}m_n)}\right).
\label{COMPLEX_ROOTS}
\end{equation}
Note that $e^{i\frac{2\pi}{M}}$ is a primitive complex $M$th root of unity.
Since all complex roots of unity
sum up to zero
the above expression vanishes.
The maximal quantum value of the left hand side of (\ref{MS_INEQUALITY}) 
equals:
\begin{equation}
\vec E^{GHZ} \cdot \vec E^{GHZ} = \frac{1}{2}M^N.
\end{equation}
If instead of $| \psi^+ \rangle$ one chooses
the state $| \psi^- \rangle = \frac{1}{\sqrt{2}}
[ |z+\rangle_1 ... |z+\rangle_N - |z-\rangle_1 ... |z-\rangle_N ]$,
for which the correlation function is given by
$E_{m_1...m_N}^{GHZ-} = - \cos(\phi_{m_1}^1 + ... + \phi_{m_N}^N)$,
one arrives at a minimal value of the Bell expression, equal to $-\frac{1}{2}M^N$,
as the vectors $\vec C$ and $\vec E^{GHZ-}$
are exactly opposite.
Since we take the modulus in the Bell expression,
both states lead to the same violation.

The Bell operator associated with the Bell expression (\ref{MS_INEQUALITY})
is defined as:
\begin{equation}
\mathcal{B'} \equiv \! \! \! \! \! \! \! 
\sum_{m_1...m_N=0}^{M-1} \! \! \! \! \! \! c_{m_1...m_N} 
\vec m_1 \cdot \vec \sigma^1 \otimes ... \otimes \vec m_N \cdot \vec \sigma^N.
\label{B}
\end{equation}
Its average in the quantum state $\rho$
is equal to the quantum prediction of the Bell expression, for this state.
We shall prove that it has only two
eigenvalues $\pm \frac{1}{2}M^N$,
and thus is of the simple form:
\begin{equation}
\mathcal{B} \equiv
\mathcal{B}(N,M) = \frac{1}{2}M^N \left[ |\psi^+ \rangle \langle \psi^+ | - 
|\psi^- \rangle \langle \psi^- |\right].
\label{BELL_OPERATOR}
\end{equation}

Both operators $\mathcal{B}$ and $\mathcal{B'}$
are defined in the Hilbert-Schmidt space
with the trace scalar product.
To prove their equivalence one should
check if the conditions:
\begin{equation}
{\rm Tr}(\mathcal{B'} \mathcal{B}) = 
{\rm Tr}(\mathcal{B} \mathcal{B}) = 
{\rm Tr}(\mathcal{B'} \mathcal{B'}),
\label{TRACES}
\end{equation}
are satisfied.
Geometrically speaking, these conditions
mean that
the ``length'' and ``direction''
of the operators are the same.

The trace ${\rm Tr}(\mathcal{B'} \mathcal{B})$
involves the traces 
${\rm Tr}\left(|\psi^{\pm} \rangle \langle \psi^{\pm} | \vec m_1 \cdot \vec \sigma^1 \otimes ... \otimes \vec m_N \cdot \vec \sigma^N \right)$.
These traces are the quantum correlation functions
(averages of the product of local results) for the GHZ states,
and thus are given by $\pm \cos(\phi_{m_1}^1 + ... + \phi_{m_N}^N)$.
Their difference doubles the cosine,
which is then multiplied by the same cosine
coming from the coefficients $c_{m_1...m_N}$.
Thus the main trace takes the form:
\begin{equation}
{\rm Tr}(\mathcal{B'} \mathcal{B}) =
M^N \! \! \! \! \! \! \sum_{m_1...m_N=0}^{M-1} \! \! \! \! \!
\cos^2(\phi_{m_1}^1 + ... + \phi_{m_N}^N) 
= \frac{1}{2} M^{2N},
\end{equation}
where the last equality
follows from the considerations below Eq. (\ref{COS_SQUARE}).

The middle trace of (\ref{TRACES})
is given by ${\rm Tr}(\mathcal{B} \mathcal{B}) = \frac{1}{2} M^{2N}$,
which directly follows from the
orthonormality of the states $| \psi^{\pm} \rangle$.

The last trace of (\ref{TRACES}) is more involved.
Inserting decomposition (\ref{B})
into ${\rm Tr}(\mathcal{B'} \mathcal{B'})$ gives:
\begin{eqnarray*}
&& \sum_{\substack{
m_1...m_N, \\
m_1'...m_N'=0}}^{M-1} 
\! \! \! \! \!
\cos(\phi_{m_1}^1 + ... + \phi_{m_N}^N) \cos(\phi_{m_1'}^1 + ... + \phi_{m_N'}^N) \\
&& \times {\rm Tr}[(\vec m_1 \cdot \vec \sigma^1)(\vec m_1' \cdot \vec \sigma^1) ]...
{\rm Tr}[(\vec m_N \cdot \vec \sigma^N)(\vec m_N' \cdot \vec \sigma^N) ]
\end{eqnarray*}
The local traces are given by:
\begin{equation}
{\rm Tr}[(\vec m_n \cdot \vec \sigma^n)(\vec m_n' \cdot \vec \sigma^n) ]
= 2 \vec m_n \cdot \vec m_n' = 2 \cos(\phi_{m_n}^n - \phi_{m_n'}^n).
\end{equation}
Thus, the factor $2^N$ appears in front of the sums.
We write all the cosines (of sums and differences) in terms of individual
angles, insert these decompositions into ${\rm Tr}(\mathcal{B'} \mathcal{B'})$,
and perform all the multiplications.
Note that whenever the final product term involves
at least one expression
$\cos\phi_{m_n}^n \sin \phi_{m_n}^n = \frac{1}{2}
\sin(2 \phi_{m_n}^n)$
(or for the primed angles)
its contribution to the trace vanishes after the summations
[for the reasons discussed in Eq. (\ref{COMPLEX_ROOTS})].
Moreover, in the decomposition of 
$\cos(\phi_{m_n}^n - \phi_{m_n'}^n) = \cos\phi_{m_n}^n \cos\phi_{m_n'}^n +
\sin\phi_{m_n}^n \sin\phi_{m_n'}^n$
only the products of the same trigonometric functions appear.
In order to contribute to the trace
they must be multiplied again by the same functions.
Since the decompositions of cosines of sums
only differ in angles (primed or unprimed)
and not in the individual trigonometric functions,
the only contributing terms come from the product
of exactly the same individual trigonometric functions
in the decomposition of $\cos(\phi_{m_1}^1 + ... + \phi_{m_N}^N)$ 
and $\cos(\phi_{m_1'}^1 + ... + \phi_{m_N'}^N)$.
There are $2^{N-1}$ such products,
as many as the number of terms in the decomposition.
Each product involves $2N$ squared 
individual trigonometric functions.
Each of these functions can be written in terms
of cosines of the double angle, e.g. $\sin^2 \phi_{m_n}^n =
\frac{1}{2}(1-\cos(2\phi_{m_n}^n))$,
and the last cosine does not contribute to the sum
[again due to (\ref{COMPLEX_ROOTS})].
Finally the trace reads:
\begin{equation}
{\rm Tr}(\mathcal{B'} \mathcal{B'}) = 2^N
\! \! \! \! \!
\sum_{ \substack{ 
m_1...m_N, \\ m_1'...m_N'=0}}^{M-1} 
\! \! \! \! \!
2^{N-1} \frac{1}{2^{2N}} = \frac{1}{2} M^{2N}.
\end{equation}
Thus, the equations (\ref{TRACES})
are all satisfied, i.e.\ 
both operators $\mathcal{B}$ and $\mathcal{B'}$ are equal.
Only the states which 
have contributions in the subspace spanned by $| \psi^{\pm} \rangle$
can violate the inequality (\ref{MS_INEQUALITY}).

\subsection{Violation of inequality with arbitrary number of settings [P3]}

Let us derive the necessary and sufficient condition
for the violation of inequality (\ref{MS_INEQUALITY}).
The expected quantum value of the Bell expression,
using Bell operator, reads:
\begin{equation}
{\rm Tr}(\mathcal{B}(N,M)\rho) = 
\frac{M^N}{2} \left[ {\rm Tr}(|\psi^+ \rangle \langle \psi^+ |\rho) - 
{\rm Tr}(|\psi^- \rangle \langle \psi^- | \rho)\right].
\label{MS_BELL_OPERATOR}
\end{equation}
The violation condition is obtained after maximization,
for a given state, over the position of the $xy$ plane, 
in which the observables lie.

Let us denote the correlation tensors of the projectors 
$|\psi^\pm \rangle \langle \psi^\pm |$ by $T_{\nu_1...\nu_N}^{\pm}$.
Using the linearity of the trace operation
and the fact that the trace of the tensor product
is given by the product of local traces,
one can write ${\rm Tr}(|\psi^\pm \rangle \langle \psi^\pm | \rho)$
in terms of correlation tensors:
\begin{equation}
{\rm Tr}(|\psi^\pm \rangle \langle \psi^\pm | \rho) = 
\frac{1}{2^{2N}} \! \! \! \! \! \! \!
\sum_{\mu_1...\mu_N,\\ \nu_1...\nu_N=0}^3 
\! \! \! \! \! \! \! \! \! \! \! \!
T_{\nu_1...\nu_N}^{\pm} T_{\mu_1...\mu_N}
{\rm Tr}(\sigma_{\mu_1} \sigma_{\nu_1})...{\rm Tr}(\sigma_{\mu_N} \sigma_{\nu_N}). \nonumber
\end{equation}
Since each of the $N$ local traces ${\rm Tr}(\sigma_{\mu_n} \sigma_{\nu_n}) = 2 \delta_{\mu_n \nu_n}$,
the global trace is given by:
\begin{equation}
{\rm Tr}(|\psi^\pm \rangle \langle \psi^\pm | \rho)
= \frac{1}{2^{N}}  \sum_{\mu_1...\mu_N=0}^3
T_{\nu_1...\nu_N}^{\pm} T_{\mu_1...\mu_N}.
\end{equation}
The nonvanishing correlation tensor components of the GHZ states $| \psi^{\pm} \rangle$
are the same in the $z$ plane: $T_{z..z0..0}^{\pm} = 1$ for even number of $z$ indices;
and are exactly opposite in the $xy$ plane:
$T_{i_1...i_N}^{+} = - T_{i_1...i_N}^{-} = (-1)^\xi$
with $2\xi$ indices equal to $y$ and all remaining equal to $x$.
Inserting the traces to the Bell operator
one finds that the components in the $z$ plane
cancel out, and components in the $xy$ plane
double themselves.
Finally, 
the necessary and sufficient condition for the violation
of the inequality is given by:
\begin{equation}
\left( \frac{M}{2}\right)^N \max \sum_{i_1...i_N \in I_\xi} (-1)^{\xi} T_{i_1...i_N} \le 
 B_{LR}(N,M),
\label{NS}
\end{equation}
where the maximization is performed over the choice of local
coordinate systems, 
$I_\xi$ includes all sets of indices $i_1...i_N$ with 2$\xi$
indices equal to $y$ and the rest equal to $x$,
and 
\begin{equation}
B_{LR}(N,M) = \left[ \sin \left(\frac{\pi}{2M}\right) \right]^{-N} \cos \left(\frac{\pi}{2M} \right)
\end{equation}
denotes the local realistic bound.

\subsubsection{Examples}

Let us present examples of states,
which violate the new inequality.
As a measure of violation, $V(N,M)$, 
we take the average (quantum) value of the Bell operator
in a given state,
divided by the local realistic bound:
\begin{equation}
V(N,M) = \frac{\langle \mathcal{B}(N,M) \rangle_{\rho}}{B_{LR}(N,M)}.
\label{VIOL_FACTOR}
\end{equation}

\emph{GHZ state}.
First, let us simply consider $| \psi^{\pm} \rangle$.
For the case of two settings per side
one recovers previously known results \cite{WW,ZB,MERMIN}:
\begin{equation}
V(N,2) = 2^{(N-1)/2}.
\end{equation}
For three settings per side
the result of \.Zukowski and Kaszlikowski is obtained \cite{ZUK_KASZ}:
\begin{equation}
V(N,3) = \frac{1}{\sqrt{3}}\left(\frac{3}{2} \right)^{N}.
\end{equation}
For the continuous range of settings one recovers \cite{ZUK93}:
\begin{equation}
V(N,\infty) = \frac{1}{2}\left(\frac{\pi}{2} \right)^{N}.
\end{equation}
In the intermediate regime one has
\begin{equation}
V(N,M) = \frac{1}{2 \cos \left(\frac{\pi}{2M} \right)}\left(M \sin \left(\frac{\pi}{2M} \right) \right)^{N}.
\end{equation}
For a fixed number of parties $N > 3$ the violation
increases with the number of local settings.
It also grows
with increasing number of parties.
Surprisingly, the inequality implies 
for the cases of $N=2$ and $N=3$
that the violation decreases when the number of local settings grows.

\emph{Generalized GHZ state.}
Consider the GHZ state with free real coefficients (\ref{GEN_GHZ})
and correlation tensor components (\ref{T_GEN_GHZ}).
All components in the $xy$ plane (there are $2^{N-1}$ of them) 
contribute to the violation condition (\ref{NS}).
The violation factor is equal to $V(N,M)=\frac{M^N}{2 B_{LR}(N,M)} \sin2\alpha$.
For $N>3$ and $M>2$
the violation is bigger than the violation of standard two-setting inequalities \cite{ZB}.
Moreover, some of the generalized GHZ states,
for small $\alpha$ and odd $N$, 
do not violate any two-setting correlation function
Bell inequality \cite{ZBLW}, and violate the multisetting
inequality.

\emph{Bound entangled state.}
The inequality can reveal non-classical
correlations of a bound entangled state introduced
by D\"ur \cite{DUR}:
\begin{equation}
\rho_N = \frac{1}{N+1} \left( |\phi \rangle \langle \phi| + \frac{1}{2} \sum_{k=1}^N (P_k + \tilde P_k) \right),
\end{equation}
where $|\phi \rangle = \frac{1}{\sqrt{2}} \left[ |z+ \rangle_1 ... | z+ \rangle_N + e^{i \alpha_N} |z- \rangle_1 ... | z- \rangle_N \right]$,
with $\alpha_N$ being an arbitrary phase factor,
and $P_k$ denoting
a projector on the state 
$|z+ \rangle_1 ... |z- \rangle_k ... |z+ \rangle_N$ with ``$z-$'' 
on the $k$th position 
($\tilde P_k$ is obtained from $P_k$ after replacing ``$z+$'' by ``$z-$'' and vice versa).
As originally shown in \cite{DUR}
this state violates the Mermin-Klyshko inequalities for $N \ge 8$.
The new inequality
predicts the violation factor of
\begin{equation}
V(N,M) = \frac{1}{N+1} \frac{M^N ~\cos \alpha_N}{2B_{LR}(N,M)},
\label{viol-bes}
\end{equation}
which comes from the contribution
of the GHZ-like state $| \phi \rangle$
to the bound entangled state.
One can follow \cite{KASZLI} 
and change the Bell-operator (\ref{MS_BELL_OPERATOR}) such that the state
$|\phi \rangle$ 
becomes its eigenstate. 
The new operator, $\tilde{\mathcal{B}} (N,M)$, 
is obtained after applying 
local unitary transformations
\begin{equation}
U = |z+\rangle \langle z+| + e^{i\alpha_N/N} |z-\rangle \langle z-|,
\end{equation}
to the operator (\ref{MS_BELL_OPERATOR}),
i.e.\ 
$\tilde{\mathcal{B}} (N,M) = U^{\otimes N} \mathcal{B} U^{\dagger \otimes N}$.
The violation factor of the new inequality 
is higher than
(\ref{viol-bes}), and equal to
\begin{equation}
\tilde{V}(N,M) = \frac{1}{N+1} \frac{M^N}{2B_{LR}(N,M)}.
\end{equation}
If one sets $M=3$ it appears
that the number of parties sufficient
to see the violation reduces to $N \ge 7$ \cite{KASZLI}.
On the other hand, the result of \cite{ADITI1}
shows that the infinite range of settings
further reduces the number of parties to $N \ge 6$.
Using the new inequality, 
$M=5$ settings per side suffice to
already violate local realism with $N \ge 6$ parties.

\subsection{Conclusions}

We have described various approaches to Bell's theorem.
Starting with elementary notions  the 
assumptions and main experimental difficulties were discussed.
Next, the whole set of Bell inequalities
for correlation experiments between $N$
parties making two local measurements on qubits was presented,
and the generalization to multisetting case was described.
The conditions for violation of these inequalities
were shown as well as examples of states which 
(do not) violate them.
Finally, using different techniques
yet another multisetting inequality was derived,
and its properties were discussed.
The ideas used to derive $4 \times 4 \times 2$ type inequalities
were recently further developed
by Chen, Albeverio, and Fei,
who showed a family of Bell inequalities
which also involve lower order correlations \cite{LOWER_ORDER_CHEN}.

\newpage

\section{Beyond Bell's theorem}

We have presented Bell's theorem,
a tool which allows a 
quantitative distinction between the
quantum and the classical (local realism).
Although there is no experiment
simultaneously closing all loopholes
which allow to describe the observed data
in local realistic way,
each individual loophole
was closed in separate experiments.
Most scientists consider
the final experimental test
as only of technical difficulty.
Therefore, it is reasonable to consider the violation of local realism a well established fact.

In this part of the thesis we go beyond Bell's theorem.
It is shown that there exists a class
of plausible \emph{nonlocal} hidden variable theories
which still give predictions incompatible
with quantum mechanics.
We present an inequality, 
similar in spirit to the CHSH inequality 
on local hidden variables, 
that allows to test the class of nonlocal hidden variable theories against quantum theory. 
The theories under test provide an explanation 
of all standard two-qubit Bell-type experiments,
and despite being nonlocal
they do not allow faster than light communication. 
The derivation to be presented is based on a recent theorem by Leggett \cite{LEGGETT_NONLOCAL}.
We extend it to apply to real experimental situations 
and to simultaneously test against all local hidden variable models. 
Finally, we perform an experiment that violates the new inequality 
and hence excludes for the first time a broad class 
of nonlocal hidden variable theories as 
possible models underlying quantum theory.
One could consider this violation
as a step towards invalidating the realism assumption.
In non-realistic theories
measurement outcomes are objectively random.\footnote{As in the Copenhagen interpretation of quantum mechanics.}
This has the practical implication
that there exist perfect random number generators,
often a crucial ingredient in communication tasks.

We also study the freedom in choosing measurement settings,
another crucial assumption of Bell.
Within a local realistic
picture the violation of Bell's inequalities can only be understood if this
freedom is denied. The minimal degree to which the
freedom has to be abandoned is determined, which allows to keep such a picture and be in
agreement with the experiment.
Furthermore, the freedom in choosing
experimental arrangements may be considered as a resource 
for quantum communication. 
Its lacking
can be used by an eavesdropper to harm the security of quantum cryptography.
This will be shown in the next Chapter.

\subsection{Plausible nonlocal realistic theories [P1]}

The logical conclusion one can draw 
from the violation of local realism 
is that at least one of its assumptions fails. 
Specifically, either locality or realism or both cannot provide a 
foundational basis for quantum theory. 
Each of the resulting possible positions 
has strong supporters and opponents in the scientific community. 
However, Bell's theorem is principally unbiased against either of these views, 
i.e.\ one cannot, even in principle, favour one over the other.
It is therefore important to ask whether incompatibility 
theorems similar to Bell's can be found, 
in which at least one of these concepts is relaxed. 
We address a class of nonlocal hidden variable theories 
that could provide an explanation for all standard Bell experiments with two qubits. 
Nevertheless we demonstrate, both in theory and experiment, 
their variance with other quantum predictions and observed measurement data. 
Nonlocal models of the considered class have been introduced by Leggett \cite{LEGGETT_NONLOCAL}.
He also has derived an inequality valid for such nonlocal theories. 
We extend Leggett's approach to apply to real experimental situations 
in such a way that it also allows for a simultaneous test of all 
local hidden variable models, 
i.e.\ the measurement data can neither be explained by 
a local realistic model nor by the given class of nonlocal models.

We focus this description 
on the polarization degree of freedom of photons. 
The theories have the following underlying assumptions: 
\begin{itemize}
\item \emph{realism} \\
Measurement outcomes are determined by pre-existing properties of the particles, independent of the measurement.

\item \emph{polarized photons}\\
Each photon separately contributes to a subensemble of experimental runs
in which the average value measured using ``polarization analyzer'' fulfils Malus' law
(this defines subensembles with definite polarization).
Different photons can contribute to different subensembles.
Finally, the expectation values actually
observed are a statistical mixture 
over subensembles with definite polarization.
\end{itemize}
A general framework of such models is the following. 
Due to the realism assumption an individual binary measurement outcome, $A = \pm 1$, 
for a polarization measurement along direction $\vec a$
(i.e.\ whether a single photon is 
transmitted or absorbed by a polarizer set at a specific angle) 
is predetermined.\footnote{All polarizations and measurement directions are represented as vectors on the Poincar{\' e} sphere.} 
One can parameterize it with hidden variables
carried by the particle.
We distinguish one of them,
a three-dimensional vector $\vec u$,
which describes to which polarization subensemble
the photon belongs.
Additionally, the outcome can depend on some other nonlocal parameters $\eta$
(e.g.\ measurement settings in space-like separated regions). 
Finally, $A = A(\lambda, \vec u, \vec a, \eta)$.
Particles with the same $\vec u$ but different $\lambda$ 
build up subensembles ``of definite polarization''. 
The expectation value 
$\overline{A}(\vec u, \vec a)$,
obtained by averaging over hidden variables $\lambda$ within the subensemble, 
is assumed to fulfill Malus' law:
\begin{equation}
\overline{A}(\vec u,\vec a) = \int d \lambda \rho_{\vec u}(\lambda) A(\lambda, \vec u, \vec a, \eta) = \vec u \cdot \vec a,
\end{equation}
where $\rho_{\vec u}(\lambda)$ describes the distribution of $\lambda$ for a given $\vec u$.
The measured expectation value 
for a general source of photons is given by 
averaging over the distribution of polarizations, $F(\vec u)$:
\begin{equation}
\langle A \rangle = \int d \vec u F(\vec u) \overline{A}(\vec u, \vec a).
\end{equation}

Consider a source which emits pairs of photons with well-defined polarizations $\vec u$ and $\vec v$.
The local polarization measurement outcomes, $A$ and $B$, are 
fully determined by the polarization vector, 
by an additional set of hidden variables $\lambda$ 
specific to the source, 
and by any set of parameters $\eta$ outside the source 
(e.g.\ the settings $\vec a$ and $\vec b$ of both measurement apparatuses). 
Each emitted pair is fully defined by the subensemble distribution $\rho_{\vec u, \vec v}(\lambda)$. 
According to the assumption of polarized photons, 
the local averages of measurements \emph{within the subensembles} satisfy:
\begin{eqnarray}
\overline{A}(\vec u, \vec a) &=& \int d \lambda \rho_{\vec u, \vec v}(\lambda) A(\lambda, \vec u, \vec a, \vec b) = \vec u \cdot \vec a, \nonumber\\
\overline{B}(\vec v, \vec b) &=& \int d \lambda \rho_{\vec u, \vec v}(\lambda) B(\lambda, \vec v, \vec b, \vec a) = \vec v \cdot \vec b.
\label{MALUS}
\end{eqnarray}
For reasons of clarity, 
we have chosen an explicit nonlocal dependence of the outcomes on the settings $\vec a$ and $\vec b$ 
of the measurement devices. 
However, this is just an example of a possible nonlocal 
dependence and one can choose any other set out of $\eta$. 
It is important to note that the validity of Malus' law 
imposes the non-signalling condition on the investigated nonlocal model.
Since the local expectation values depend only on local parameters,
changing the accessible parameters in one lab does not influence
statistics in the other lab.
The correlation function of measurement results for 
a source emitting well-polarized photons 
is defined as the average of the products of the local measurement outcomes:
\begin{equation}
\overline{AB}(\vec u, \vec a, \vec v,\vec b) = \int d \lambda 
\rho_{\vec u, \vec v}(\lambda) A(\lambda, \vec u, \vec a, \vec b) B(\lambda, \vec v, \vec b, \vec a).
\end{equation}
For a general source producing mixtures of polarized photons 
the observable correlations are averaged over a distribution of the polarizations $F(\vec u, \vec v)$, 
and the general correlation function $E_{\vec a \vec b}$ is given by:
\begin{equation}
E_{\vec a \vec b} \equiv \langle AB \rangle = 
\int d \vec u d \vec v F(\vec u, \vec v) \overline{AB}(\vec u, \vec a, \vec v,\vec b).
\end{equation}
It is a crucial trait of this model 
that predictions for the subensembles 
of definite polarization agree with Malus' law. 
It is clear that other classes of nonlocal theories 
may exist that do not have this property 
when reproducing entangled states 
and are fully compliant with all quantum mechanical predictions. 
For example, in Bohm's theory \cite{BOHM1,BOHM2}
realistic "spin vectors" of individual particles are strictly zero
just after emission from the source,
clearly violating the assumption of definite polarization.
Holland makes the following comment
concerning bohmian spin vectors of individual particles \cite{HOLLAND_BOOK}:
\begin{quote}
The initial conditions 
[$\vec v_1 = \vec v_2 = \vec s_1 = \vec s_2 = 0$, where $\vec v_n$
describe velocities of the particles and $\vec s_n$ their spins]\footnote{Present author's comments are given in square brackets.}
provide a good example of how analogous quantities in the one- and two-body theories
have quite different properties.
The spin vectors are strictly zero, something that is not possible in the one-body case.
Notice in particular that the spins are not determined by either of the addends in [the singlet state],
i.e., the particles are not in an initial state in which the spin of one is up (down)
while the other is down (up), as one might expect in the analogous classical case.
The usual informal way of speaking about the singlet state in terms of 'antiparallel spins'
is, according to this model, misleading.
\end{quote}

\subsubsection{Explicit nonlocal model}

We construct an explicit nonlocal model compliant with the
class of hidden variables considered.
One deals with well-polarized photons
which carry predetermined outcomes of all possible measurements.
The model perfectly simulates all quantum
mechanical predictions for measurements performed in an 
arbitrary 
plane of the local Poincar\`e sphere: we model the correlation function
$E_{\vec a \vec b}^{QM} = -\vec a \cdot \vec b$, for which all
local averages $\langle A \rangle$ and $\langle B \rangle$
vanish. In particular, in this way one explains a violation of any CHSH
inequality. The model also rebuilds all perfect
correlations of the singlet state obtained for all measurements
performed along the
same directions.
The validity condition for the model is derived,
which expresses the conflict between
modelling all quantum predictions and
satisfying Malus' law on the level of subensembles
with definite polarizations.

Let us start with a source that emits photons with a well-defined
polarization. Polarisation $\vec u$ is sent to Alice and $\vec v$ to
Bob. Alice sets her measuring device to $\vec a$ and Bob to $\vec
b$. The random hidden real number $\lambda \in [0,1]$ is carried by both
particles and predetermines the individual measurement result as
follows:
\begin{eqnarray}
A \equiv A(\vec a, \vec u, \lambda) &=& \Big\{
\begin{array}{ccc}
+1 & \textrm{ for } & \lambda \in [0,\lambda_A], \\
-1 & \textrm{ for } & \lambda \in (\lambda_A,1],
\end{array}
\textrm{ with } \lambda_A = \frac{1}{2} (1 + \vec u \cdot \vec a),
\label{A_DEF}
\end{eqnarray}
where $A$ is the outcome of Alice.
This means, whenever $\lambda\leq\lambda_A$ the result of the
measurement $A$ is $+1$, and for $\lambda >\lambda_A$ the
result is $-1$. Note that the measurement settings only
enter in $\lambda_A$ and are hence independent of the hidden
variable $\lambda$ of the source. The outcome of Bob is given by
\begin{eqnarray*}
B & \equiv & B(\vec a, \vec b, \vec u, \vec v, \lambda) = \Big\{
\begin{array}{ccc}
+1 & \textrm{ for } & \lambda \in [x_1,x_2], \\
-1 & \textrm{ for } & \lambda \in [0,x_1) \cup (x_2,1],
\end{array} \\
&& \textrm{ with } x_1,x_2 \in [0,1] \textrm{ arbitrary but } x_2-x_1 =
\frac{1}{2}(1+\vec v \cdot \vec b).
\end{eqnarray*}
All nonlocal dependencies are put on the side of Bob. His measuring
device has the information about the setting of Alice, $\vec a$, and
her polarization $\vec u$. The requirement of the nonlocal models
discussed here is that the local averages performed on the
subensemble of definite (but arbitrary) polarizations $\vec u$ and
$\vec v$ obey Malus' law, i.e.\ $\overline{A_{\vec u}} = \vec u \cdot
\vec a$ for Alice, and $\overline{B_{\vec v}} = \vec v \cdot \vec b$
for Bob. Indeed, a straight-forward calculation shows that this
requirement is fulfilled for both Alice and Bob:
\begin{eqnarray*}
\overline{A}(\vec u, \vec a) & = & \int_0^{\lambda_A} d \lambda -
\int_{\lambda_A}^1 d \lambda = 2 \lambda_A - 1 = \vec u \cdot \vec a, \\
\overline{B}(\vec v, \vec b) & = & \int_{x_1}^{x_2} d \lambda - \int_{0}^{x_1} d \lambda - \int_{x_2}^{1} d \lambda
= 2 (x_2-x_1) - 1 = \vec v \cdot \vec b.
\end{eqnarray*}
In order to get the correct formula for correlated
counts one can fix the values of $x_1$ and $x_2$ in the following
way:
\begin{eqnarray}
x_1 &=& \frac{1}{4}[1 + \vec u \cdot \vec a - \vec v \cdot \vec b + \vec a \cdot \vec b], \nonumber \\
x_2 &=& \frac{1}{4}[3 + \vec u \cdot \vec a + \vec v \cdot \vec b
+ \vec a \cdot \vec b]. \label{xes}
\end{eqnarray}
With these definitions and whenever $x_1 \le \lambda_A \le x_2$ the
expectation value for measurements on the subensembles reproduces
quantum correlations:
\begin{equation}
\overline{AB}(\vec u, \vec a, \vec v, \vec b) = - \int_0^{x_1} d \lambda +
\int_{x_1}^{\lambda_A} d \lambda - \int_{\lambda_A}^{x_2} d
\lambda + \int_{x_2}^{1} d \lambda = 2(\lambda_A - x_1 - x_2) +
1=-\vec a \cdot \vec b.
\end{equation}
Therefore, in the next step, one must find the conditions for which
both $x_1$ and $x_2$ take values from $[0,1]$ and $x_1 \le \lambda_A \le x_2$.

Using the definitions (\ref{xes}) one finds that the first
condition is equivalent to a set of four inequalities:
\begin{eqnarray}
-1 + \vec v \cdot \vec b \quad \le & \vec a \cdot \vec b + \vec u \cdot \vec a & \le \quad 3 + \vec v \cdot \vec b, \nonumber \\
-3 - \vec v \cdot \vec b \quad \le & \vec a \cdot \vec b + \vec u \cdot \vec a & \le \quad 1 - \vec v \cdot \vec b.
\end{eqnarray}
Note that the upper bound, $3 + \vec v \cdot \vec b$,
cannot be exceeded by the middle term,
as well as the lower bound, $-3 - \vec v \cdot \vec b$.
Thus, this set of four inequalities is equivalent
to a single one:
\begin{equation}
|\vec a \cdot \vec b + \vec u \cdot \vec a|  \le 1 - \vec v \cdot \vec b.
\label{COMPLEMENTARITY1}
\end{equation}
Similarly, the second condition can be reexpressed as:
\begin{equation}
|\vec a \cdot \vec b - \vec u \cdot \vec a|  \le 1 + \vec v \cdot \vec b.
\label{COMPLEMENTARITY2}
\end{equation}
Finally, the validity condition for the model
is a conjunction of (\ref{COMPLEMENTARITY1}) and (\ref{COMPLEMENTARITY2}):
\begin{equation}
|\vec a \cdot \vec b \pm \vec u \cdot \vec a|  \le 1 \mp \vec v \cdot \vec b.
\label{COMPLEMENTARITY}
\end{equation}
If this relation  is not satisfied the model does not recover
quantum correlations. Either it becomes inconsistent since $x_1$ or
$x_2$ leave their range or the necessary relation $x_1 \le \lambda_A
\le x_2$ is not satisfied, or both. This is the origin of the
incompatibility with general quantum predictions. Nevertheless the
model can explain all perfect correlations and the violation of CHSH inequalities.

Consider a source producing pairs with the following property:
whenever polarization $\vec u$ is sent to Alice polarization $\vec v
= - \vec u$ is sent to Bob. Both parties locally observe random
polarizations. For Alice, the local average over different
polarizations yields
\begin{equation}
\langle A \rangle = \frac{1}{2} \overline{A}(\vec u, \vec a) +  
\frac{1}{2}\overline{A}(-\vec u, \vec a) = \frac{1}{2}\vec u \cdot \vec a - \frac{1}{2}\vec u \cdot \vec a = 0,
\end{equation}
as it should be for the singlet state. The same result holds for
Bob. In this way, we have reproduced the randomness of local
measurement outcomes, typical for measurements on entangled
states.

With the same source one can explain perfect correlations for
measurements along the same basis, i.e.\ $\vec b = \pm \vec a$. To
see how the model works take $\vec v = - \vec u$ and $\vec b = \vec a$. 
In this case $x_1 = \frac{1}{2}[1 + \vec u \cdot \vec a] =
\lambda_A$ and $x_2 = 1$. As it should be, Bob's outcomes are always
opposite to Alice's:
\begin{eqnarray}
B \equiv B(\vec a, \vec a, \vec u, - \vec u, \lambda) &=& \Big\{
\begin{array}{ccc}
+1 & \textrm{ for } & \lambda \in [\lambda_A,1], \\
-1 & \textrm{ for } & \lambda \in [0,\lambda_A).
\end{array}
\end{eqnarray}
If in the same subensemble one takes $\vec b = - \vec a$, one obtains
$x_1 = 0$ and $x_2 = \lambda_A$, which results in $B=A$, again in
full agreement with quantum mechanics. Note that for these
measurement settings condition (\ref{COMPLEMENTARITY}) imposes no additional restrictions. 
For example, if $\vec u$ is sent to
Alice and $\vec b = -\vec a$ one obtains $|-1 \pm \vec u \cdot \vec
a| \le 1 \mp \vec u \cdot \vec a$, which always holds. The same
argument applies to the other subensemble and other measurement
possibilities $\vec b = \pm \vec a$.

Finally, the full predictions of quantum theory are recovered if
Alice and Bob restrict their measurements to lie in the planes
orthogonal to the vectors $\vec u$ and $\vec v$, respectively, i.e.\ 
$\vec u \cdot \vec a = \vec v \cdot \vec b = 0$. In this case,
condition (\ref{COMPLEMENTARITY}) is satisfied for all the settings,
as $|\vec a \cdot \vec b| \le 1$. In general, if condition
(\ref{COMPLEMENTARITY}) is satisfied, i.e.\ for a consistent set
of parameters, our model reproduces quantum correlations since
they are already reproduced in every subensemble and hence
averaging over different polarizations does not change this result:
\begin{equation}
\langle AB \rangle = \overline{AB}(\vec u, \vec a, -\vec u, \vec b) =
\overline{AB}(- \vec u, \vec a, \vec u, \vec b) = - \vec a \cdot \vec b.
\end{equation}
Therefore, every experimental violation of any CHSH
inequality can be explained by the presented nonlocal model.

\subsubsection{Incompatibility}

The theories described are incompatible with quantum theory.
All of them satisfy certain inequality
which is violated by suitable quantum predictions.
A detailed derivation of the inequality will now be presented.
It is an extension of the work by Leggett~\cite{LEGGETT_NONLOCAL}.

For any dichotomic measurement results, $A=\pm 1$ and $B=\pm 1$,
the following identity holds\ \cite{LEGGETT_NONLOCAL}:
\begin{equation}
-1 + |A+B| = A B = 1 - |A - B|. \label{NL_IDENTITY}
\end{equation}
If the signs of $A$ and $B$ are the same $|A+B| = 2$ and $|A-B|=0$. 
If $A = -B$ then $|A+B| = 0$ and $|A-B|=2$. Any
kind of nonlocal dependencies is allowed. Taking the average
over the subensemble with definite polarizations one obtains:
\begin{equation}
-1 + \int d \lambda \rho_{\vec u, \vec v}(\lambda) |A+B| = \int d
\lambda \rho_{\vec u, \vec v}(\lambda) AB = 1 - \int d \lambda
\rho_{\vec u, \vec v}(\lambda)|A-B|,
\end{equation}
which in an abbreviated notation, with the averages denoted
by bars, is
\begin{equation}
-1 + \overline{|A + B|} = \overline{AB} = 1 - \overline{|A - B|}.
\end{equation}
Since the average of the modulus is greater or equal to the
modulus of the averages one gets the set of inequalities
\begin{equation}
-1 + |\overline A + \overline B| \le \overline{AB} \le 1 -
|\overline A - \overline B|. \label{SET_OF_INEQ}
\end{equation}
From now on only the upper bound will be considered. 
However, all the steps apply to the lower bound as well. 
We will discuss the point in which the lower bound 
becomes equal to the negative upper
bound and the modulus appears in the inequality.

With the assumption that photons with well defined polarization
obey Malus' law:
\begin{eqnarray}
\overline A &=& \vec u \cdot \vec a, \nonumber \\
\overline B &=& \vec v \cdot \vec b, \label{LOCAL_AVERAGES}
\end{eqnarray}
the upper bound of Eq.\ (\ref{SET_OF_INEQ}) becomes:
\begin{equation}
\overline{AB} \le 1 - |\vec u \cdot \vec a_k - \vec v \cdot \vec
b_l|,
\end{equation}
where $\vec a_k$ and $\vec b_l$ are unit vectors associated with
the measurement settings of Alice and Bob, respectively.

Taking the average over arbitrary polarizations one obtains:
\begin{equation}
E_{kl} \le 1 - \int_0^\pi \sin \theta_u d \theta_u \int_0^{2 \pi}
d \phi_u \int_0^\pi \sin \theta_v d \theta_v \int_0^{2 \pi} d
\phi_v F(\theta_u,\phi_u,\theta_v,\phi_v)
 |\vec u \cdot \vec a_k - \vec v \cdot \vec b_l|,
\end{equation}
where all the vectors and the weight function
$F(\theta_u,\phi_u,\theta_v,\phi_v)$ are written in the spherical
coordinate system. 
We stress that the correlation function
$E_{kl}$ can be
 experimentally measured.
Let us denote the plane spanned by $\vec a_k$ and $\vec b_l$
as the $xy$ plane and the angle relative to the
$\hat z$ axis as $\theta$. 
In this coordinate system the vectors $\vec a_k$ and $\vec b_l$
are parameterized by the angles within
the $xy$ plane, $\phi_{a_k}$ and $\phi_{b_l}$,
respectively.
The scalar products read:
\begin{eqnarray}
\vec u \cdot \vec a_k &=& \sin \theta_u \cos(\phi_{a_k} - \phi_u), \\
\vec v \cdot \vec b_l &=& \sin \theta_v \cos(\phi_{b_l} - \phi_v),
\end{eqnarray}
and the inequality transforms to:
\begin{eqnarray*}
E_{kl} & \le & 1 - \int_0^\pi \sin \theta_u d \theta_u \int_0^{2 \pi}
d \phi_u \int_0^\pi \sin \theta_v d \theta_v \int_0^{2 \pi} d
\phi_v F(\theta_u,\phi_u,\theta_v,\phi_v)  \\
&& \times
|\sin \theta_u \cos(\phi_{a_k} - \phi_u) 
- \sin \theta_v \cos(\phi_{b_l} - \phi_v)|.
\end{eqnarray*}
The sines $\sin \theta_u$ and $\sin \theta_v$ describe the
magnitude of the projection of $\vec u$ and $\vec v$ onto the
$xy$ plane, respectively. These magnitudes can always be
decomposed into a sum and difference of the other two real
numbers:
\begin{eqnarray}
\sin \theta_u &=& n_1 + n_2, \\
\sin \theta_v &=& n_1 - n_2.
\end{eqnarray}
Note that both $n_1$ and $n_2$ are functions of the $\theta$s
only. We insert this decomposition into the last inequality. 
The terms multiplied by $n_1$ and $n_2$ respectively read:
\begin{eqnarray*}
\cos(\phi_{a_k} - \phi_u) - \cos(\phi_{b_l} - \phi_v) & = & 2 \sin \frac{\phi_{a_k} + \phi_{b_l} - (\phi_u + \phi_v)}{2} \sin\frac{-(\phi_{a_k} - \phi_{b_l})+\phi_u - \phi_v}{2}, \\
\cos(\phi_{a_k} - \phi_u) +\cos(\phi_{b_l} - \phi_v)  & = & 2 \cos \frac{\phi_{a_k} + \phi_{b_l} - (\phi_u + \phi_v)}{2}
\cos \frac{\phi_{a_k} - \phi_{b_l} - (\phi_u - \phi_v)}{2}.
\end{eqnarray*}
One can make
the following substitution for the measurement angles:
\begin{eqnarray}
\xi = \frac{\phi_{a_k} + \phi_{b_l}}{2}, \qquad && \varphi =
\phi_{a_k} - \phi_{b_l},
\end{eqnarray}
and change the integration variables $\phi_u,\phi_v$ to $\psi,
\chi$:
\begin{eqnarray}
\psi = \frac{\phi_u + \phi_v}{2}, \qquad && \chi = \phi_u -
\phi_v.
\end{eqnarray}
The absolute value of the Jacobian of this transformation equals
one, thus it does not introduce any new factors to the integral.
Within these new variables one arrives at:
\begin{eqnarray*}
E_{kl}(\xi,\varphi) & \le & 1- 2\int_0^\pi \sin \theta_u d \theta_u
\int_0^{2 \pi} d \psi \int_0^\pi \sin \theta_v d \theta_v
\int_0^{2 \pi} d \chi F(\theta_u,\theta_v,\psi, \chi) \\
&& \times |n_2
\cos\frac{\varphi - \chi}{2} \cos (\xi-\psi) - n_1 \sin
\frac{\varphi - \chi}{2} \sin (\xi-\psi)|,
\end{eqnarray*}
where in the correlation function
$E_{kl}(\xi,\varphi)$ we explicitly
state the angles it is dependent on.
The expression
within the modulus is a linear combination of two harmonic
functions of $\xi - \psi$, and therefore it is a harmonic function itself.
Its amplitude reads $\sqrt{n_2^2 \cos^2(\frac{\varphi - \chi}{2})
+ n_1^2 \sin^2(\frac{\varphi - \chi}{2})}$, and the phase is some
fixed real number $\alpha$:
\begin{eqnarray}
E_{kl}(\xi, \varphi) \le 1 - 2
\int_0^\pi \sin \theta_u d \theta_u \int_0^{2 \pi} d \psi
\int_0^\pi \sin \theta_v d \theta_v \int_0^{2 \pi} d \chi
F(\theta_u,\theta_v,\psi, \chi) \nonumber \\
\times \sqrt{n_2^2
\cos^2(\frac{\varphi - \chi}{2}) + n_1^2 \sin^2(\frac{\varphi -
\chi}{2})} |\cos(\xi - \psi + \alpha)|.
\label{XI_PHI_CORR}
\end{eqnarray}
In the next step one averages both sides of this inequality
over the
\emph{measurement angle} $\xi = \frac{\phi_{a_k} + \phi_{b_l}}{2}$.
This means an integration over $\xi \in [0,2\pi)$
and multiplying by $\frac{1}{2\pi}$.
Experimentally one should perform a series of measurements in which the
angle between the observables is kept constant, $\varphi =
\textrm{const}$, and the two measurement vectors are rotated in
their plane. 
The integral of the $\xi$-dependent part 
of the right-hand side of (\ref{XI_PHI_CORR}) reads:
\begin{equation}
\int_0^{2 \pi} \frac{d \xi}{2 \pi} |\cos(\xi - \psi + \alpha)| =
\frac{2}{\pi}.
\end{equation}
If one denotes the average of the correlation function over the angle
$\xi$ as:
\begin{equation}
\overline{E}_{kl}(\varphi) 
\equiv 
\int_0^{2 \pi} \frac{d \xi}{2 \pi} E_{kl}(\xi, \varphi),
\end{equation}
one can write (\ref{XI_PHI_CORR}) in the form:
\begin{eqnarray*}
\overline{E}_{kl}(\varphi) & \le & 1 - \frac{4}{\pi} 
\int_0^\pi \sin \theta_u d \theta_u \int_0^{2 \pi} d \psi
\int_0^\pi \sin \theta_v d \theta_v \int_0^{2 \pi} d
\chi F(\theta_u,\theta_v,\psi,\chi) \\
&& \times
\sqrt{n_2^2 \cos^2 \frac{\varphi -
\chi}{2} + n_1^2 \sin^2 \frac{\varphi - \chi}{2}}.
\end{eqnarray*}
Further, the integrand is no longer dependent on $\psi$,
and the $\psi$ integration results in the marginal
weight function:
\begin{equation}
F(\theta_u,\theta_v,\chi) = \int_0^{2 \pi} d \psi F(\theta_u,\theta_v,\psi,\chi).
\end{equation}
The last inequality can thus be slightly simplified to:
\begin{eqnarray*}
\overline{E}_{kl}(\varphi) & \le & 1 - \frac{4}{\pi} 
\int_0^\pi \sin \theta_u d \theta_u 
\int_0^\pi \sin \theta_v d \theta_v \int_0^{2 \pi} d
\chi F(\theta_u,\theta_v,\chi)
\sqrt{n_2^2 \cos^2 \frac{\varphi -
\chi}{2} + n_1^2 \sin^2 \frac{\varphi - \chi}{2}}.
\end{eqnarray*}
This inequality is valid for any choice of observables in the
plane defined by $\vec a_k$ and $\vec b_l$. One can introduce two
new observable vectors in this plane and write the inequality for
the averaged correlation function of these new observables,
$\overline{E}_{k'l'}(\varphi')$. 
Let us consider the sum of these two inequalities:
\begin{eqnarray*}
\overline{E}_{kl}(\varphi) + \overline{E}_{k'l'}(\varphi')\le 2- \frac{4}{\pi}
\int_0^\pi \sin \theta_u d \theta_u \int_0^\pi \sin \theta_v d
\theta_v \int_0^{2 \pi} d \chi
F(\theta_u,\theta_v,\chi) \\
\times \left( \sqrt{n_2^2 \cos^2 \frac{\varphi - \chi}{2} + n_1^2
\sin^2 \frac{\varphi - \chi}{2}} +\sqrt{n_2^2 \cos^2
\frac{\varphi' - \chi}{2} + n_1^2 \sin^2 \frac{\varphi' -
\chi}{2}} \right).
\end{eqnarray*}
One can use the triangle inequality:
\begin{eqnarray}
||\vec x + \vec y|| & \le & ||\vec x || + || \vec y ||, \\
\sqrt{(x_1+y_1)^2 + (x_2 + y_2)^2} & \le & \sqrt{x_1^2 + x_2^2} + \sqrt{y_1^2 + y_2^2}
\label{TRIANGLE_TRICK}
\end{eqnarray}
for the two-dimensional vectors $\vec x = (x_1,x_2)$
and $\vec y = (y_1,y_2)$, with components defined by:
\begin{eqnarray*}
x_1 = |n_2 \cos\frac{\varphi - \chi}{2}|, & \quad & 
y_1 = |n_2 \cos\frac{\varphi' - \chi}{2}|, \\
x_2 = |n_1 \sin\frac{\varphi - \chi}{2}|, & \quad & 
y_2 = |n_1 \sin\frac{\varphi' - \chi}{2}|.
\end{eqnarray*}
This implies that integrand is bounded from below by:
\begin{eqnarray*}
\sqrt{n_2^2 \cos^2 \frac{\varphi - \chi}{2} + n_1^2
\sin^2 \frac{\varphi - \chi}{2}} +\sqrt{n_2^2 \cos^2
\frac{\varphi' - \chi}{2} + n_1^2 \sin^2 \frac{\varphi' -
\chi}{2}} \\
\ge
\sqrt{n_2^2 \Big( |\cos\frac{\varphi - \chi}{2}| + |\cos\frac{\varphi' - \chi}{2}| \Big)^2
+ n_1^2 \Big( |\sin\frac{\varphi - \chi}{2}| + |\sin \frac{\varphi' - \chi}{2}| \Big)^2 }
\end{eqnarray*}
The bound can be simplified by noting that:
\begin{eqnarray}
|\cos(\frac{\varphi - \chi}{2})| + |\cos(\frac{\varphi' - \chi}{2})| & \ge & |\sin\frac{\varphi-\varphi'}{2}|, \nonumber \\
|\sin(\frac{\varphi - \chi}{2})| + |\sin(\frac{\varphi' - \chi}{2})| & \ge & |\sin\frac{\varphi-\varphi'}{2}|,
\label{NL_APPROX_BOUND}
\end{eqnarray}
which follows
after using the formula for the sine of the difference angle,
$\frac{\varphi-\varphi'}{2} = \frac{\varphi-\chi}{2}-\frac{\varphi'-\chi}{2}$,
to the right-hand side of these inequalities:
\begin{eqnarray*}
|\sin\frac{\varphi-\varphi'}{2}| & = &
|\sin\frac{\varphi - \chi}{2} \cos\frac{\varphi' - \chi}{2} - 
\cos\frac{\varphi - \chi}{2} \sin\frac{\varphi' - \chi}{2}| \\
& \le & |\sin\frac{\varphi - \chi}{2}| |\cos\frac{\varphi' - \chi}{2}| + 
|\cos\frac{\varphi - \chi}{2}| |\sin\frac{\varphi' - \chi}{2}|.
\end{eqnarray*}
After these estimations the lower bound equals the negative upper
bound, and one can shortly write the modulus:
\begin{eqnarray*}
| \overline{E}_{kl}(\varphi) +  \overline{E}_{k'l'}(\varphi') | & \le &  2 - \frac{4}{\pi}
|\sin(\frac{\varphi - \varphi'}{2})| \int_0^\pi \sin \theta_u d \theta_u \int_0^\pi \sin \theta_v d \theta_v \\
&& \times \int_0^{2 \pi} d \chi
F(\theta_u,\theta_v,\chi) \sqrt{n_2^2 + n_1^2} .
\end{eqnarray*}
Recall that the numbers $n_1$ and $n_2$ are functions of
$\theta_{u}$ and $\theta_{v}$ only. Thus one can perform the
integration over $\chi$,
which results in yet another marginal weight function:
\begin{equation}
F(\theta_u,\theta_v) =
\int_0^{2\pi} d \chi F(\theta_u,\theta_v,\chi).
\end{equation}
Going back to the magnitudes:
\begin{eqnarray}
| \overline{E}_{kl}(\varphi) +  \overline{E}_{k'l'}(\varphi') | & \le & 2 - \frac{2
\sqrt{2}}{\pi} |\sin(\frac{\varphi - \varphi'}{2})| \int_0^\pi
\sin \theta_u d \theta_u \int_0^\pi \sin \theta_v d \theta_v \nonumber \\
&& \times F(\theta_u,\theta_v) \sqrt{\sin^2 \theta_u + \sin^2 \theta_v}.
\label{XY}
\end{eqnarray}
This inequality is valid for any set of four observables in one
plane and for any choice of the plane. The bound involves only
the angles of vectors $\vec u$ and $\vec v$ relative to the axis
orthogonal to the plane of observables. For a plane orthogonal to
the initial one, e.g.\ the $xz$ plane, the inequality therefore
reads:
\begin{eqnarray}
|\overline{E}_{mn}(\varphi_y) + \overline{E}_{m'n'}(\varphi'_y)| & \le & 2 - \frac{2
\sqrt{2}}{\pi} |\sin \frac{\varphi_y-\varphi_y'}{2}| \int_0^\pi
\sin \theta_u d \theta_u \int_0^\pi \sin \theta_v d \theta_v \nonumber \\
&& \times F(\theta_u,\theta_v) \sqrt{\sin^2 \theta'_u + \sin^2 \theta'_v},
\label{XZ}
\end{eqnarray}
where the primed angles $\theta'$ under the square root are now
relative to the $y$ axis (the distribution of vectors is still
the same), and $\varphi_y $ is the angle between in the $xz$ plane. We add the inequalities for orthogonal observation
planes, (\ref{XY}) and (\ref{XZ}), choose $\varphi' = \varphi_y' =
0$ and $\varphi = \varphi_z = \varphi_y$ to obtain:
\begin{eqnarray*}
|\overline{E}_{kl}(\varphi_z) + \overline{E}_{k'k'}(0)| + 
|\overline{E}_{mn}(\varphi_y) + \overline{E}_{m'm'}(0)| \le 4 - \frac{2\sqrt{2}}{\pi}|\sin\frac{\varphi}{2}| \\
\times \int_0^\pi \sin \theta_u d \theta_u
\int_0^\pi \sin \theta_v d \theta_v \nonumber 
F(\theta_u,\theta_v) \left(\sqrt{\sin^2 \theta_u + \sin^2
\theta_v} + \sqrt{\sin^2 \theta'_u + \sin^2 \theta'_v} \right).
\end{eqnarray*}
On the left-hand side we use the notation $\varphi_z$ and
$\varphi_y$ to stress that the averaged correlations in the
moduli are valid for observables from orthogonal planes. 
To the expression within the bracket
one can apply the trick
with the triangle inequality for two-dimensional vectors (\ref{TRIANGLE_TRICK}).
This time the components of vectors $\vec x$ are $\vec y$ read:
\begin{eqnarray}
x_1 = \sin \theta_u, & \quad & y_1 = \sin \theta'_u, \\
x_2 = \sin \theta_v, & \quad & y_1 = \sin \theta'_v.
\end{eqnarray}
The integrand is lower-bounded by
\begin{equation}
\sqrt{\sin^2 \theta_u + \sin^2
\theta_v} + \sqrt{\sin^2 \theta'_u + \sin^2 \theta'_v}
\ge
\sqrt{(\sin \theta_u + \sin \theta'_u)^2 + (\sin \theta_v + \sin \theta'_v)^2}.
\label{PROJ_BOUND_NL_DERIV}
\end{equation}
Let us consider the term involving vector $\vec u$ only.
Since both $0 \le \theta_u \le \pi$
and $0 \le \theta'_u \le \pi$ their sines 
are always non-negative.
This implies
\begin{equation}
(\sin \theta_u + \sin \theta'_u)^2 \ge \sin^2 \theta_u + \sin^2 \theta'_u.
\end{equation}
Recall that angles $\theta_u$ and $\theta'_u$ (of two spherical coordinate systems)
are relative to orthogonal Cartesian axes $z$ and $y$, respectively.
Thus, the vector $\vec u$ has the following
components in the Cartesian coordinate system:
\begin{equation}
\vec u = (\delta, \cos\theta'_u, \cos\theta_u),
\quad {\rm with} \quad
\delta^2 + \cos^2 \theta'_u + \cos^2 \theta_u = 1,
\end{equation}
The normalization implies that 
$\cos^2 \theta'_u + \cos^2 \theta_u \le 1$,
which is equivalent to:
\begin{equation}
\sin^2 \theta_u + \sin^2 \theta'_u \ge 1.
\end{equation}
The same steps obviously apply to vector $\vec v$
and one finds the bound of (\ref{PROJ_BOUND_NL_DERIV}) to be equal to:
\begin{equation}
\sqrt{\sin^2 \theta_u + \sin^2
\theta_v} + \sqrt{\sin^2 \theta'_u + \sin^2 \theta'_v} \ge \sqrt{2}.
\label{PROJECTIONS_BOUND}
\end{equation}
Since the $F(\theta_u,\theta_v)$ function is normalized the final
inequality reads:
\begin{equation}
S_{NLHV} \equiv |\overline{E}_{kl}(\varphi_z) + \overline{E}_{k'k'}(0)| 
+ | \overline{E}_{mn}(\varphi_y) + \overline{E}_{m'm'}(0)| 
\le 4 - \frac{4}{\pi}|\sin\frac{\varphi}{2}|.
\label{NL_INEQUALITY}
\end{equation}
To conclude, this inequality has to be satisfied by all the nonlocal theories in question.
The \emph{averaged} correlation functions 
for observables from orthogonal planes enter the inequality.
This contrasts the standard experimental configuration 
to test the CHSH inequality, which is maximally violated for settings in \emph{one} plane.

Quantum theory predicts violation of the inequality (\ref{NL_INEQUALITY}). 
Consider the polarization singlet state of two photons:
\begin{equation}
| \psi^- \rangle = \frac{1}{\sqrt{2}}
\Big[ | H \rangle_1 | V \rangle_2 - | V \rangle_1 | H \rangle_2 \Big],
\label{POL_SINGLET}
\end{equation}
where e.g.\ $| H \rangle_1$ denotes a horizontally polarized photon propagating towards detector ``1''. 
The quantum correlation function for the measurements $\vec a_k$ and $\vec b_l$ 
performed on the photons depends only on the relative angles between these vectors (Appendix A):
\begin{equation}
E_{\vec a \vec b} = - \vec a \cdot \vec b = - \cos \varphi.
\end{equation}
Thus the left hand side of (\ref{NL_INEQUALITY}), 
for quantum predictions, reads  $2|\cos \varphi+1|$. 
The maximal violation of inequality (\ref{NL_INEQUALITY}) is for $\varphi_{max} = 18.8^{\circ}$. 
For this difference angle the bound 
equals $3.792$ and the quantum value is $3.893$.

If we additionally assume that the correlations predicted by the nonlocal models
depend only on a difference angle between observables,
as it is predicted by quantum mechanics and can possibly be experimentally verified,
the averaged correlations $\overline{E}_{kl}$
are given by correlations for one pair of settings only, $E_{kl}$.
In this case the inequality (\ref{NL_INEQUALITY})
involves two settings for Alice and three settings for Bob.
The possible correlations obtained 
with the chosen settings could still have a local realistic model.\footnote{Note, however, that such local realistic theories 
must not be constrained by Malus' law
-- all local realistic theories with additional constraints 
are special cases of nonlocal theories with the same constraints.}
In order to avoid that, 
we have to exclude both local and 
nonlocal hidden-variable theories. 
The violation of a CHSH inequality invalidates all local realistic models
irrespectively of the number of alternative local settings.
If one takes:
\begin{equation}
S_{CHSH} \equiv |E_{11}+E_{12}-E_{21}+E_{22}| \le 2,
\label{NL_CHSH}
\end{equation}
and the settings 
used to maximally violate the inequality (\ref{NL_INEQUALITY})
in our experiment
(see the spheres in Fig. \ref{NL_SETUP}):
\begin{eqnarray*}
\vec a_1 = (1,0,0), & \quad & \vec a_2 = (0,0,1), \\
\vec b_1 = (\cos \varphi_{max}, 0,-\sin \varphi_{max}), & \quad &
\vec b_2 = (0,\sin\varphi_{max},\cos\varphi_{max}), \quad
\vec b_3 = \vec a_2,
\end{eqnarray*}
the quantum value of the left-hand side is $2.215$.

The $2 \times 3$ scenario considered
is the simplest one in which one can
simultaneously disprove possibility of any local and the nonocal
hidden variable description.
We show that a violation
of local realism with two measurement settings per side
which are suitable for a violation of the nonlocal inequality,
is impossible.
We even make a bit more general proof
and allow settings from orthogonal planes
to be freely rotated.
In the derivation of the nonlocal inequality
we have assumed settings in orthogonal planes
have the same difference angle.
However, even if one generalizes the nonlocal inequality
to the case of freely rotated observables in orthogonal
planes, this will not help
to simultaneously disprove local realism.

Suppose Alice and Bob both choose between two settings.
The nonlocal inequality requires
Alice and Bob to measure along the same direction.
Say that $\vec a_1 = \vec b_1 = \hat x$.
The remaining settings have to lie in orthogonal planes.
Take $\vec a_2$ is rotated in $xz$ plane
and $\vec b_2$ in $xy$ plane.
In the usual spherical coordinate system
the observable vectors have the following components:
\begin{equation}
\vec a_1 = \vec b_1 = (1,0,0), \qquad
\vec a_2 = (\sin\theta,0,\cos\theta), \qquad
\vec b_2 = (\cos\phi, \sin \phi, 0).
\end{equation}
If the singlet state
is measured with these settings,
the appropriate correlations read:
\begin{equation}
E_{11} = -1, \qquad
E_{12} = -\sin\theta, \qquad
E_{21} = -\cos\phi, \qquad
E_{22} = -\sin\theta \cos\phi.
\end{equation}
Consider the CHSH inequality in the form
\begin{equation}
|E_{11} + E_{12} + E_{21} - E_{22}| \le 2.
\end{equation}
Under the chosen settings the left-hand side equals
\begin{equation}
|E_{11} + E_{12} + E_{21} - E_{22}| =
|(1+\cos\phi) + \sin\theta (1 - \cos\phi)|.
\end{equation}
Using the following trigonometric identities
\begin{equation}
1+\cos\phi = 2 \cos^2 \frac{\phi}{2}, \qquad
1-\cos\phi = 2 \sin^2 \frac{\phi}{2},
\end{equation}
and the Pythagorean trigonometric identity, 
$\sin^2 \frac{\phi}{2} + \cos^2 \frac{\phi}{2} = 1$,
one can rewrite the CHSH expression to the form:
\begin{equation}
|E_{11} + E_{12} + E_{21} - E_{22}|
= 2 |1 + \sin^2 \frac{\phi}{2}(\sin\theta - 1)|.
\end{equation}
Since $-2 \le \sin\theta - 1 \le 0$, the above
expression cannot exceed the bound of two:
\begin{equation}
2 |1 + \sin^2 \frac{\phi}{2}(\sin\theta - 1)| \le 2,
\end{equation}
i.e.\ local realism cannot be violated.

Similar techniques disprove the violation of other CHSH inequalities.
For example, $|E_{11} + E_{12} - E_{21} + E_{22}|$
can be written as $2|\sin^2 \frac{\phi}{2} + \sin\theta \cos^2 \frac{\phi}{2}|$,
which transforms to $2|1+\cos^2 \frac{\phi}{2}(\sin\theta-1)|$.
For the same reason as before one cannot expect a violation of local realism.

\subsubsection{Experiment}

The correlation function determined in an actual experiment 
is typically reduced by a visibility factor, $V$, to:
\begin{equation}
E^{exp} = - V \cos \varphi,
\end{equation}
due to noise and imperfections. 
Thus to observe in the experiment violation 
of inequality (\ref{NL_INEQUALITY}) [and (\ref{NL_CHSH})]
one must have a sufficiently high experimental 
visibility of the quantum interference.
For the optimal difference angle $\varphi_{max} = 18.8^{\circ}$
the minimum required visibility to see the violation of inequality (\ref{NL_INEQUALITY})
is given by the ratio of 
its bound [$3.792$] 
and the quantum value [$3.893$],
or approx. $97.4\%$. 
We remind that in standard Bell-type experiment
to have a minimum visibility of only $\frac{2}{2\sqrt{2}} \approx 71\%$ 
is sufficient to violate the CHSH inequality (\ref{NL_CHSH}) 
at its optimal settings.
For the settings used here the critical 
visibility reads $\frac{2}{2.215} \approx 90.3 \%$,
which is much lower than $97.4\%$.

Quantum mechanics predicts the violation of inequality (\ref{NL_INEQUALITY}).
We experimentally demonstrate this violation
and hence exclude the class of nonlocal hidden variable theories discussed.
Our results are in very good agreement with quantum calculations.

\begin{figure}
\begin{center}
\includegraphics[scale=0.6]{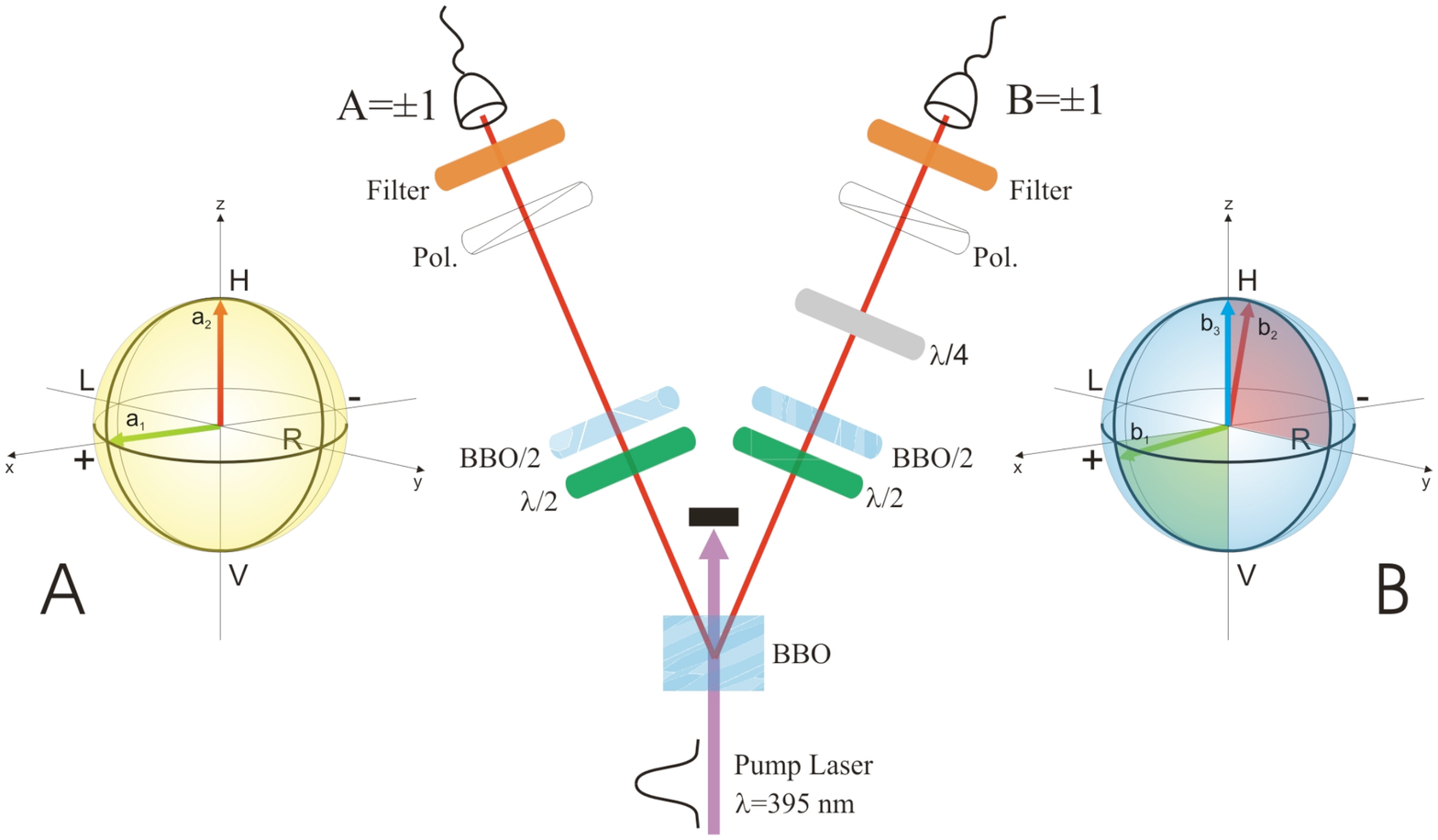}
\end{center}
\caption{Experimental setup.
A 2 mm thick type-II  $\beta$-barium-borate (BBO) crystal 
is pumped with a pulsed frequency-doubled Ti:SA laser 
($180$ fs) at $\lambda=395$ nm wavelength and approx. $150$~mW optical cw-power. 
The crystal is aligned to produce the polarization-entangled singlet state (Appendix C). 
Spatial and temporal distinguishability of the produced photons 
(induced by birefringence in the BBO) 
are compensated by a combination of half-wave plates ($\lambda/2$) 
and additional BBO-crystals (BBO/2).
Spectral distinguishability 
(due to the broad spectrum of the pulsed pump) 
is eliminated by narrow spectral filtering of $1$ nm 
bandwidth in front of each detector. 
In addition, the reduced pump power 
diminishes higher-order SPDC-emissions of multiple photon pairs. 
This allows to achieve a two-photon interference visibility 
of about $99\%$.
The arrows in the Poincar{\' e} spheres indicate 
the measurement settings of Alice's and Bob's polarizers 
for the maximal violation of inequality (\ref{NL_INEQUALITY}). 
Note that setting $\vec b_2$ lies in the $yz$ plane 
and therefore a quarter-wave plate has to be introduced on Bob's side. 
The coloured planes indicate actually measured 
settings.}
\label{NL_SETUP}
\end{figure}

In the experiment (see Fig.\ \ref{NL_SETUP}), 
pairs of polarization entangled photons are generated
via spontaneous parametric down-conversion (SPDC) 
(Appendix C). 
The photon source is aligned to produce pairs 
described in quantum mechanics by the
polarization singlet state (\ref{POL_SINGLET}). 
The experiment consists of a series of measurements
in which one sets the polarizer angle (and inserts a quarter-wave plate if necessary)
and registers the number of coincidences
within a ten seconds time slot.
We observe maximal coincidence count rates, 
in the $H/V$ basis, of around 3500 with single count rates of 95000 (Alice) 
and 105000 (Bob), 3300 coincidences in the $\pm 45^{\circ}$ basis 
(75000 singles at Alice and 90000 at Bob), 
and 2400 coincidences in the $R/L$ basis 
(70000 singles at Alice and 70000 at Bob). 
The reduced count rates in the $R/L$ basis are 
due to additional retarding elements in the beam path. 
The two-photon visibilities are approximately 
$99.0 \pm 1.2 \%$ in the $H/V$ basis, 
$99.2 \pm 1.6 \%$ in the $\pm 45$ basis 
and $98.9 \pm 1.7 \%$ in the $R/L$ basis.

In terms of experimental count rates the correlation function for a given pair of measurement settings
$(\vec a, \vec b)$ is given by:
\begin{equation}
E(\vec a, \vec b) = \frac{N_{++} + N_{--}-N_{+-}-N_{-+}}{N_{++} + N_{--}+N_{+-}+N_{-+}},
\end{equation}
where $N_{AB}$ denotes the number of coincident detection events 
between Alice's and Bob's measurements within the integration time. 
We ascribe the number $+1$, if Alice (Bob) detects a photon 
after a polarizer set along  $\vec a$ ($\vec b$), 
and $-1$ for the orthogonal direction $\vec a^{\perp}$ ($\vec b^{\perp}$). 
For example, $N_{+-}$ 
denotes the number of coincidences in the experimental runs 
in which Alice sets $\vec a$ and Bob sets $\vec b^{\perp}$. 
Recall that the difference angle $\varphi$ 
between measurement settings is calculated for vectors on the Poincar{\' e} sphere
(it is a double angle with respect to that between the polarizers). 
We introduce the notation $\overline{E}_{kl}(\varphi) = E(\vec a_k, \vec b_l)$,
which encodes the assumption of rotational invariance.

To test inequality (\ref{NL_INEQUALITY}) three correlation 
functions [$E(\vec a_1,\vec b_1)$, $E(\vec a_2,\vec b_2)$, $E(\vec a_2,\vec b_3)$]
have to be extracted from the measured data. 
We choose observables $\vec a_1$ and $\vec b_1$ as 
linear polarization measurements 
(in the $xz$ plane on the Poincar{\' e} sphere; Fig.\ \ref{NL_SETUP}) 
and $\vec a_2$ and $\vec b_2$ as elliptical polarization 
measurements in the $yz$ plane. 
Two further correlation functions [$E(\vec a_1,\vec b_2)$ and $E(\vec a_2,\vec b_1)$]
are extracted to test the CHSH inequality (\ref{NL_CHSH}). 

\begin{figure}
\hspace{-1.7cm}
\includegraphics[scale=0.83]{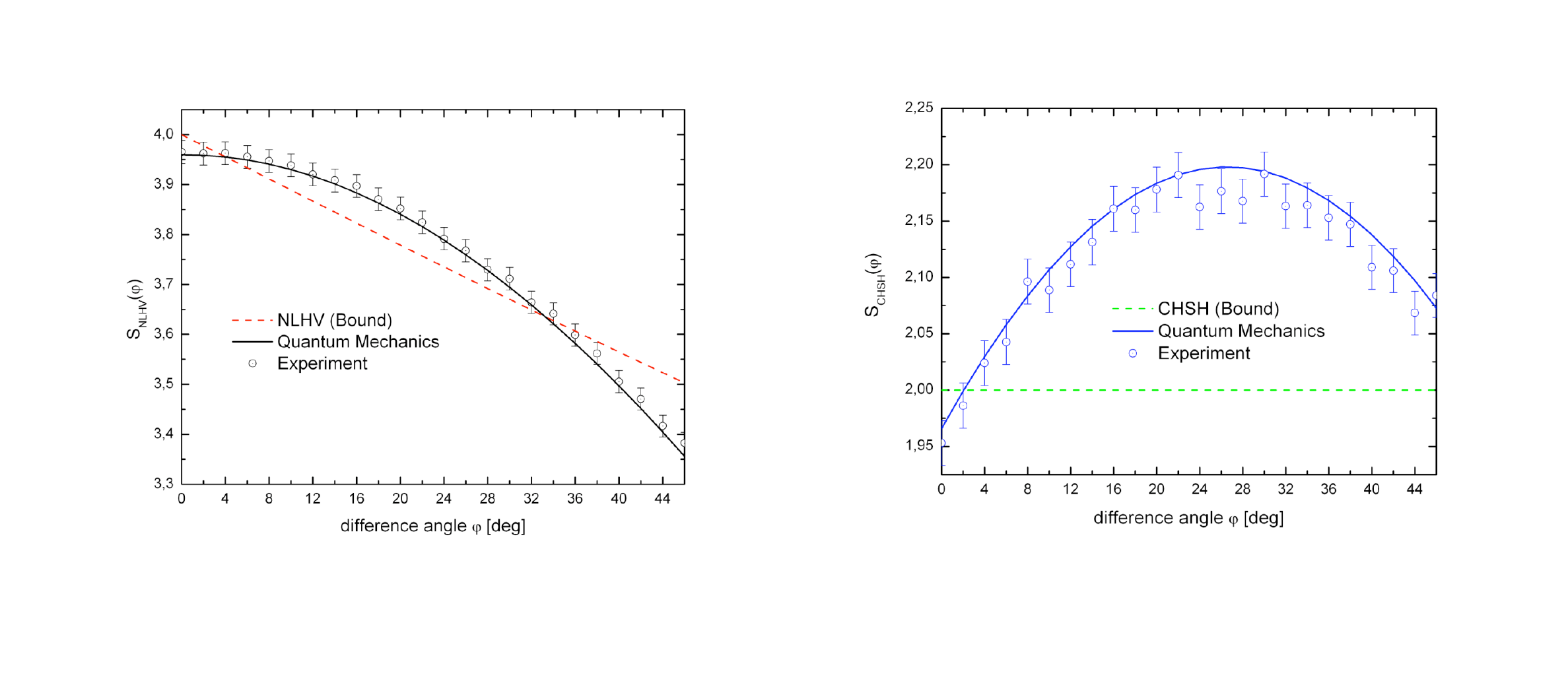}
\caption{
Experimental violation of the inequalities 
for the nonlocal hidden variable theories (NLHV) 
and for local realistic theories (CHSH).
{\bf Left panel:} 
Dashed lines indicate the bound of inequality (\ref{NL_INEQUALITY}) 
for the investigated class of nonlocal hidden variable theories (see text). 
The solid line is the quantum theoretical prediction 
\emph{including} the experimental visibility. 
The shown experimental data was taken for 
various difference angles $\varphi$ 
of local measurement settings. 
The bound is clearly violated for $4^{\circ} \le \varphi \le 32^{\circ}$. 
Maximum violation is observed for $\varphi_{max} \approx 20^{\circ}$.
{\bf Right panel:}
At the same time, no local realistic theory 
can model the correlations for 
the investigated settings as 
the same set of data also 
violates the CHSH inequality (\ref{NL_CHSH}). 
The bound (doted line) is overcome for 
all values $\varphi$ around $\varphi_{max}$ and hence 
excludes any local realistic explanation 
of the observed correlations. 
Again, the solid line gives the quantum 
prediction for the observed experimental visibility.}
\label{NL_RESULTS}
\end{figure}

The first set of correlations, in the $xz$ plane, 
is obtained by using linear polarizers set to $\alpha_1$ and $\beta_1$ 
(these are polarizer angles)
at Alice's and Bob's location, respectively. 
In particular, $\alpha_1 = \pm 45^{\circ}$ while $\beta_1$ 
is chosen to lie between $45^{\circ}$ and $160^{\circ}$ 
(green arrows in Fig.\ \ref{NL_SETUP}). 
The second set of correlations (necessary for CHSH) 
is obtained in the same plane for $\alpha_2 = 0^{\circ}/90^{\circ}$ 
and $\beta_1$ between $45^{\circ}$ and $160^{\circ}$. 
The set of correlations for measurements in the $yz$ plane 
is obtained by introducing a quarter-wave plate 
with the fast axis aligned along the (horizontal) $0^{\circ}$-direction 
at Bob's site, which effectively rotates the 
polarization state by $90^{\circ}$ around the $z$-axis 
on the Poincar{\' e} sphere. 
The polarizer angles are then set to $\alpha_2 = 0^{\circ}/90^{\circ}$ 
and $\beta_2$ between $0^{\circ}$ and $115^{\circ}$ (red arrows in Fig. \ref{NL_SETUP}). 
With the same $\beta_2$ and $\alpha_1 = \pm 45^{\circ}$ 
the last expectation values for the CHSH case are measured. 
The remaining one for inequality (\ref{NL_INEQUALITY}) 
is the check of perfect correlations, 
for which we choose $\alpha_2 = \beta_3 = 0^{\circ}$, 
i.e.\ the intersection of the two orthogonal planes. 
Fig.\ \ref{NL_RESULTS} shows the experimental violation of (\ref{NL_INEQUALITY}) 
and (\ref{NL_CHSH}) for various difference angles. 

We finally obtain the following expectation values 
for the optimal settings for a test of (\ref{NL_INEQUALITY}) 
(the errors are calculated assuming that 
the counts follow a Poissonian distribution):
\begin{eqnarray*}
E(\vec a_1,\vec b_1) & = & - 0.9298 \pm 0.0105, \\
E(\vec a_2,\vec b_2) & = & - 0.942 \pm 0.0112, \\
E(\vec a_2,\vec b_3) & = & - 0.9902 \pm 0.0118.
\end{eqnarray*}
This results in
\begin{equation}
S_{NLHV} = 3.8521 \pm 0.0227,
\end{equation}
which violates inequality (\ref{NL_INEQUALITY}) 
by $3.2$ standard deviations. 
At the same time, one can extract the additional correlation 
functions:
\begin{eqnarray*}
E(\vec a_1,\vec b_2) & = & 0.0374 \pm 0.0091, \\
E(\vec a_2,\vec b_1) & = & 0.3436 \pm 0.0088,
\end{eqnarray*}
required for the CHSH inequality. 
One obtains:
\begin{equation}
S_{CHSH} = 2.178 \pm 0.0199,
\end{equation}
which is a violation by approximately $9$ standard deviations. 
The stronger violation of (\ref{NL_CHSH}) 
is due to the relaxed visibility requirements 
on the probed entangled state.

\subsubsection{Removing assumption of rotational invariance}

Let us make few remarks on the averaged correlation
functions, which enter inequality (\ref{NL_INEQUALITY}).
In principle, to measure these averaged correlations
one needs to perform infinite series of measurements
in which the angle between the observables, $\varphi$, is kept constant,
and the angle $\xi$, describing the position of the two vectors in the plane, 
is rotated.
A more physical attempt
is to perform a finite number of measurements and approximate
the value of $\overline{E}_{kl}(\varphi)$.
Note that, according to quantum mechanics,
the correlation function of a singlet state
is a function of a difference angle between observables only.
Thus, quantum mechanics predicts
that averaging correlations over rotations which keep
the difference angle constant does not change the correlations, i.e.
$E_{kl}^{QM}(\varphi) = \overline{E}_{kl}^{QM}(\varphi)$.
This suggests another way to deal with
additional averaging,
which we have followed in the experiment.
Simply, one replaces the averaged correlations in inequality (\ref{NL_INEQUALITY})
with correlations measured for one pair of settings.
This is equivalent to making the \emph{additional assumption}
of rotational invariance
which here means that $E_{kl}(\varphi) = \overline{E}_{kl}(\varphi)$.
This can be justified from the rotational invariance 
of the correlations predicted by quantum mechanics, 
which is confirmed experimentally when measuring the visibility of the setup.
Moreover, no experimental evidence against the rotational invariance
has been found.
Yet, this is an additional assumption.

We show an inequality,
violated by quantum predictions,
which can experimentally be tested 
and which involves no extra assumptions
to realism and polarized photons.
We mainly follow the derivation of the nonlocal inequality
given above.
Briefly, in the proof the correlation function
is written in the spherical coordinate system.
The equatorial plane of the system is spanned by observable
vectors $\vec a_k$ and $\vec b_l$.
Thus the correlation function depends on the
spherical angle  $\phi$ only:
\begin{equation}
E_{kl} \to E(\phi_k,\phi_l)
\end{equation}
Next, one changes the variables to
the difference angle, $\varphi \equiv \phi_k - \phi_l$, 
and angle to the center between the vectors, $\xi \equiv \frac{\phi_k + \phi_l}{2}$,
\begin{equation}
E(\phi_k,\phi_l) \to E_{kl}(\xi,\varphi),
\end{equation}
and shows that these correlations satisfy inequality (\ref{XI_PHI_CORR}):
\begin{eqnarray*}
E_{kl}(\xi,\varphi) \le 1 - 2 
\int_0^{\pi} d\theta_u \sin\theta_u \int_0^{2 \pi} d \psi
\int_0^{\pi} d\theta_v \sin\theta_v \int_0^{2 \pi} d \chi
F(\theta_u,\theta_v,\psi,\chi) N |\cos(\xi - \psi + \alpha)|,
\end{eqnarray*}
where we define the abbreviation
\begin{equation}
N \equiv \sqrt{n_2^2 \cos^2\frac{\varphi - \chi}{2} + n_1^2
\sin^2 \frac{\varphi - \chi}{2}}.
\label{N}
\end{equation}
Instead of the integration over $\xi$,
consider a sum of \emph{two elements} only:
\begin{eqnarray}
E_{\varphi} & \equiv & E_{kl}(0,\varphi)+E_{kl}(\pi/2,\varphi) \nonumber \\
& \le &
2-2
\int_0^\pi \sin \theta_u d \theta_u \int_0^{2 \pi} d \psi
\int_0^\pi \sin \theta_v d \theta_v \int_0^{2 \pi} d \chi
F(\theta_u,\theta_v,\psi, \chi) \nonumber \\
&& \times N
\Big[
|\cos(0 - \psi + \alpha)| + |\cos(\pi/2 - \psi + \alpha)|
\Big]
\label{ROT_INV_FREE_N_COSINES}
\end{eqnarray}
Since for arbitrary argument $x$ one has:
\begin{equation}
|\cos(x)|+|\cos(\pi/2+x)|\ge 1,
\end{equation}
the expression in the square bracket in (\ref{ROT_INV_FREE_N_COSINES}) 
is always grater or equal to one,
independently of $\psi$ and $\alpha$. 
One can perform integration over $\psi$
and all other steps as before.
Note already here that the sum of the two terms introduces a 
factor of one
in front of the integral,
whereas the whole integration over $\xi$ 
before (without normalization) introduced a factor of four.
Thus, one can expect that the new inequality 
will require higher visibility to be violated.
As before, one introduces 
two new observable vectors in the same plane and adds appropriate
inequalities integrated over $\psi$:
\begin{eqnarray*}
E_{\varphi} + E_{\varphi'}
\le
4-2
\int_0^\pi \sin \theta_u d \theta_u 
\int_0^\pi \sin \theta_v d \theta_v \int_0^{2 \pi} d \chi
F(\theta_u,\theta_v, \chi)
[N+N'],
\end{eqnarray*}
where $N'$ is obtained after replacement of angle $\varphi$ by $\varphi'$ in
(\ref{N}).
With the triangle inequality for two-dimensional vectors
$\vec x$ and $\vec y$ one obtains:
\begin{eqnarray*}
|E_{\varphi} + E_{\varphi'}| \le
4-2|\sin\frac{\varphi-\varphi'}{2}|
\int_0^\pi \sin \theta_u d \theta_u 
\int_0^\pi \sin \theta_v d \theta_v
F(\theta_u,\theta_v)
\sqrt{n_2^2 + n_1^2},
\end{eqnarray*}
where we have already integrated over $\chi$.
One introduces the analogical inequality
for \emph{four} observables from an orthogonal plane, say the $yz$ plane:
\begin{eqnarray*}
|E_{\varphi}^{yz} + E_{\varphi'}^{yz}| \le
4-2|\sin\frac{\varphi_{yz}-\varphi'_{yz}}{2}|
\int_0^\pi \sin \theta_u d \theta_u 
\int_0^\pi \sin \theta_v d \theta_v
F(\theta_u,\theta_v)
\sqrt{n_2'^2 + n_1'^2},
\end{eqnarray*}
where now $n_1'$ and $n_2'$
describe projections of vectors $\vec u$ and $\vec v$
onto the $yz$ plane.
Summing up the last two inequalities
with $\varphi = \varphi_{yz}$
and $\varphi' = \varphi'_{yz} = 0$,
and using the bound on the length
of projections onto orthogonal planes (\ref{PROJECTIONS_BOUND}),
one gets the final inequality:
\begin{equation}
|E_{\varphi}^{xy} + E_{0}^{xy}| + |E_{\varphi}^{yz} + E_{0}^{yz}| \le
8 - 2 |\sin\frac{\varphi}{2}|.
\label{RI_FREE_INEQ}
\end{equation}
The left-hand side of this expression
involves eight correlation functions.
Namely, to obtain $E_{\varphi}^{xy}$
one needs to measure both $E^{xy}(0,\varphi)$ and $E^{xy}(\frac{\pi}{2},\varphi)$.
The same holds for all other correlations.

Quantum predictions violate this inequality.
For the singlet state the left-hand side equals $2|2\cos\varphi+2|$.
The optimal angle, for which the violation is
maximal, equals $\varphi_{opt} \approx 14.6^{\circ}$.
For this angle the bound is given by $7.746$
and quantum mechanics predicts for the left-hand side
the value $7.871$.
Thus, the 
visibility required to see the violation is approximately $98.41\%$
at the optimal angle.

\begin{figure}
\begin{center}
\includegraphics{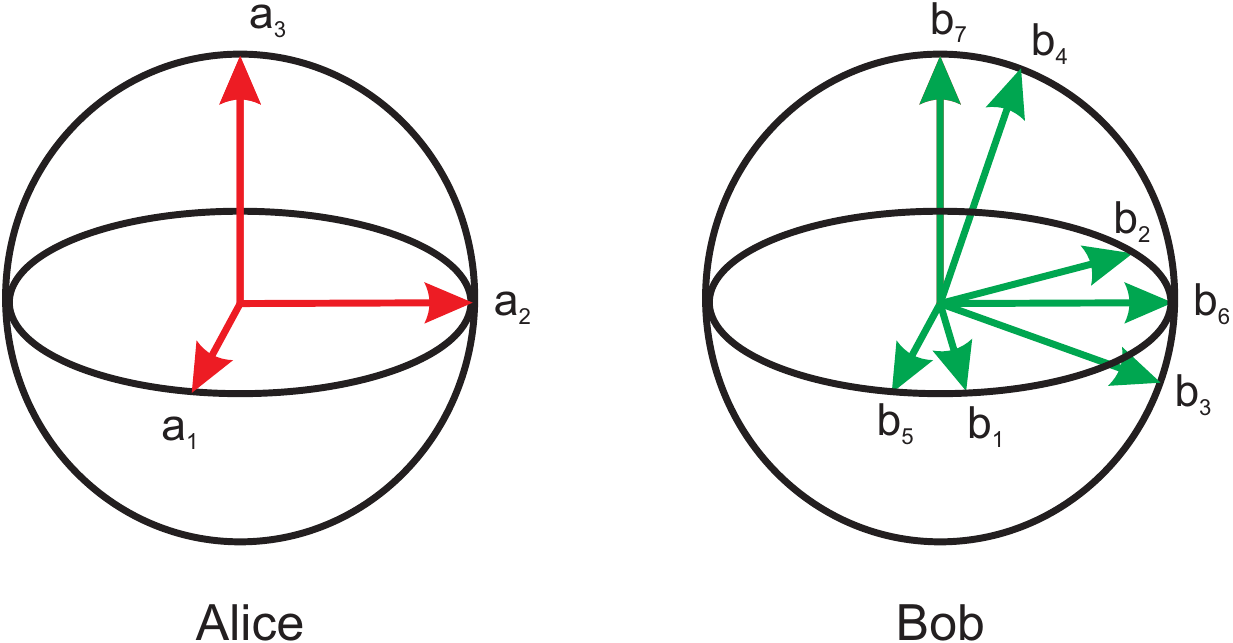}
\end{center}
\caption{Exemplary measurements required to violate the nonlocal 
inequality without the rotational invariance assumption.}
\label{ROT_INV_FREE_FIG}
\end{figure}

Finally, we give a particular choice of settings,
there are $3 \times 7$ of them,
which allow to measure seven
correlation functions and fully define
the left-hand side of (\ref{RI_FREE_INEQ}).
Let Alice measure in all three orthogonal directions:
\begin{equation}
\vec a_1 = (1,0,0), \quad
\vec a_2 = (0,1,0), \quad
\vec a_3 = (0,0,1).
\end{equation}
The settings of Bob are the following:
\begin{equation}
\begin{array}{lll}
\vec b_1 = (\cos\varphi_{opt},\sin\varphi_{opt},0), & \vec b_2 = (-\sin\varphi_{opt},\cos\varphi_{opt},0), & \\
\vec b_3 = (0,\cos\varphi_{opt},-\sin\varphi_{opt}), & \vec b_4 = (0,\sin\varphi_{opt},\cos\varphi_{opt}), & \\
\vec b_5 = \vec a_1 = (1,0,0), & \vec b_6 = \vec a_2 = (0,1,0), & \vec b_7 = \vec a_3 = (0,0,1).
\end{array}
\end{equation}
Using this notation, inequality (\ref{RI_FREE_INEQ}) reads:
\begin{equation}
|E_{11} + E_{22} + E_{15} + E_{26}|
+ |E_{23} + E_{34} + E_{26} + E_{37}|
\le 8 - 2 |\sin\frac{\varphi_{opt}}{2}|.
\end{equation}
Note that the correlation $E_{26}$ appears twice,
thus it is enough to measure seven correlations
in order to acquire the value of the left-hand side.
All the vectors are pictured on the Bloch sphere
in Fig.\ \ref{ROT_INV_FREE_FIG}.

To conclude, there can be sceptics
who might think that the averaging of correlations
present in the inequality (\ref{NL_INEQUALITY})
is essential to the experimental verification
of the incompatibility of quantum mechanics
and the nonlocal theories.
Experimental verification of the violation of the inequality
(\ref{RI_FREE_INEQ})
will convince them that this is not the case.

\subsubsection{Conclusion}

We have for the first time experimentally excluded a class of plausible nonlocal hidden variable theories. 
The theories under consideration assume realism, classical mixtures of polarizations 
(for which the Malus' law is valid) and 
arbitrary nonlocal dependencies via the measurement devices in 
order to model quantum correlations of entangled states. 
This class of theories is relevant insofar as it allows to model both perfect correlations 
of entangled states and the violation of CHSH inequalities. 
In order to simplify the experiment, the additional requirement of rotational invariance
of the  correlation functions was assumed for the nonlocal models.
We also discussed how this assumption can be relaxed
and proposed a feasible test [inequality (\ref{RI_FREE_INEQ})]
which can exclude a broader (not necessarily rotationally invariant) class of nonlocal theories.

\newpage

\subsection{Reduced experimenter's freedom [P4]}

One of the assumptions behind Bell's theorem is
the freedom to choose different experimental arrangements.
In this section we describe an approach to quantify this freedom
within a local realistic picture.
We show that the experimentally observed degree of violation of
Bell's inequalities sets a minimal degree to which the free choice has to be
abandoned if one insists on a local realistic explanation. 

Let us set the stage.
\textit{Realism} supposes that measurement results are determined by
hidden variables which exist prior to and independent of observation.
\textit{Locality} supposes that the results obtained at one location are
independent of any measurements or actions performed at space-like separated
regions. Finally, \textit{``freedom of choice''} assumes that the
experimenter's choice of the measurement setting is independent of the local
realistic mechanism which determines the measurement results. In what follows
we pursue the approach of Gill \textit{et al.}~\cite{GILL1,GILL2} in formulating
these concepts in a mathematically rigorous way.

Consider two spatially separated partners, Alice and Bob, performing
space-like separated experiments on particles which are pairwise emitted by
some source. Let $X$ and $Y$ denote the \textit{actual} measurement outcomes
obtained, and $k$ and $l$ the actual measurement settings chosen by Alice and
Bob, respectively. The outcomes $X$ and $Y$ can take values $+1$ or $-1$, and
the settings $k$ and $l$ values $1$ or $2$. The probability to observe the two
outcomes to be equal, $X\!=\!Y$, under the chosen setting $k$ (Alice) and $l$ (Bob) is
denoted by $P(X\!=\!Y|kl)$.

Local realism assumes the existence of a quadruple of variables $\{X_{1}%
,X_{2},Y_{1},Y_{2}\}$, each taking values $+1$ or $-1$, which represents the
\textit{potential} measurement outcomes in a thought experiment, under any of
the possible measurement settings. This quadruple exists independently of
whether any or which experiment is actually performed on either side. Because
of locality the variables on Alice's side do not depend on the choice of
setting on Bob's side, and vice versa. Thus, local realism requires
$X\in\{X_{1},X_{2}\}$ and $Y\in\{Y_{1},Y_{2}\}.$

The freedom assumption expresses the independence between the choice $k,l$ of
the measurement settings and the local realistic mechanism which finally
selects the actual outcomes ${X,Y}$ from the potential ones ${X_{1}%
,X_{2},Y_{1},Y_{2}}$. Gill \textit{et al.}~\cite{GILL1,GILL2} put this formally in
the requirement that $\{k,l\}$ are statistically independent of $\{X_{1}%
,X_{2},Y_{1},Y_{2}\}$. This means that in many thought repetitions of the
experiment the probabilities with which the quadruple $\{X_{1},X_{2}%
,Y_{1},Y_{2}\}$ takes on any of its $2^{4}$ possible values remain the same
within each subensemble defined by the four possible combinations of $k$ and
$l$. In particular, one has $P(X_{k}\!=\!Y_{l})\!=\!P(X\!=\!Y|kl)$, where
$P(X_{k}\!=\!Y_{l})$ is the (mathematical) probability for having $X_{k}%
=Y_{l}$.

What if the experimenter's freedom is just an illusion? Imagine that the
choices of experimental settings and experimental results are both
consequences of some common local realistic mechanism. In such a case the two
probabilities $P(X_{k}\!=\!Y_{l})$ and $P(X\!=\!Y|kl)$ may differ from each
other and we use their difference:
\begin{equation}
\Delta_{kl}\equiv P(X\!=\!Y|kl)-P(X_{k}\!=\!Y_{l}), \label{measure}%
\end{equation}
to measure the lack of freedom. This measure can acquire values from $-1$ to
$1$, and the freedom case corresponds to all $\Delta_{kl}=0$. It is important
to note that while the probabilities $P(X\!=\!Y|kl)$ can directly be measured,
the $P(X_{k}\!=\!Y_{l})$ are only mathematical entities of the local realistic
theory without a direct operational meaning. Nevertheless, they satisfy a
set-theoretical constraint which is mathematically equivalent to the
Clauser--Horne--Shimony--Holt (CHSH) inequality~\cite{CHSH}. The product of
local realistic results $X_{2}Y_{2}$ is always equal to the multiplication of
$(X_{1}Y_{1})(X_{1}Y_{2})(X_{2}Y_{1})$, because the square of a dichotomic
variable is equal to $+1$. This implies that the
following expression can attain only one of two values \cite{GILL1,GILL2}:
\begin{eqnarray}
\openone \{X_{1}=Y_{1}\}+\openone \{X_{1}=Y_{2}\}
+ \openone \{X_{2}=Y_{1}\} - \openone \{X_{2}=Y_{2}\}=0\textrm{ or }2\,,
\end{eqnarray}
where $\openone \{X_{k}=Y_{l}\}$ is the indicator of the event $X_{k}=Y_{l}$,
i.e.,\ it is equal to $1$ if it happens and $0$ if it does not happen. The
expectation value of the indicator variable is the probability for the event
to happen, $P(X_{k}\!=\!Y_{l})$. Finally, the expectation value of the
left-hand side cannot be greater than the maximum value of the averaged
expression:%
\begin{eqnarray}
S_{\textrm{CHSH}} \equiv P(X_{1}\!=\!Y_{1})+P(X_{1}\!=\!Y_{2})
  +P(X_{2}\!=\!Y_{1})-P(X_{2}\!=\!Y_{2})\leq2\,. \label{kids}%
\end{eqnarray}
The equivalence to the CHSH inequality is evident as soon as one recalls that
the correlation function of dichotomic variables equals $E_{kl}=2P(X_{k}%
\!=\!Y_{l})-1$. The above inequality, in turn, implies a new bound on the set
of probabilities that can experimentally be measured:%
\begin{eqnarray}
S_{\Delta}  \equiv P(X\!=\!Y|11)+P(X\!=\!Y|12)
+P(X\!=\!Y|21)-P(X\!=\!Y|22)\leq2+\Delta_{\textrm{CHSH}},
\label{S_DELTA}
\end{eqnarray}
where $\Delta_{\textrm{CHSH}}\equiv\Delta_{11}+\Delta_{12}+\Delta_{21}%
-\Delta_{22}$. Note that on the basis of measured probabilities (relative
frequencies) one cannot make statements about the individual measures
$\Delta_{kl}$ but rather on their combination as given in $\Delta
_{\textrm{CHSH}}$. In particular, it is possible that $\Delta_{\textrm{CHSH}%
}\!=\!0$, although all the individual $\Delta_{kl}\!\neq\!0$, and it may also
be negative. However, only the case of positive $\Delta_{\textrm{CHSH}}$ ---
implying at least one individual $\Delta_{kl}$ to be unequal to zero --- makes
the freedom assumption within a local realistic model experimentally testable,
as the bound on the right-hand side of (\ref{S_DELTA}) is increased. As well one
could study the lower bounds of $S_{\textrm{CHSH}}$ and $S_{\Delta}$.

To give an example of the lack of freedom model, imagine a local realistic
mechanism in which the source ''knows'' in advance the settings ``to be
chosen'' by Alice and Bob. The source can arbitrarily manipulate the value
of $S_{\Delta}$ in this case. Even the algebraic (logical) bound of
$S_{\Delta}=3$ can be reached: whenever Alice and Bob both measure the second
setting, the source sends (local realistic) correlated pairs such that the
measurement results anticoincide, i.e.\ $P(X\!=\!Y|22)=0$, and in all other
measurements it produces pairs for which the results coincide,
i.e.\ $P(X\!=\!Y|11)=P(X\!=\!Y|12)=P(X\!=\!Y|21)=1$, and thus $S_{\Delta}=3$.
For this local realistic model (without freedom) inequality (\ref{S_DELTA}) is
satisfied, but only because of the adapted bound $2+\Delta_{\textrm{CHSH}}=3$.
Imagine another experiment, in which the observers (freely) choose their
settings independently from the local realistic source. Then
$P(X\!=\!Y|kl)=P(X_{k}\!=\!Y_{l})$, i.e.\ $\Delta_{\textrm{CHSH}}=0$, and
inequality (\ref{S_DELTA}) is fulfilled with the bound of 2, as it becomes the
CHSH inequality (\ref{kids}).

The value of $\Delta_{\textrm{CHSH}}$ for which the inequality is still
satisfied, defines the minimal extent to which the experimenter's freedom has
to be abandoned such that a local realistic explanation of the experiment is
still possible. Denote the left-hand side of inequality~(\ref{S_DELTA}) as the
CHSH expression. The maximal possible quantum value of this expression,
$S_{\textrm{QM}}=1+\sqrt{2}$, can be observed for the maximally entangled state,
for example, the singlet state $|\psi^{-}\rangle=(|0\rangle|1\rangle
-|1\rangle|0\rangle)/\sqrt{2}$, where $|0\rangle$ and $|1\rangle$ are two
orthogonal quantum states, and for an appropriate choice of possible settings
$\{k,l\}$. This quantum value requires an abandonment of the experimentalist's
freedom to the extent of at least $\Delta_{\textrm{CHSH}}=\sqrt{2}%
-1\approx0.414$.

Since, basing on the experiment, we can only make statements about
$\Delta_{\textrm{CHSH}}$, a large number of local realistic theories are
possible that deny the experimenter's freedom and are in agreement with
quantum mechanical predictions and experiments. In order to be able to make
further statements about these theories we need to impose some structure on
them. In what follows we restrict ourselves to the case in which the degree to
which the freedom is abandoned --- that is the absolute value of the measure
$\Delta_{kl}$ --- is independent of the actual experiment performed,
i.e.\ $|\Delta_{kl}|\!=\!\Delta$ is the same for all $k,l$. Roughly speaking,
the level of conspiracy is assumed to be the same for all experimental
situations. Choosing $\Delta_{11}\!=\!\Delta_{12}\!=\!\Delta_{21}%
\!=\!-\Delta_{22}\!\equiv\!\Delta$ one obtains $\Delta=\frac{1}{4}(\sqrt
{2}-1)\approx0.104$ for the minimal degree required to explain the quantum
value of the CHSH expression by a local realistic model. If all the
$\Delta_{kl}$ are positive (i.e.\ if $P(X\!=\!Y|kl)>P(X_{k}\!=\!Y_{l})$ for
all $k,l$), one finds the even higher value $\Delta=\frac{1}{2}(\sqrt
{2}-1)\approx0.207$.

It is known that with an increasing number of parties, $N$, the discrepancy
between the results of Bell tests and local realistic predictions that respect
the experimenter's freedom increases rapidly (exponentially) with
$N$~\cite{MERMIN}. We now determine how the degree of the lack of freedom
needs to scale with $N$ in a local realistic theory that agrees with these tests.

Consider $N$ space-like separated parties who can each choose between two
possible measurement settings. Let $X^{(j)}\!\in\!\{1,-1\}$ denote the actual
measurement result obtained and $k_{j}\!\in\!\{1,2\}$ the actual measurement
setting chosen by party $j$. The probability to observe correlation, i.e.\ the
probability that the product of local results is equal to $1$ under settings
$k_{1},...,k_{N}$, is denoted by $P(\prod_{j=1}^{N}\!X^{(j)}
\!=\!1|k_{1}...k_{N})$. Local realism assumes the existence of $2\,N$ numbers
$\{X_{1}^{(1)},X_{2}^{(1)},...,X_{1}^{(N)},X_{2}^{(N)}\}$, each taking values
$+1$ or $-1$ and representing the potential measurement outcomes of $N$
parties under any possible combination of their measurement settings. The
(mathematical) probability that the product of the potential outcomes is equal
to $1$ is denoted by $P(\prod_{j=1}^{N}\!X_{k_{j}}^{(j)}\!=\!1)$. Note again
that this probability cannot be measured experimentally.

We apply the approach used above to the present case of $N$ parties. We
introduce the difference%
\begin{equation}
\Delta_{k_{1}...k_{N}}\equiv P(\textstyle\prod_{j=1}^{N}\!X^{(j)}%
\!=\!1|k_{1}...k_{N})-P(\textstyle\prod_{j=1}^{N}\!X_{k_{j}}^{(j)}\!=\!1)
\label{cold}%
\end{equation}
to measure the lack of freedom of $N$ experimenters. The probabilities
$P(\prod_{j=1}^{N}\!X_{k_{j}}^{(j)}\!=\!1)$ satisfy a set-theoretical
constraint that is mathematically equivalent to the Mermin
inequality~\cite{MERMIN}:
\begin{equation}
M\equiv\!\sum_{k_{1},...,k_{N}=1}^{2}\!S(k_{1},...,k_{N})\,P(\textstyle
\prod_{j=1}^{N}\!X_{k_{j}}^{(j)}\!=\!1)\leq B(N)\,, \label{cave}%
\end{equation}
where $S(k_{1},...,k_{N})\!=\!\sin\!\left[  (k_{1}+...+k_{N})\frac{\pi}%
{2}\right]  $ are coefficients taking values $0$, $+1$ or $-1$. The inequality
is bounded by $B(N)\!=\!\frac{1}{2}\!\left[  2^{\lfloor N/2\rfloor
}\!+\!2^{N/2}\sin(\frac{N\pi}{4})\right]  $, where $\lfloor x\rfloor$ is the
greatest integer less or equal to $x$. Using inequality~(\ref{cave}) and
definition~(\ref{cold}), one obtains a new inequality:%
\begin{eqnarray}
M_{\Delta}  \equiv \!\sum_{k_{1},...,k_{N}=1}^{2}\!S(k_{1},...,k_{N}%
)\,P(\textstyle\prod_{j=1}^{N}\!X^{(j)}\!=\!1|k_{1}...k_{N})
\leq B(N)+\Delta_{\textrm{Merm}}, \label{you}%
\end{eqnarray}
where $\Delta_{\textrm{Merm}}\!=\!\sum_{k_{1},...,k_{N}=1}^{2}S(k_{1}%
,...,k_{N})\,\Delta_{k_{1}...k_{N}}$. Importantly, the probabilities entering
this inequality are measurable.

In a Bell experiment involving the maximally entangled $N$-party (GHZ) state
one observes $M_{\textrm{QM}}=\frac{1}{2}\left[  2^{N-1}+2^{N/2}\sin(\frac{N\pi
}{4})\right]  $ for the maximal possible value of the left-hand side of
inequality~(\ref{you}). This implies $2^{N-2}-2^{\lfloor(N-2)/2\rfloor}$ for
the minimal value of $\Delta_{\textrm{Merm}}=M_{\textrm{QM}}-B(N)$ that still
allows a local realistic explanation of the experiment. Suppose again that the
degree of the lack of freedom is independent of the measurement setting. With
an adequate choice of signs one has $\Delta_{k_{1}...k_{N}}\!=\!\Delta_{N}$
for $k$'s for which $S(k_{1},...,k_{N})\!=\!1$ and $\Delta_{k_{1}...k_{N}
}\!=\!-\Delta_{N}$ for $k$'s for which $S(k_{1},...,k_{N})\!=\!-1$. This
results in $\Delta_{\textrm{Merm}}\!=\!2^{N-1}\Delta_{N}$. 
Finally, one obtains
that the degree to which the experimenter's freedom has to be abandoned in
order to have an agreement between local realism and Bell's experiments with
$N$ parties \textit{saturates exponentially fast} with $N$ as $\Delta
_{N}=\frac{1}{2}-\frac{1}{2^{\lfloor(N+1)/2\rfloor}}$. In the limit of
infinitely many partners $\Delta_{N}$ reaches the value of $\frac{1}{2}$. It
is remarkable that if the sign of all $\Delta_{k_{1}...k_{N}}$ is chosen
positive, there will be no way to obtain agreement between local realism and
the experimental results, since $\Delta_{N}$ would have to leave the range
from $-1$ to $+1$ in the limit of large $N$. The other argument which
invalidates all $\Delta_{k_{1}...k_{N}}$ to be positive involves only four
parties. In this case in the expression defined in (\ref{you}) the number of
probabilities with a positive sign is equal to the number of probabilities
with a negative sign. Thus, if all $\Delta_{k_{1}...k_{N}}$ are positive and
have the same value they cancel each other, i.e.\ $\Delta_{\textrm{Merm}}=0$,
and no explanation of the violation of the bound $B(N\!=\!4)$ is possible.

In this section we showed that quantum correlations for $N$ partners can be
explained within local realism only if both the number of measurement settings
in which the experimenter's freedom is abandoned increases exponentially
(all $2^{N-1}$ combinations of local settings entering the Mermin inequality)
and the degree of this abandonment saturates exponentially fast with $N$.

\chapter{Quantum communication}

With the emergence of this new sub-branch of physics (and information theory)
Bell's theorem and Bell inequalities found
applications far away from the foundations of quantum physics.
The security analysis of the first entanglement based quantum cryptography
scheme involves Bell inequalities \cite{EKERT}.
This now is strengthened by the analysis of Scarani and Gisin,
who showed that the violation of Bell's inequality is indeed
a valid security criterion in quantum crypto-key distribution \cite{SCARANIGISIN}.
It was shown that with every Bell inequality one can associate
a specific communication complexity problem.
The solution to the problem making use 
of quantum states which violate the Bell inequality
outperforms all possible classical solutions \cite{BZPZ}.
Furthermore, Bell's theorem was identified in the non-classical
part of the quantum teleportation procedure \cite{TELEPORTATION,ZUK_TELEPORT}.

In this chapter we review
famous examples of quantum communication:
\emph{quantum dense coding} 
allows to transfer two bits with the 
exchange of a single qubit;
\emph{quantum teleportation}
allows to transfer a quantum state
to a distant location;
\emph{quantum cryptography}
allows for secret communication due to the laws of physics;
and
\emph{quantum communication complexity}
reduces the amount of communication
needed to perform certain tasks.
All of them were experimentally realized
and some even appeared on the market.

Next, we present new results in the fields of
quantum cryptography and quantum communication complexity.
In the field of quantum cryptography,
using the results obtained when studying Bell inequalities
with restricted freedom,
we show that to a certain extend one can allow information leakage
from the laboratory, and still extract a secret key.
It is also shown that it is reasonable
to realize quantum cryptography with higher-dimensional 
systems using qudits composed of two subsystems.
This is preceded by a general solution to the eigenproblem
of the unitary generalizations of Pauli operators.
In the field of communication complexity
we present problems and their solutions linked with one of the multisetting inequalities.
In this case the quantum solutions are better than all classical solutions.
The other protocols, utilizing higher-dimensional entangled systems,
are shown to be better than a broad class of classical protocols.
The
discrepancy
between quantum and the class of classical protocols
grows with dimensionality.

\section{Brief review of basic ideas}

\subsection{Quantum dense coding (superdense coding)}

Quantum dense coding allows transmision of
two bits of information while exchanging a single qubit.
It was invented by Bennett and Wiesner in 1992 \cite{DENSE_CODING}.
Since the information storage capacity of a single qubit is
limited to a single bit (the Holevo bound),
one needs to measure two qubits to read two bits.
The trick of Bennett and Wiesner is to use entanglement
and certain ``entangled'' measurements on two qubits.

In the protocol Alice prepares a pair of qubits in a maximally
entangled state, say:
\begin{equation}
| \phi^+ \rangle_{12} = \frac{1}{\sqrt{2}} \Big[ |z+ \rangle_1 |z+ \rangle_2 + |z- \rangle_1 |z- \rangle_2 \Big],
\label{PHI+}
\end{equation}
where $| z \pm \rangle$ are the eigenstates of the local $\sigma_z$ operator.
Alice and Bob have previously agreed on the same reference frame.
Alice sends \emph{one} qubit from the pair to Bob,
who performs one of the four encoding operations (encodes two bits):
\begin{eqnarray}
U_0 &=& | z+ \rangle \langle z+ | + | z- \rangle \langle z- | = \hat 1, \nonumber \\
U_1 &=& | z+ \rangle \langle z+ | - | z- \rangle \langle z- | = \sigma_z, \nonumber \\
U_2 &=& | z- \rangle \langle z+ | + | z+ \rangle \langle z- | = \sigma_x, \nonumber \\
U_3 &=& | z- \rangle \langle z+ | - | z+ \rangle \langle z- | = \sigma_x \sigma_z = -i \sigma_y.
\end{eqnarray}
His actions evolve the initial state into the four orthogonal Bell states:
\begin{eqnarray}
U_0 | \phi^+ \rangle_{12} & = & | \phi^+ \rangle_{12}, \nonumber \\
U_1 | \phi^+ \rangle_{12} & = & | \phi^- \rangle_{12} = \frac{1}{\sqrt{2}} \Big[ |z+ \rangle_1 |z+ \rangle_2 - |z- \rangle_1 |z- \rangle_2 \Big], \nonumber \\
U_2 | \phi^+ \rangle_{12} & = & | \psi^+ \rangle_{12} = \frac{1}{\sqrt{2}} \Big[ |z+ \rangle_1 |z- \rangle_2 + |z- \rangle_1 |z+ \rangle_2 \Big], \nonumber \\
U_3 | \phi^+ \rangle_{12} & = & | \psi^- \rangle_{12} = \frac{1}{\sqrt{2}} \Big[ |z+ \rangle_1 |z- \rangle_2 - |z- \rangle_1 |z+ \rangle_2 \Big].
\label{BELL_BASIS}
\end{eqnarray}
Next, he sends the qubit back to Alice,
who performs the measurement in the Bell basis,
defined by equations (\ref{BELL_BASIS}),
and thus can perfectly decode the action of Bob.

Classically this task is impossible
because all one can do to a classical bit
is either to keep its value or flip it.
Two classical bits prepared in an arbitrary initial state
always end up in one of two different final states,
depending on the action of Bob.
Even the encoding of two bits
via acting on a single bit from a pair
cannot be defined classically.

The exemplary realisations of 
the dense coding scheme can be found in \cite{DC1,DC2}.
The main experimental challenge is to realize the full Bell measurement.

\subsection{Quantum teleportation}

Quantum teleportation uses entangled particles and two classical bits of communication
to transmit an unknown quantum state from one location to another \cite{TELEPORTATION}.
Initially, Alice and Bob share a pair of qubits in a maximally entangled state.
Alice has an extra qubit the state of which she wants to teleport to Bob.
She transmits to Bob the result of the Bell measurement
on her qubits, and after a suitable local operation
the state of Bob's particle is the same 
as the initial state of Alice's extra qubit.

Let us now present the protocol in detail.
Assume that the initial maximally entangled state between Alice and Bob is the $| \phi^+ \rangle_{AB}$ state.
This is their quantum channel.
Alice wants to teleport an arbitrary (unknown) state of her extra particle:
\begin{equation}
|\psi \rangle_E = \alpha |z+ \rangle_E + \beta |z- \rangle_E,
\label{TO_TELEPORT}
\end{equation}
with $|\alpha|^2 + |\beta|^2 = 1$.
The initial three-particle state between Alice and Bob reads:
\begin{equation}
|\Psi \rangle_{ABE} = | \phi^+ \rangle_{AB} |\psi \rangle_E.
\end{equation}
We insert decomposition (\ref{TO_TELEPORT}) for $|\psi \rangle_E$
and (\ref{PHI+}) for $| \phi^+ \rangle_{AB}$
and write the total state in terms
of Bell states of particles labelled by $A$ and $E$.
After noting that:
\begin{eqnarray}
| z+ \rangle_A | z+ \rangle_E & = & \frac{1}{\sqrt{2}}\Big[ | \phi^+ \rangle_{AE} + | \phi^+ \rangle_{AE} \Big], \nonumber \\
| z+ \rangle_A | z- \rangle_E & = & \frac{1}{\sqrt{2}}\Big[ | \psi^+ \rangle_{AE} + | \psi^+ \rangle_{AE} \Big], \nonumber \\
| z- \rangle_A | z+ \rangle_E & = & \frac{1}{\sqrt{2}}\Big[ | \psi^+ \rangle_{AE} - | \psi^+ \rangle_{AE} \Big], \nonumber \\
| z- \rangle_A | z- \rangle_E & = & \frac{1}{\sqrt{2}}\Big[ | \phi^+ \rangle_{AE} - | \phi^- \rangle_{AE} \Big],
\end{eqnarray}
the initial state before Alice measures her particle in the Bell basis reads:
\begin{eqnarray*}
|\Psi \rangle_{ABE} &=& 
\frac{1}{2} \Big[ | \phi^+ \rangle_{AE} \Big( \alpha |z+ \rangle_B + \beta |z- \rangle_B \Big)
+ | \phi^- \rangle_{AE} \Big( \alpha |z+ \rangle_B - \beta |z- \rangle_B \Big) \\
&&+ | \psi^+ \rangle_{AE} \Big( \alpha |z- \rangle_B + \beta |z+ \rangle_B \Big)
+ | \psi^- \rangle_{AE} \Big( -\alpha |z- \rangle_B + \beta |z+ \rangle_B \Big)
\Big].
\end{eqnarray*}
This relation lies at the heart of quantum teleportation.
Note that if the result of Alice corresponds to the $| \phi^+ \rangle_{AE}$ state,
the particle of Bob has collapsed to exactly the same state
as the original one given by Eq.\ (\ref{TO_TELEPORT}).
If her result corresponds to $| \phi^- \rangle_{AE}$,
Bob should flip the phase in front of the $| z- \rangle_B$
component of his qubit, i.e.\ perform the local $\sigma_z$ operation.
If the result of Alice corresponds to $| \psi^+ \rangle_{AE}$,
Bob should flip his qubit.
Finally, if her result is linked with $| \psi^- \rangle_{AE}$ state,
Bob has to change the phase in front of $| z- \rangle_B$
and then flip the qubit.

As soon as Bob knows the outcome
of Alice's Bell measurement (two classical bits have to be transmitted to Bob) 
he can locally bring the state of his qubit
to exactly the state of the initial extra qubit.

Note the following features of the teleportation scheme.
Since two classical bits have to be communicated,
teleportation does not allow sending signals faster than light.
The no-cloning theorem \cite{NOCLONING} is not violated 
because the Bell measurement changes the state
of the extra qubit (to the completely mixed one).

Quantum teleportation is widely studied experimentally,
both with a full Bell state analyzer and without it.
Some milestones can be found in \cite{EXP_TELEPORT1,EXP_TELEPORT2,EXP_TELEPORT3,EXP_TELEPORT4}.

\subsection{Quantum cryptography}

Quantum cryptography, or more precisely quantum-key distribution,
allows for secure communication. 
The message sent using quantum cryptography protocols
is not accessible to third parties.
Security of classical cryptography relies on algorithms
which are based on tasks believed to be computationally hard.
Factorization is an example of such a task:
given a number, find its prime factors.
Classically, there is no known efficient solution to this problem.
However, there exists a quantum algorithm
which solves this problem efficiently,
i.e.\ the number of resources required by the algorithm scales
polynomially with the number of digits in the integer to be factored \cite{SHOR}.
Thus, classical cryptography is, in principle, vulnerable.
Quantum cryptography is secure as long as quantum mechanics
is a correct description of nature.
Any action of an eavesdropper
causes disturbance of a quantum system,
which can be detected by the legitimate partners.

Let us first describe the role of quantum key distribution
in the process of secret communication.
Suppose Alice wants to send a message, a string of $N$ bits $M_i$, to Bob.
They can follow the so-called one-time-pad procedure.
The security of the one-time-pad requires
Alice and Bob to initially share 
a random string of bits, $K_i$, the key
which is known only to them.
Alice encrypts her information by adding
to the $i$th bit of her message string
the $i$th bit of the key, $E_i = [M_i+K_i]_2$ where $[x]_2$ denotes $x$ modulo $2$
and $i = 1,...,N$.
Since, by assumption, only Alice and Bob know the key
the cryptogram (encrypted message)
can be given to anyone without compromising the security.
After Bob receives the cryptogram
he can 
decode a message by applying the inverse operation.
In this case this is again addition modulo $2$.
This protocol was invented by Vernam in 1926 \cite{VERNAM},
and it was proven to be unconditionally secure by Shannon \cite{SHANNON_CRYPTO}
under the following circumstances:
(i) the key is truly random,
(ii) it is never reused,
(iii) it is as long as the message,
and of course (iv) it is known to Alice and Bob only.
In this way the difficulty of direct secure communication
is moved to the difficulty the key generation.
This problem is solved using quantum cryptography.

The first well established quantum cryptography protocol
is due to Bennett and Brassard \cite{BB84}. 
Since it was invented in 1984
it is often cited as the BB84 protocol.
In this protocol Alice 
sends to Bob a sequence of suitably prepared qubits.
She chooses at random
one of two conjugated bases, say the basis of $\sigma_x$ or $\sigma_y$.
The next random choice she makes
is either to send the state which corresponds to the eigenvalue $+1$ or the one which corresponds to $-1$.
Thus, she randomly sends to Bob one of the states
$|x\pm \rangle$ or $|y \pm \rangle$.
She records the choice of the basis and the choice of the eigenstate.
At his side Bob randomly measures either in  $\sigma_x$ basis or in $\sigma_y$ basis.
He records the choice of the basis
as well as the result he has obtained.
Next, both parties publicly announce their bases (but not the results!).
This public channel does not have to be confidential.
It has to be authentic -- the information which enters the channel 
cannot be modified by anybody (including the eavesdropper).
If the bases coincide, the parties have arrived at the correlated data
called a sifted key.
They have to sacrifice a part of the sifted key in order to check for eavesdropping.
If the check shows no eavesdropper
all that remains is a secret random key,
that can be used in the one-time-pad procedure.

Alternatively, one can use an equivalent entanglement-based schemes,
introduced by Ekert \cite{EKERT} and Bennett, Brassard and Mermin (BBM92) \cite{BBM92}.
In these schemes the initial randomness of the preparation of the eigenstates
is hidden in the properties of entangled states.
Take the $| \phi^+ \rangle_{12}$ state.
It has the following properties, ideal for cryptography:
local results are random, but they are always perfectly correlated
with the results of the same measurement on the other side.
In the BBM92 protocol Alice prepares a $| \phi^+ \rangle_{12}$ state
and sends one particle to Bob.
Both parties randomly measure either $\sigma_x$ or $\sigma_y$.
If the $| \phi^+ \rangle_{12}$ state is defined in the $\sigma_z$ basis,
its correlation function for measurements
in the $xy$ plane reads $\cos(\phi_1+\phi_2)$,
where $\phi_1$ and $\phi_2$ denote
the angles of observables within the plane.
Next, Alice and Bob publicly announce their bases.
Whenever the bases coincide parties obtained perfectly correlated (identical) results.
These results are used to check for a possible eavesdropper
and finally a part of it forms the key.\footnote{In the Ekert's protocol violation
of the CHSH inequality indicates the absence of an eavesdropper.}

Quantum cryptography is already at the stage of being available on the market.
The experimental progress concerning this field is impressive.
The first demonstration of quantum cryptography
was performed at IBM in early $1990$s \cite{QC_EXP1}.
Alice and Bob were separated by 30 cm.
In recent experiments this distance is far beyond 100 km \cite{QC_EXP2}.
There are prospects to set a satellite-based quantum cryptography.
Finally, bank transfers were already secured 
in the quantum way \cite{BANK_TRANSFER_CRYPTO}.

\section{Leaking labs and security [P4]}

The standard assumption in cryptography is that
the laboratories of the authorized parties are safe
and no information is allowed to leak out of them.
We first take this assumption
and present already known results concerning quantum cryptography.
Next, we prove that to some extend one can
relax it and still keep security of the quantum protocol.

\subsubsection{Secure labs}

Remarkably, the security of quantum crytography 
is linked with the violation of Bell inequalities.
We will only sketch the proof, details can be found in \cite{ACIN_IJQI,SCARANIGISIN}.
In a more practical scenario, after revealing the bases,
Alice and Bob never obtain perfectly correlated data
not only due to eavesdropping but also due to experimental
imperfections.
Nevertheless, they can efficiently extract a secret key
despite of these perturbations,\footnote{By running some extra data processing called error correction and privacy amplification.} 
if and only if \cite{CSISZAR}:
\begin{equation}
I_{AB} > \min[I_{AE},I_{BE}],
\label{SECURITY_NS}
\end{equation}
where $I_{XY} = H(X) - H(X|Y)$ is the mutual information,
which measures how much knowledge about the outcomes of 
one of the parties
reduces the uncertainty about the outcomes of the other;
$H(X) = -p(x=1) \lg p(x=1) - p(x=-1) \lg p(x=-1)$ is the Shannon entropy (or Shannon information),
where $p(x = \pm 1)$ denotes the probability of a certain outcome $x$ of party $X$
(we consider only binary outcomes) and $\lg$ is
the logarithm with base $2$;
$H(X|Y) = \sum_{y=\pm 1} P(y) H(X|y)$ is the conditional entropy
with $H(X|y)$ being the Shannon entropy of 
the conditional probability distribution $p(x|y)$.

From now on only individual attacks are considered,
i.e.\ an eavesdropper Eve operates on individual qubits transmitted to Bob.
The best individual attack Eve can do
uses a single ancillary qubit initially prepared in the state $| z+ \rangle_E$
and the following unitary transformation \cite{NIU_OPT_EVE}:
\begin{eqnarray}
U_{BE} | z+ \rangle_B | z+ \rangle_E & = & | z+ \rangle_B | z+ \rangle_E, \nonumber \\
U_{BE} | z- \rangle_B | z+ \rangle_E & = & \cos \varphi | z- \rangle_B | z+ \rangle_E + \sin\varphi | z+ \rangle_B | z- \rangle_E,
\end{eqnarray}
with $\varphi \in [0,\pi/2]$.
Explicit calculation of the mutual informations which enter condition (\ref{SECURITY_NS})
was performed in \cite{SG_MUT_INF}
with the conclusion that the protocol is secure (equivalently (\ref{SECURITY_NS}) is fulfilled)
if and only if $\varphi < \frac{\pi}{4}$.

On the other hand one checks the violation of CHSH inequality,
as given by the Horodeckis criterion \cite{CHSH_NS},
between any pair of parties.
Assume Alice prepares the $| \phi^+ \rangle$ state.
The three-particle state after Eve's attack reads:
\begin{eqnarray}
| \Psi \rangle_{ABE} &=& \frac{1}{\sqrt{2}} \Big[ |z+ \rangle_A |z+ \rangle_B |z+ \rangle_C \nonumber \\
&& + \cos\varphi |z- \rangle_A |z- \rangle_B |z+ \rangle_C
+ \sin\varphi |z- \rangle_A |z+ \rangle_B |z- \rangle_C  \Big].
\end{eqnarray}
The two-particle states $\rho_{AB}$, $\rho_{AE}$ and $\rho_{BE}$
are obtained after tracing out the appropriate subsystem.
Finally, $\rho_{BE}$ does not violate the CHSH inequality,
the maximal value of the CHSH expression for $\rho_{AB}$ equals
\begin{equation}
S_{AB} = 2 \sqrt{2} \cos\varphi,
\end{equation}
and a similar expression for $\rho_{AE}$ is given by
\begin{equation}
S_{AE} = 2 \sqrt{2} \sin\varphi.
\end{equation}
The CHSH inequality between Alice and Bob is violated,
$S_{AB} > 2$, if and only if $S_{AE} < 2$,
and this happens precisely for $\varphi < \frac{\pi}{4}$.
Thus, whenever Alice and Bob
observe violation of any CHSH inequality
they are sure they can extract a secret key 
(under the restriction of individual attacks).

\begin{figure}
\begin{center}
\includegraphics{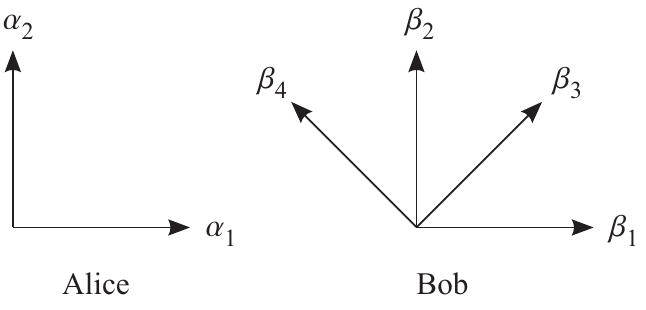}
\end{center}
\caption{Settings in the BBM--CHSH protocol. Alice chooses between two
orthogonal measurement directions $\alpha_{1}$ and $\alpha_{2}$, whereas Bob
has four different possibilities, namely the same directions as Alice, i.e.,
$\beta_{1}$ and $\beta_{2}$, as well as two directions rotated by $\frac{\pi}{4}$, 
i.e. $\beta_{3}$ and $\beta_{4}$.}
\label{BBMCHSH_SETTINGS}
\end{figure}

This proof suggests a practical protocol
for quantum-key distribution,
which we refer to as the BBM-CHSH protocol,
as it combines the BBM92 settings
and the settings for the check of the CHSH inequality.
Alice chooses her settings between $\sigma_x$ ($\alpha_1$) and $\sigma_y$ ($\alpha_2$),
and Bob chooses one of four angles:
two of which are the same as those of Alice (for perfect correlations)
and the remaining two are used for the check of violation of 
the CHSH inequality (Fig. \ref{BBMCHSH_SETTINGS}).

\subsubsection{Leaking labs}

The violation of a Bell inequality by the legitimate parties was found to be a
necessary and sufficient condition for the efficient extraction of a quantum
secret key as described in the previous section.
We show that this link disappears
as soon as one takes into account that some information
can leak out of the laboratories of Alice and Bob.
However, if the amount of leaked information
is known, one can adapt a new bound
in the Bell inequality and recover security again.

Apart from its fundamental meaning, the freedom to choose between different
measurement settings can be regarded as an important resource in quantum
secret key distribution. In particular,
 if the freedom in choosing the settings
by the legitimate partners is abandoned
an eavesdropper can both simulate the violation of a Bell
inequality and successfully eavesdrop,
 as recently shown by
Hwang~\cite{HWANG}. Effectively, one can assume that each
measurement device chooses its setting according to a pseudo-random sequence
that is installed in the device beforehand. Such a model of lack of freedom
allows the eavesdropper to know the algorithm generating pseudo-random
numbers, at least to some extent, and correspondingly predict the future
measurement settings.

In what follows we will consider the BBM-CHSH protocol
and
analyze both the violation of the CHSH inequality and the security of the key
distribution as a function of the amount of knowledge that the eavesdropper
Eve (E) has about the settings chosen by the legitimate parties Alice (A) and
Bob (B). 

Consider a source that emits pairs of spin-$\frac{1}{2}$ particles in the
singlet state 
$|\psi^{-}\rangle=(|z+\rangle_1 |z-\rangle_2 - |z-\rangle_1
|z+\rangle_2)/\sqrt{2}$, where $|z+\rangle$ and $|z-\rangle$ denote spin-up and
spin-down along the $z$ direction, respectively. The legitimate parties
measure the incoming particles in the $xy$ plane. Alice can choose between
two orthogonal settings, characterized by the azimuthal angles $\alpha
_{1}\equiv0$ and $\alpha_{2}\equiv\frac{\pi}{2}$, whereas Bob has four possible
measurement directions, namely $\beta_{1}\equiv\alpha_{1}\equiv0$, $\beta
_{2}\equiv\alpha_{2}\equiv\frac{\pi}{2}$, $\beta_{3}\equiv\frac{\pi}{4}$, and
$\beta_{4}\equiv\frac{3\pi}{4}$ (note that the $\beta_{j}$ are not numbered in
ascending order). Therefore, depending on their choice of settings, they
sometimes measure correlations for determining the violation of the CHSH
inequality, namely with the four settings $(\alpha_{1},\beta_{3})$, $(\alpha
_{1},\beta_{4})$, $(\alpha_{2},\beta_{3})$, and $(\alpha_{2},\beta_{4})$, or
they can establish a key, since their outcomes are perfectly anti-correlated
for measurements along $(\alpha_{1},\beta_{1})$ and $(\alpha_{2},\beta_{2})$.
If they choose $(\alpha_{1},\beta_{2})$ or $(\alpha_{2},\beta_{1})$,
i.e.\ orthogonal directions, they discard their results. A schematic of the
measurement directions is shown in Fig.\ \ref{BBMCHSH_SETTINGS}.

Let $P(X\!=\!-Y|ij)$ denote the probability that Alice and Bob obtain
anti-correlated results if they measure along $\alpha_{i}$ and $\beta_{j}$,
respectively, where $i=1,2$ and $j=1,2,3,4$. The (measured) CHSH expression
has the form
\begin{equation}
S  \equiv P(X\!=\!-Y|13)+P(X\!=\!-Y|23)
+P(X\!=\!-Y|24)-P(X\!=\!-Y|14)\leq2\,. \label{chsh
(bbm)}%
\end{equation}
For the (maximally entangled) singlet state it is equal to $1+\sqrt{2}$. The
classical bound is 2, whereas the logical bound is equal to 3.

Let us now assume that an eavesdropper has some knowledge about the choice
of settings of Alice and Bob, for instance by having some insight into their
random number generators. We \textit{model} this knowledge in the following
way: In each run, i.e., for each singlet pair, Eve knows that the combination
of local settings $(\alpha_{i},\beta_{j})$ will happen with probability
$q_{ij}$. For simplicity we assume that one out of the 8 joint settings will
happen with (high) probability $Q\geq\tfrac{1}{8}$, whereas all the other 7
have equal (low) probability $\frac{1-Q}{7}$ to be manifested. The number $Q$
shall be the same for all runs; the setting which it indicates to be most
probable of course changes from run to run. The case $Q=1$ corresponds to
perfect knowledge of the eavesdropper and to the complete lack of free will of
Alice and Bob, whereas $Q= \tfrac{1}{8}$ means that Eve has no knowledge at all.

We impose the following \textit{attack algorithm}: If Eve believes one of
the CHSH settings to be most likely, she sends the corresponding optimal
product state. In general, if $q_{ij}=Q$, which means that the setting
$(\alpha_{i},\beta_{j})$ is most probable from Eve's viewpoint, she intercepts
and sends either $\left|  \alpha_{i}\right\rangle _{\text{A}}\left|  \beta
_{j}+\pi\right\rangle _{\text{B}}$ or $\left|  \alpha_{i}+\pi\right\rangle
_{\text{A}}\left|  \beta_{j}\right\rangle _{\text{B}}$ (by tossing a fair
coin, such that the local results of Alice and Bob are always totally random).
Only in the special case $q_{14}=Q$, Eve sends $\left|  \alpha_{1}%
\right\rangle _{\text{A}}\left|  \beta_{4}\right\rangle _{\text{B}}$ or
$\left|  \alpha_{1}+\pi\right\rangle _{\text{A}}\left|  \beta_{4}%
+\pi\right\rangle _{\text{B}}$. This is the CHSH setting where the probability
of anti-correlation should be minimized, since $P(X\!=\!-Y|14)$ appears with a
minus sign in the CHSH inequality. Therefore, she attacks the CHSH
measurements in order to achieve a maximal violation ($S=3$) \textit{and} the
key establishing measurements to find the key (or rather produce it herself).

To further motivate the choice of this attack algorithm note that (i)
it is canonical in the way that Eve attacks all events in the same way, namely
with the appropriate product state. (ii) The attack is already good enough to
show that the connection between violation of local realism and secure key
distribution is lost in the case in which the eavesdropper has partial knowledge
about the settings. (iii) Eve sends a product state for each pair that is
generated by the source. Hence, Alice and Bob are faced with measurement
results that can be described by local realism but nevertheless can violate
the CHSH inequality (\ref{chsh (bbm)}) due to restricted freedom.

\begin{figure}
\begin{center}
\includegraphics{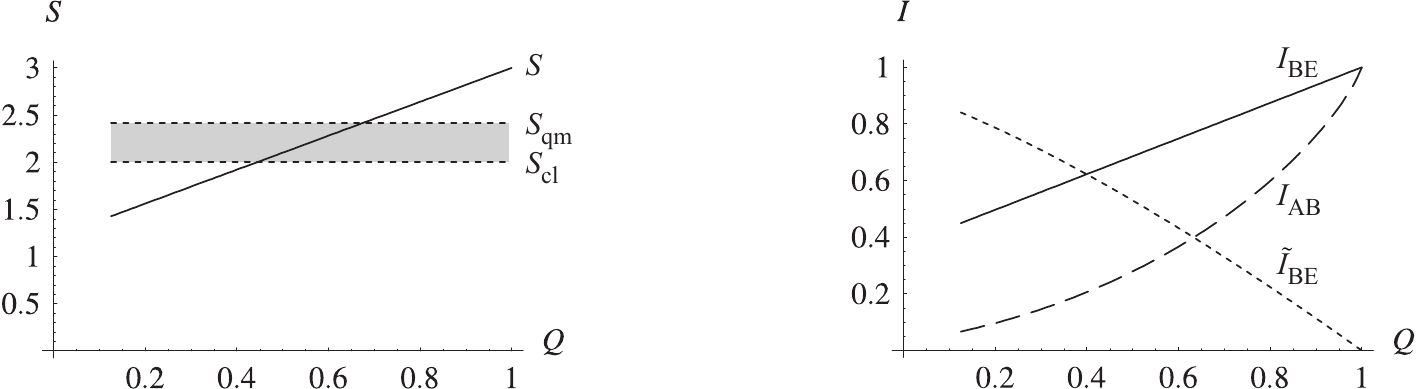}
\end{center}
\caption{\textbf{Left:} The (measured) CHSH expression $S$ as a function of Eve's setting
knowledge $Q$ (solid line). The CHSH inequality with classical bound
$S_{\text{cl}}=2$ (dotted line) is violated for every setting knowledge
$Q>Q_{\text{cl}}\approx0.44$. The quantum bound $S_{\text{qm}}=1+\sqrt{2}$ is
also indicated. \textbf{Right:} The mutual information between Alice and Bob
$I_{\text{AB}}$ (dashed line) and the actual mutual information between Bob
and Eve $I_{\text{BE}}$ (solid line), which is always smaller than (or equal
to) the Alice--Eve mutual information. For every setting knowledge $Q$ one has
$I_{\text{BE}}\geq I_{\text{AB}}$ and thus Alice and Bob can never extract a
secret key. An optimal attack without setting knowledge leads to
$\tilde{I}_{\text{BE}}$ (dotted line). Only for $Q\leq Q_{0}\approx0.63$ the
BBM--CHSH protocol is secure, because Alice and Bob find $I_{\text{AB}}%
\leq\tilde{I}_{\text{BE}}$ and they will not use their key.}
\label{FIG_EVE}
\end{figure}

According to Eve's setting knowledge and the attack strategy, one can compute
the value for the CHSH expression as measured by Alice and Bob. \textit{In the
subensemble} of cases in which, e.g.\ Alice measures along $\alpha_{1}$ and Bob
along $\beta_{3}$, Eve sends with probability $Q$ the product states resulting
in anti-correlations $P(X\!=\!-Y|13)=1$. In the rest of the cases she sends 7
possible ''wrong guesses'' which each happen with probability $\frac{1-Q}{7}$
and for each of them the probability for anti-correlations takes values
between $\frac{1}{2}$ and $\cos^{2}\frac{\pi}{8}\approx0.854$, depending on
the specific wrong attack. The measured probability $P(X\!=\!-Y|13)$ is the
expectation value of all 8 sets of anti-correlated results weighted with their
probabilities to happen. Analogously, the other probabilities for
anti-correlation are calculated and we find:
\begin{align}
P(X\! &  =\!-Y|13)=Q+\tfrac{1-Q}{7}\left(  \tfrac{5}{2}+2\,\cos^{2}\tfrac{\pi
}{8}\right)  \!,\\
P(X\! &  =\!-Y|23)=P(X\!=\!-Y|13)\,,\\
P(X\! &  =\!-Y|24)=Q+\tfrac{1-Q}{7}\left(  \tfrac{5}{2}+\cos^{2}\tfrac{\pi}%
{8}+\sin^{2}\tfrac{\pi}{8}\right)  \!,\\
P(X\! &  =\!-Y|14)=\tfrac{1-Q}{7}\left(  \tfrac{5}{2}+\cos^{2}\tfrac{\pi}%
{8}+\sin^{2}\tfrac{\pi}{8}\right)  \!.
\end{align}
The CHSH expression finally results in:
\begin{equation}
S=3\,Q+\tfrac{1-Q}{7}\left(  5+4\,\cos^{2}\tfrac{\pi}{8}\right)
\approx1.2+1.8\,Q\,.\label{chsh}%
\end{equation}
Thus, the logical bound $S_{\text{log}}\equiv3$ is reached in the limit
$Q\rightarrow1$. The classical bound of $S_{\text{cl}}\equiv2$ is beaten for
all $Q>Q_{\text{cl}}\approx0.44$ and the quantum mechanics (Cirel'son) bound
$S_{\text{qm}}\equiv1+\sqrt{2}\approx2.41$ is beaten for setting knowledge
$Q>Q_{\text{qm}}\approx0.67$. If $Q$ is larger than $Q_{\text{qm}}$, Eve
should reduce the strength of her attack, e.g.\ by mixing some noise into her
product states, for otherwise even the quantum bound would be broken. The CHSH
expression (\ref{chsh}) and the bounds are shown in the left panel of Fig.\ \ref{FIG_EVE}.

When is Eve's knowledge about the settings also sufficient to find out the key
which is established by Alice and Bob? To answer this question, we have to
compute mutual informations between the parties. The mutual information
between Alice and Bob is determined by the bit error rate~\cite{CRYPTO_REVIEW} which
they can compute in the \textit{subensembles} where they measured along
$\alpha_{1}\!=\!\beta_{1}\!=\!0$ or $\alpha_{2}\!=\!\beta_{2}\!=\!\frac{\pi
}{2}$. Let us consider the first; the error rate in the second is the same for
symmetry reasons. The bit error rate $D$ is given by the sum of 8 terms
corresponding to the 8 settings that were potentially possible from Eve's
point of view. Each term is the probability with which Eve believed this event
would happen --- $Q$ for the event $(\alpha_{1},\beta_{1})$ itself and
$\frac{1-Q}{7}$ for all the others (the wrong guesses), corresponding to our
definition of the setting knowledge --- multiplied with the probability that
the attack $(\alpha_{i},\beta_{j})$ leads to a correlation (error) rather than
an anti-correlation as for the original singlet state. This ''destruction
probability'' is $0$ for the ''correct'' event $(\alpha_{1},\beta_{1})$. It is
$\sin^{2}\frac{\pi}{8}$ for both $(\alpha_{1},\beta_{3})$ and $(\alpha
_{1},\beta_{4})$, and $\frac{1}{2}$ for all the others (where an orthogonal
state was sent to Alice or Bob). Finally, we find the bit error rate
\begin{equation}
D=\tfrac{1-Q}{7}\left(  \tfrac{5}{2}+2\,\sin^{2}\tfrac{\pi}{8}\right)
\approx0.4\,(1-Q)\,.\label{ber}%
\end{equation}

The mutual information between Alice and Bob reads \cite{CRYPTO_REVIEW}:
\begin{equation}
I_{\text{AB}}\equiv1-H(D)\,,\label{IAB}
\end{equation}
The maximal mutual information between Alice (or Bob for symmetry reasons) and
Eve from Alice's and Bob's viewpoint, which can be attained by an optimal
attack of Eve for a given error rate $D$ and \textit{under the condition that
Eve has no setting knowledge} is given by~\cite{CRYPTO_REVIEW}:
\begin{equation}
\tilde{I}_{\text{AE}}=\tilde{I}_{\text{BE}}=1-H(\tfrac{1}{2}\!+\!\sqrt
{D-D^{2}})\,.\label{IAE-tilde}%
\end{equation}

The \textit{actual} mutual information between Alice and Eve, $I_{\text{AE}}$,
can be computed from the conditional entropy $H(A|E)$ by
$I_{\text{AE}}=H(A) - H(A|E)$. 
Since the outcomes of Alice are
locally random for all possible attacks
the Shannon information of Alice is $H(A)=1$. 
As all chosen settings are publicly
revealed after the measurements, Eve can compute $H(A|E)$ in
the \textit{subensemble} of the key establishing measurement $(\alpha
_{1},\beta_{1})$. (If Alice and Bob measure along $(\alpha_{2},\beta_{2})$,
the result does not change.) The calculation itself is straightforward, once
one realizes that $H(A|e)=0$ for all 4 events $e$ in which Eve
(justly) believed that Alice would choose $\alpha_{1}$, as Eve knows her
result in this case. 
If Eve made the (wrong) guess $\alpha_{2}$ then
$H(A|e)=1$ for these 4 possible events, for Alice measures in
the orthogonal direction $\alpha_{1}$. Thus, $H(A|E)
=4\,\frac{1-Q}{7}$ and
\begin{equation}
I_{\text{AE}}=1-4\,\tfrac{1-Q}{7}=\tfrac{3}{7}+\tfrac{4}{7}\,Q\,.\label{IAE}%
\end{equation}
Analogously, one can find the \textit{actual} mutual information between Bob
and Eve:
\begin{equation}
I_{\text{BE}}=1-\tfrac{1-Q}{7}\left(  2+4\,H(\cos^{2}\tfrac{\pi}{8})\right)
\approx0.37+0.63\,Q\,,
\end{equation}
which is always smaller than (or equal to) $I_{\text{AE}}$. We have
\begin{equation}
I_{\text{AB}}\leq I_{\text{BE}}%
\end{equation}
for all $Q$ and equality only holds for $Q=1$. Alice and Bob can never extract
a secret key, since the condition $I_{\text{AB}}>I_{\text{AE}}$ is never
fulfilled (right panel of Fig.\ \ref{FIG_EVE}).

If Eve has no setting knowledge the quantum protocol
is secure if and only if quantum bit error rate
is below $D < D_0 =\tfrac{1}{2}\,(1-\tfrac{1}{\sqrt{2}})\approx0.15$,
which in turn is equivalent to $S>2$ \cite{CRYPTO_REVIEW}.
In the present case the 
critical error rate $D_{0}$ corresponds, according to (\ref{ber}), to a setting knowledge $Q_{0}\approx0.63$.
For this knowledge $I_{\text{AB}}=\tilde{I}_{\text{BE}}$.
If $Q>Q_{0}$, the BBM--CHSH protocol is insecure, since $I_{\text{AB}} \le I_{\text{BE}}$ 
\textit{and} Alice and Bob find both their error rate to be
sufficiently small (below $D_{0}$) and the CHSH inequality (\ref{chsh (bbm)})
to be violated, which makes them think they are safe.
In fact, for the CHSH expression not to exceed the quantum value
the setting knowledge should be below $Q_{\text{qm}}\approx0.67$.
 For $Q\leq Q_{0}$ Eve's
setting knowledge is ''insufficient'' and the protocol becomes secure: Alice
and Bob cannot extract a secret key because still $I_{\text{AB}}<I_{\text{BE}%
}$, but they find $I_{\text{AB}}\leq\tilde{ I}_{\text{BE}}$ and know that
there might be an eavesdropper and thus they will not use the key. 
For $0.44\approx Q_{\text{cl}}<Q\leq Q_{0}\approx0.63$ Alice and Bob find the CHSH
inequality (\ref{chsh (bbm)}) to be violated ($S>2$) and nonetheless they
cannot extract a secret key ($I_{\text{AB}}\leq\tilde{I }_{\text{BE}}$, $D\geq D_{0}$). 
Therefore, we deduce that the equivalence between the violation of
Bell's inequality (with complete freedom) and the secure key distribution
(without freedom) is lost.

If Alice and Bob knew $Q$, which means they knew to which extent their freedom
is restricted, and if they calculated the maximal $\Delta_{\text{CHSH}}$ under
the constraint of an insecure key, $I_{\text{AB}}\leq\min\{I_{\text{AE}%
},I_{\text{BE}}\}$, \textit{for all possible attacks}, then a violation of the
CHSH inequality with the new bound $2+\Delta_{\text{CHSH}}$ would be
equivalent to the possibility of efficient secret key extraction (unless the
new bound is larger than $S_{\text{qm}}=1+\sqrt{2}$). A violation of this new
bound is equivalent to statement that the classical bound 2 is violated in the
case of total freedom and for this situation there exists a complete
equivalence between the CHSH inequality violation and the security of the BBM
protocol~\cite{ACIN_IJQI,SCARANIGISIN}.

To conclude,
if some information leaks out of the authorized parties' laboratories
the violation of the standard CHSH inequality is not equivalent to a secure key
distribution. Nevertheless, one can define a new (higher) bound whose
violation indeed guarantees the security of the key. Therefore, one can keep
the security while, to some extent, relaxing the assumption that no
information about the measurement settings is revealed to an eavesdropper, as long as the amount of this
information is known.

\section{Qudit quantum cryptography with composite systems [P2]}

Quantum cryptography as just described uses two-level quantum systems.
If higher-dimensional systems (qudits) are at disposal
one can increase the security of the quantum protocol,
i.e.\ the level of allowed errors can be bigger \cite{MOHAMED,BPP,BRUS}.
The question to be answered in this section
is whether such cryptosystems are feasible.
It will be shown that two-bases quantum cryptography with qudits
composed of two lower-dimensional subsystems
can be realized with individual measurements on
subsystems accompanied with classical communication.
We will use the description of a $d$-level system
in terms of unitary generalizations of Pauli operators: $S_{kl} = S_x^k S_z^l$
with $k,l=0,...,d-1$ (Appendix B).

Generally, an arbitrary measurement can be viewed as a unitary
evolution of the system which transforms the eigenvectors of the observable 
into the eigenvectors which can be distinguished by the measurement apparatus.
Thus, we solve the eigenproblem of the generalized
Pauli operators, and next apply it to the measurements of quantum cryptography.

\subsection{Eigenproblem of the generalized Pauli operators}

The matrix of any $S_{kl}$ operator, written in the $S_z$ basis $| \kappa \rangle_z$,
has only $d$ non-vanishing entries, one per column and row:
\begin{equation}
S_{kl} = 
\left(
\begin{array}{ccccccc}
0      & 0        &        & 0                 & \alpha_d^{(d-k)l} &        & 0 \\
\vdots & \vdots   &        & \vdots            & \vdots          &        & \vdots \\
0      & 0        &        & 0                 & 0               &        & \alpha_d^{(d-1)l} \\
1      & 0        & \ldots & 0                 & 0               & \ldots & 0\\
0      & \alpha_d^l &        & 0                 & 0               &        & 0\\
\vdots & \vdots   &        & \vdots            & \vdots          &        & \vdots \\
0      & 0        &        & \alpha_d^{(d-k-1)l} & 0               &        & 0
\end{array}
\right). \nonumber
\end{equation}
The only non-vanishing element of the first column, a ``1'',
appears in the $k$th row (recall that $k=0,1,...,d-1$).
Generally, the matrix elements of the $S_{kl}$ operator,
$[S_{kl}]_{rm}$, are given by 
$[S_{kl}]_{rm} = \delta_{r-k,m} \alpha_d^{ml}$,
where $\delta_{x,y}$ is the Kronecker delta.
Since every $S_{kl}$ is unitary it can be diagonalized:
\begin{equation}
S_{kl} = V D V^{\dagger},
\label{DIAGONALIZATION}
\end{equation}
where $V$ 
is a unitary matrix the columns of which are eigenstates of $S_{kl}$, 
$V = (|0 \rangle, ..., |d-1 \rangle)$,
and $D$ is a diagonal matrix with entries being eigenvalues of $S_{kl}$, denoted by $\lambda_{j}$.
The form of $[S_{kl}]_{rm}$ and  (\ref{DIAGONALIZATION}) imply 
conditions, which must be satisfied by the eigenvectors $|j \rangle$:
\begin{equation}
\sum_{j = 0}^{d-1} \lambda_{j} v_{k+m,j} v_{m,j}^{*} = \alpha_d^{ml}, \quad \textrm{for all} \quad m=0,...,d-1,
\label{CONDITION}
\end{equation}
where $v_{i,j}$ is the element of the matrix $V$ in the $i$th row and $j$th column, 
i.e.\ $i$th coefficient of the eigenvector $|j \rangle$.
A study of this condition allows one to construct the eigenbasis.

We first present the result, that is give a candidate for an eigenbasis,
and then prove that this is indeed the eigenbasis.
Depending on $k$ 
the eigenstates of $S_{kl}$ are given 
by superposition of different number of states $| \kappa \rangle_z$.
Let us denote by $f$ the smallest multiple of $k$ modulo $d$,
i.e. $f = \min_{x=1,2,...} [k x]_d$.
The values of $f$ are taken to be strictly positive, i.e. $f=1,2,...$.
Within this definition $k=wf$ is a multiple of $f$.
Eigenstates $| j \rangle$
involve every $f$th state of the $S_z$ basis:
\begin{equation}
| \kappa \rangle_z = |a + \eta' f \rangle_z = |a + \eta k \rangle_z,
\label{STATES_INVOLVED}
\end{equation}
where $\eta'=0,...,d/f-1$ and $a=0,...,f-1$,
and of course $\eta' = w \eta$.
Both $\eta'$ and $\eta$ enumerate different states $| \kappa \rangle_z$
into which $| j \rangle$ is decomposed, i.e. $\eta=0,...,d/f-1$.
All other coefficients vanish.
The whole eigenbasis splits into $f$
groups of eigenvectors which are superposition of vectors $|\kappa \rangle_z$
with fixed $a$.
There are $d/f$ eigenvectors within each group.
To uniquely identify the eigenvector $| j \rangle$
one needs to specify $a$, and additionally an integer $g=0,...,d/f-1$, i.e. $j = j_{g,a}$.
With these definitions we can present
the form of eigenvectors (a candidate):
\begin{equation}
|j \rangle = |j_{g,a} \rangle = \sum_{\eta=0}^{d/f-1} v_{\eta k,j_{g,a}} | a + \eta k \rangle_z,
\label{EIGENVECTORS}
\end{equation}
with
\begin{equation}
v_{\eta k,j_{g,a}} = \frac{1}{\sqrt{d/f}} \lambda_{j_{g,0}}^{- \eta} \alpha_d^{\frac{ \eta (\eta -1)}{2}kl},
\label{EIGEN_COEFFICIENTS}
\end{equation}
where generally the eigenvalues $\lambda_{j_{g,a}}$ are given by:
\begin{equation}
\lambda_{j_{g,a}} = e^{i \varphi} \alpha_d^{gf+al},
\label{EIGENVALUES}
\end{equation}
and $e^{i \varphi}$ is a phase factor common to all the eigenvalues.\footnote{To get rid of this phase, instead of $S_{kl}$
one can consider an operator $e^{-i \varphi} S_{kl}$.}
We will show below how to compute this phase.
Note that the coefficients (\ref{EIGEN_COEFFICIENTS}) are independent of $a$.
This can be intuitively explained by noting
that for different $a$'s the eigenvectors $|j_{g,a} \rangle$
are orthogonal just due to the fact that they involve 
orthogonal vectors $| a + \eta k\rangle_z$.
For a fixed $a$, but different $g$'s, the vectors (\ref{EIGENVECTORS}) with coefficients (\ref{EIGEN_COEFFICIENTS})
are also orthogonal.
Their scalar product
$\langle j_{g',a} | j_{g,a} \rangle = (d/f)^{-1} \sum_{\eta = 0}^{d/f-1}(\lambda_{j_{g',0}} \lambda_{j_{g,0}}^{-1})^{\eta}$
involves the product of 
$\lambda_{j_{g',0}} \lambda_{j_{g,0}}^{-1} = \alpha_d^{(g'-g)f} = \alpha_{d/f}^{(g'-g)}$,
and the whole sum is equal to the Kronecker delta $\delta_{g',g}$.
Thus, the vectors $| j_{g,a} \rangle$ form an orthonormal basis.

To prove that this basis is the eigenbasis one needs to check whether
\begin{equation}
S_{kl} |j_{g,a} \rangle = \lambda_{j_{g,a}} |j_{g,a} \rangle.
\label{EIGEN_JA}
\end{equation}
The action of $S_{kl}$, defined in the Appendix B by (\ref{APPB_SZSX_DEF}), on the state $|j_{g,a} \rangle$ is given by:
\begin{equation}
S_{kl} |j_{g,a} \rangle =
\frac{1}{\sqrt{d/f}} \sum_{\eta =0}^{d/f-1} \lambda_{j_{g,0}}^{- \eta} \alpha_d^{\frac{ \eta (\eta -1)}{2}kl} 
\alpha_d^{l(a+ \eta k)}| a+(\eta+1)k\rangle_z, \nonumber
\end{equation}
Changing the summation index to $\eta_1 = \eta+1$
one finds:
\begin{equation}
S_{kl} |j_{g,a} \rangle = 
\lambda_{j_{g,0}} \alpha_d^{al} 
\frac{1}{\sqrt{d/f}} \sum_{\eta_1=1}^{d/f} \lambda_{j_{g,0}}^{-\eta_1} \alpha_d^{\frac{\eta_1(\eta_1-1)}{2}kl} 
| a+\eta_1 k\rangle_z.
\label{S_J}
\end{equation}
The coefficients within the sum are equal to the coefficients of the initial
$| j_{g,a} \rangle$ state, (\ref{EIGEN_COEFFICIENTS}), 
if for the last term in (\ref{S_J}),
for which $\eta_1 = d/f$, one has:
\begin{equation}
\lambda_{j_{g,0}}^{-d/f} = \alpha_d^{-\frac{1}{2}\frac{d}{f}(\frac{d}{f}-1)kl}.
\label{COMPUTE_EIGENVALUES}
\end{equation}
This equation gives the eigenvalues $\lambda_{j_{g,0}}$.
If one takes one of the solutions to (\ref{COMPUTE_EIGENVALUES}), say $\lambda_{j_{0,0}}$,
in the form $\lambda_{j_{0,0}} = e^{i \varphi}$, then the remaining solutions
are given by $\lambda_{j_{g,0}} = e^{i \varphi} \alpha_{d/f}^g$.
Indeed, if $\lambda_{j_{0,0}}$ satisfies (\ref{COMPUTE_EIGENVALUES}),
then also $\lambda_{j_{g,0}}$ satisfy it.
The eigenvalues for other $a$'s are given by:
\begin{equation}
\lambda_{j_{g,a}} = \lambda_{j_{g,0}} \alpha_{d}^{al}.
\label{LAMBDA_A_DEF}
\end{equation}
Note that degeneracies in the eigenproblem can only appear for $f \ne 1$
(since for $f=1$ one has only $a=0$, and $g$ takes all $d$ different values).

Practically, to compute the eigenvectors one should find the value of $f$.
If it is different than unity, set $a=0$ and compute
the coefficients according to Eq.\ (\ref{EIGEN_COEFFICIENTS}).
For other values of $a$ the coefficients are the same,
but now they are multiplied with orthogonal vectors $| a + \eta k\rangle_z$.
To compute the eigenvalues one needs to solve Eq.\ (\ref{COMPUTE_EIGENVALUES}).
Moreover, once $\lambda_{j_{g,0}}$ has been found for some $g$
the other eigenvalues for $a=0$ 
are obtained by multiplication of $\alpha_{d/f}$: 
$\lambda_{j_{g',0}} = \lambda_{j_{g,0}} \alpha_{d/f}^{g'-g}$.
The eigenvalues for $a \ne 0$ can be found from (\ref{LAMBDA_A_DEF}).

\emph{Example}.
Take $S_{43}$ for $d=6$, i.e.\ $k=4, l=3$ and one finds $f=2$.
Put $a=0$.
From (\ref{COMPUTE_EIGENVALUES}) one has 
$\lambda_{j_{g,0}} = e^{ig\frac{2\pi}{3}} = \alpha_6^{2g}=\alpha_3^{g}$ ($e^{i \varphi}=1$).
According to (\ref{LAMBDA_A_DEF}),
the eigenvalues $\lambda_{j_{g,1}}$ are equal to $\lambda_{j_{g,1}} = -\lambda_{j_{g,0}}$.
This can be summarized in the eigenbasis:
\begin{equation}
\begin{array}{lcl}
|0 \rangle = \frac{1}{\sqrt{3}} \Big(|0\rangle_z + |2\rangle_z + |4\rangle_z \Big), & \quad &
|1 \rangle = \frac{1}{\sqrt{3}} \Big(|1\rangle_z + \alpha_3^2 |3\rangle_z + \alpha_3 |5\rangle_z \Big), \\
|2 \rangle = \frac{1}{\sqrt{3}} \Big(|0\rangle_z + \alpha_3 |2\rangle_z + \alpha_3^2 |4\rangle_z \Big), & &
|3 \rangle = \frac{1}{\sqrt{3}} \Big(|1\rangle_z + |3\rangle_z + |5\rangle_z \Big), \\
|4 \rangle = \frac{1}{\sqrt{3}} \Big(|0\rangle_z + \alpha_3^2 |2\rangle_z + \alpha_3 |4\rangle_z \Big), & &
|5 \rangle = \frac{1}{\sqrt{3}} \Big(|1\rangle_z + \alpha_3 |3\rangle_z + \alpha_3^2 |5\rangle_z \Big).
\end{array}
\nonumber
\end{equation}

\subsection{Cryptography}

Consider the two-bases quantum cryptography protocol
with $d$-level systems, as described in \cite{MOHAMED}. 
One has a qudit randomly prepared in a state of a certain basis, or of another basis,  
which is unbiased with respect to the first one, 
i.e. every state form the first basis has equal overlap with all the states of another basis
\cite{MUBS,MUBS2}.
The measurement basis is also randomly chosen between these two.\footnote{Interestingly, the performance of the two-bases protocol
is only slightly worse than the performance of the many-bases protocol
(compare Table I of \cite{MOHAMED}).}
The two mutually unbiased bases 
can be chosen as the eigenbases of the $S_z$ and $S_x$ generalized Pauli operators.
Applying relation (\ref{EIGEN_COEFFICIENTS}) to $S_x = S_{10}$ 
one finds, for arbitrary dimension, 
the well-known Fourier relation between the $S_z$ and $S_x$ eigenbases:
\begin{equation}
|j\rangle_x = \frac{1}{\sqrt{d}} \sum_{\kappa = 0}^{d-1} \alpha_d^{- \kappa j} |\kappa \rangle_z,
\end{equation}
i.e. $| _x \langle j | \kappa \rangle_z| = 1/\sqrt{d}$ for all $j$ and $\kappa$ (indeed the bases are mutually unbiased).

Consider a $d$-level system encoded in two subsystems.
Let us define the eigenbasis of a global $S_z$ operator as:
\begin{equation}
| \kappa \rangle_z = | d_0 \kappa_1 + \kappa_0 \rangle_z \equiv |\kappa_1 \rangle_1 |\kappa_0 \rangle_0,
\label{DEFINITION}
\end{equation}
where $\kappa_i = 0,...,d_i - 1$, and $|\kappa_0 \rangle_0$, $|\kappa_1 \rangle_1$
denote the states of subsystems ``0'' and ``1'', respectively.
Within this definition a measurement of the global observable $S_z$
is equivalent to individual measurements on the components.
These individual measurements reveal the values of $\kappa_0$ and $\kappa_1$,
and the eigenvalue of $S_z$ is $\alpha_d^{d_0 \kappa_1 + \kappa_0}$.

To measure $S_x$ one uses the definition (\ref{DEFINITION}) and the fact that 
the dimension of a global system, $d$,
is the product of dimensions of subsystems,
$d=d_1d_0$,
and finds that:
\begin{equation}
| j \rangle_x = \frac{1}{\sqrt{d_1}} 
\sum_{\kappa_1=0}^{d_1-1} \alpha_d^{-d_0 \kappa_1 j} |\kappa_1 \rangle_1 \otimes
\frac{1}{\sqrt{d_0}} 
\sum_{\kappa_0=0}^{d_0-1} \alpha_d^{-\kappa_0 j}|\kappa_0 \rangle_0,
\label{SEPARATED}
\end{equation}
where we have used the symbol $\otimes$
to stress the factorization of this state.
For $j = j_1 + d_1 j_0$
the state of subsystem ``1'' reads:
\begin{equation}
\frac{1}{\sqrt{d_1}}  \sum_{\kappa_1=0}^{d_1-1} \alpha_d^{-d_0 \kappa_1 j_1-d_0 \kappa_1 d_1 j_0} |\kappa_1 \rangle_1.
\end{equation}
Since $\alpha_d^{d_0} = \alpha_{d_{1}}$, see (\ref{ALPHA}),
and $e^{-i 2\pi \kappa_1 j_0} = 1$ a
measurement on this subsystem in the basis
\begin{equation}
|\phi_{j_1} \rangle_1 = \frac{1}{\sqrt{d_1}}  \sum_{\kappa_1=0}^{d_1-1} \alpha_{d_1}^{-\kappa_1 j_1} |\kappa_1 \rangle_1
\end{equation}
reveals the value of $j_1$.
The value of $j_0$ can be measured once $j_1$ is known.
A measurement in the basis
\begin{equation}
|\psi_{j_0} \rangle_0 = \frac{1}{\sqrt{d_0}}  \sum_{\kappa_0=0}^{d_0-1} \alpha_{d}^{-(j_1 + d_1 j_0) \kappa_0} |\kappa_0 \rangle_0
\end{equation}
on the subsystem ``0''
reveals the value of $j_0$.
In this way all values of $j$ can be measured
using individual measurements and classical communication.
Since the measurement on subsystem ``0'' depends on 
the outcome of the measurement on subsystem ``1'',
the (classical) information about the outcome needs to be 
fed-forward to the device measuring subsystem ``0''.

\section{Quantum communication complexity}

In a communication complexity problem (CCP) \cite{YAO},
separated parties performing \emph{local} computations
exchange information in order to accomplish a \emph{globally}
defined task,
which is impossible to solve singlehandedly.
Two types of CCPs can be distinguished:
in the first type one asks for a minimal amount of information exchange
necessary to solve a task with certainty \cite{QCCP2,QCCP3,QCCP4};
in the second type one maximizes the probability
of successfully solving a task with
a restricted amount of communication \cite{BZPZ,QCCP4,QCCP5,QCCP_QUTRIT_PRL}.
Such studies aim, e.g., at a speedup of a distributed computation
of very large scale integrated circuits
and data structures \cite{CC_BOOK}.

Communication complexity was introduced in 1979 by Yao \cite{YAO},
who was also the first to introduce the quantum version.
However, only recently it was noticed that the problems
are ultimately linked with violation of local realism.
It was shown that one can link a CCP with every Bell
inequality for qubits \cite{BZPZ}. 
Quantum protocols for the CCPs,
utilizing entangled states which violate the inequalities,
outperform the best classical protocol \cite{EXP_CCP}.

Interestingly, it is possible to recast some entanglement-based CCPs
in terms of a single qubit sequentially transmitted
between the participants \cite{GALVAO}.
Also in this case the limits of performance
of classical protocols 
are described by a form of ``Bell inequality'' \cite{EXP_CCP}.
This type of communication complexity
was experimentally realized in the group of Weinfurter \cite{EXP_CCP}.

In this section we follow the general link 
between violation of Bell inequality and CCPs \cite{BZPZ}
for the case of the
inequality with an arbitrary number of settings [P3].
Next, we give quantum communication complexity protocols 
using higher-dimensional entangled systems [P7]
that are linked with the Bell inequality for multilevel systems \cite{CGLMP}.

\subsection{Qubits [P3]}

Let us focus on a variant of a CCP,
in which each of $N$ separated partners
receives arguments, $y_n=\pm 1$ and $x_n = 0,...,M-1$,
of some globally defined function, $\mathcal{F} \equiv \mathcal{F}(y_1,x_1,...,y_N,x_N)$.
The inputs of the $n$th party are not known to any other party.
Assume the bits $y_n$ are randomly distributed,
and inputs $x_n$ (representing $\textrm{lg} M$ bits of information) 
can in general be distributed
according to a weight $\mathcal{W}(x_1,...,x_2)$.
The goal is to maximize the probability
that Alice arrives at the correct value of the function,
under the restriction of $N-1$ bits of overall
communication.
Before participants receive their inputs
they are allowed to do anything
from which they can derive benefit.
In particular, they can share some correlated strings of numbers
in the classical scenario or entangled states in the quantum case.

\emph{The problem.}
Following \cite{BZPZ} one chooses for a task-function:
\begin{eqnarray}
\mathcal{F} &=& y_1...y_N {\rm Sign}[\cos(\phi_{x_1}^1 + ... + \phi_{x_N}^N)] = \pm 1,
\label{FF}
\end{eqnarray}
with the angles defined by (\ref{ANGLES}).
Additionally, the $x_n$ inputs are distributed with the weight
\begin{equation}
\mathcal{W}(x_1,...,x_2) = (1/\mathcal{N}) |\cos(\phi_{x_1}^1 + ... + \phi_{x_N}^N)|,
\label{WW}
\end{equation}
with the normalization factor 
$\mathcal{N} = \sum_{x_1...x_N=0}^{M-1}|\cos(\phi_{x_1}^1 + ... + \phi_{x_N}^N)|$.
After the communication takes place,
if Alice misses some of the random variables $y_n$,
her ``answer'' can only be random.
Thus, in an optimal protocol 
each party must communicate one bit.
Essentially, there are two communication structures
which lead to a non-random answer:
(i) each party transmits one bit directly to Alice,
and (ii) sequence of a peer-to-peer exchanges
with Alice at the end.
The task is to maximize the probability
of correct answer 
$\mathcal{A} \equiv \mathcal{A}(y_1,x_1,...,y_N,x_N)$.
Since both $\mathcal{A}$ and $\mathcal{F}$
are dichotomic variables this amounts in maximizing:
\begin{equation}
P_{{\rm correct}}
= \frac{1}{2^N}
\sum_{{\bf y},{\bf x}} \mathcal{W}(x_1,...,x_2)
P_{{\bf y},{\bf x}}(\mathcal{A} \mathcal{F} = 1),
\end{equation}
where $\frac{1}{2^N}$ describes (random) distribution of $y_n$'s,
and $P_{{\bf y},{\bf x}}(\mathcal{A} \mathcal{F} = 1)$
is a probability that $\mathcal{A} = \mathcal{F}$
for given inputs ${\bf y} \equiv (y_1,...,y_N)$ and ${\bf x} = (x_1,...,x_N)$.
It is useful to express the last probability
in terms of an average value (over many runs of the protocol) 
of a product $\langle \mathcal{A} \mathcal{F} \rangle_{{\bf y},{\bf x}}$:
\begin{equation}
P_{{\bf y},{\bf x}}(\mathcal{A} \mathcal{F} = 1) = \frac{1}{2} \Big[ 1 + \langle \mathcal{A} \mathcal{F} \rangle_{{\bf y},{\bf x}} \Big].
\end{equation}
Since $\mathcal{F}$ is independent of $\mathcal{A}$,
and for given inputs it is constant, one has
$\langle \mathcal{A} \mathcal{F} \rangle_{{\bf y},{\bf x}} =  \mathcal{F} \langle \mathcal{A} \rangle_{{\bf y},{\bf x}}$.
Finally the probability of a correct answer reads 
$P_{{\rm correct}} = \frac{1}{2}[1 + (\mathcal{F},\mathcal{A})]$,
and it is in one-to-one correspondence with
a ``weighted'' scalar product (average success):
\begin{equation}
(\mathcal{F},\mathcal{A})
= \frac{1}{2^N}\sum_{{\bf y},{\bf x}} 
 \mathcal{W}(x_1,...,x_2) \mathcal{F} \langle \mathcal{A} \rangle_{{\bf y},{\bf x}}.
\end{equation}
Using the definitions (\ref{WW}) for 
$\mathcal{W}$ and (\ref{FF}) for $\mathcal{F}$
one gets:
\begin{equation}
(\mathcal{F},\mathcal{A})
= \frac{1}{2^N} \frac{1}{\mathcal{N}} \sum_{{\bf y},{\bf x}} 
 y_1...y_N \cos(\phi_{x_1}^1 + ... + \phi_{x_N}^N) \langle \mathcal{A} \rangle_{{\bf y},{\bf x}},
 \label{QUBIT_CCP_SUCCESS}
\end{equation}
with angles given by (\ref{ANGLES}).
We focus our attention on maximization of this quantity.

\emph{Classical scenario.}
In the \emph{best} classical protocol
each party locally computes a bit function $e_n = y_n f(x_n,\lambda)$,
with $f(x_n,\lambda) = \pm 1$,
where $\lambda$ denotes some previously shared
classical resources.
Next, the bit is
sent to Alice, who puts as an answer
the product
$\mathcal{A}_c = y_1 f(x_1,\lambda) e_2...e_N = y_1 ... y_N f(x_1,\lambda)...f(x_N,\lambda)$.
The same answer can be reached
in the peer-to-peer strategy,
simply the $n$th party sends $e_n = y_n f(x_n,\lambda) e_{n-1}$.
For the given inputs the procedure is always the same,
i.e. $\langle \mathcal{A}_c \rangle_{{\bf y},{\bf x}} = \mathcal{A}_c$.
To prove the optimality of this protocol,
one follows the proof of \cite{EXP_CCP},
with the only difference that $x_n$ is a $M$-valued variable now.
This, however, does not invalidate any of the steps
of \cite{EXP_CCP}, and we will not repeat that proof.

Inserting the product form of $\mathcal{A}_c$
into the average success (\ref{QUBIT_CCP_SUCCESS}),
using the fact that $y_n^2=1$, and summing over all $y_n$'s
one obtains
\begin{equation}
(\mathcal{F},\mathcal{A}_c) = 
\frac{1}{\mathcal{N}} \sum_{x_1...x_N=0}^{M-1} \! \! \! \! \! 
\cos(\phi_{x_1}^1 + ... + \phi_{x_N}^N)
f(x_1,\lambda)...f(x_N,\lambda),
\end{equation}
which has the same structure as the local realistic expression (\ref{INEQ_DETER}).
Thus, the highest classically achievable average success
is given by a local realistic bound:
$\max (\mathcal{F},\mathcal{A}_c) = (1/\mathcal{N}) B_{LR}(N,M)$.

\emph{Quantum scenario.}
In the quantum case participants share a $N$-party entangled state $\rho$ before delivery of the inputs.
After receiving inputs each party measures the $x_n$th observable on the state,
where the observables are enumerated as in the Bell inequality (\ref{MS_INEQUALITY}).
This results in a measurement outcome, $f_n$.
Each party sends $e_n = y_n f_n$ to Alice,
who then puts as an answer the product
$\mathcal{A}_{q} = y_1...y_N f_1 ... f_N$.
For the given inputs the average answer
reads $\langle \mathcal{A}_{q} \rangle_{{\bf y},{\bf x}} = y_1...y_N  \langle f_1 ... f_N \rangle
= y_1...y_N E_{x_1...x_N}^{\rho}$,
and the maximal average success is given by a quantum value:
\begin{equation}
(\mathcal{F},\mathcal{A}_q) = 
\frac{1}{\mathcal{N}} \sum_{x_1...x_N=0}^{M-1} \! \! \! \! \! 
\cos(\phi_{x_1}^1 + ... + \phi_{x_N}^N)
E_{x_1...x_N}^{\rho}.
\end{equation}
The average advantage of quantum versus classical protocol
can be quantified by $(\mathcal{F},\mathcal{A}_q)/(\mathcal{F},\mathcal{A}_c)$
which is equal to a violation factor, $V(N,M)$, introduced before in Eq.\ (\ref{VIOL_FACTOR}).
Thus, 
all the states which violate the Bell inequality (including bound entangled states)
are a useful resource for the communication complexity task.
Optimally one should use the GHZ states $| \psi^{\pm} \rangle$,
as they maximally violate the inequality.

Alternatively, one can compare the probabilities of success, $P_{\textrm{correct}}$,
in the quantum and 
classical case.
In Table \ref{TABLE_ADV}
we gather the ratios between quantum
and classical success probabilities
for small number of participants.
Clearly, one outperforms classical protocols
for every $N$ and every $M$. 
\begin{table}
\begin{center}
\begin{tabular}{c  c c c c c}
\hline \hline 
$N \backslash M$  &   2     &   3     &   4   &   5   &   $\infty$   \\ 
\hline
2 \qquad & 1.1381  \qquad     & 1.1196 \qquad         &1.1009 \qquad   &1.1002  \qquad      &1.0909 \\
3 \qquad & 1.3333  \qquad     & 1.2919  \qquad        &1.2815 \qquad   &1.2773  \qquad       &1.2709 \\
4 \qquad & 1.3657  \qquad    & 1.4395    \qquad      &1.4038   \qquad  &1.4258   \qquad      &1.4192 \\
5 \qquad & 1.6000  \qquad     & 1.5582   \qquad       &1.5467  \qquad   &1.5418   \qquad      &1.5336  \\
\hline \hline
\end{tabular}
\caption{The ratio between probabilities of success in quantum and
(optimal) classical protocol $P_{\textrm{correct}}^{qm} / P_{\textrm{correct}}^{cl}$ for
the communication complexity problem with $N$ observers and $M$ settings.
Quantum protocol uses GHZ state.
}
\label{TABLE_ADV}
\end{center}
\end{table}

\subsection{Qudits [P7]}

In this section we present CCPs connected with Bell inequalities
for higher-dimensional systems \cite{CGLMP}.
For a wide class of classical protocols we find 
an increase in the separation between the efficiency of the quantum and classical strategies, 
which grows with the dimensionality of the entangled systems. 
We show that the quantum protocol is more efficient than
the classical ones if and only if the protocol participants share a state that
 violates the CGLMP inequalities for higher-dimensional systems \cite{CGLMP}. 
The results form a
generalization of those of \cite{QCCP_QUTRIT_PRL} 
to arbitrarily high-dimensional systems.

\emph{The problem.}
Two parties are asked to give a single answer to 
$2 \lfloor d/2 \rfloor$ questions,
where $\lfloor x \rfloor$ stands for the integer part of $x$.
The integer $d$ describes the number of possible answers to each question.
Each party locally receives two inputs,
one bit and one dit,\footnote{A dit is a generalization of a bit, to a unit of information which can have $d$ values.}
but is restricted to communicate only a dit to the other party.
Further, the
parties are not allowed to differ in their answers. That is, they
must produce two identical answers each time.

Formally, the $2 \lfloor d/2 \rfloor$ questions will be formulated as a problem
of computation of 
$\lfloor d/2 \rfloor$ functions $f_k^+$ and
$\lfloor d/2 \rfloor$ functions $f_k^-$, with $k=0,..., \lfloor d/2 \rfloor - 1$.
The parties give one answer to the question about the values of
all $2 \lfloor d/2 \rfloor$ functions and their goal is to give the correct
value of $\lfloor d/2 \rfloor$ functions $f_k^+$, with the highest possible
probability, and \textit{at the same time}, the correct value of $\lfloor d/2 \rfloor$
functions $f_k^-$ with the lowest possible probability.
The questions are not treated equally.
The importance of questions changes with the
weight $1-\frac{2k}{d-1}$.

We now introduce the two-party task in detail and give all the
functions explicitly: 
Alice receives a data string $\alpha =
(a_{bit},a_{dit})$ and Bob a string $\beta = (b_{bit},b_{dit})$.
Alice's string is a combination of a bit $a_{bit} \in \{0,1\}$
and a dit $a_{dit} \in \{1,\alpha_d,\alpha_d^2, \dots ,
\alpha_d^{d-1}\}$ where $\alpha_d = e^{i(2\pi/d)}$. 
 Similarly, Bob's string is a
combination of a bit $b_{bit} \in \{0,1\}$ and a dit $b_{dit} \in
\{1,\alpha_d,\alpha_d^2, \dots , \alpha_d^{d-1}\}$. All possible input
strings are distributed randomly.
Before they give their answers, Alice and Bob are allowed to
exchange two dits of information. 
The answers are in the form of one dit. 
The task of
Alice and Bob is to maximize (having in mind the weight of the
questions) all differences between the probabilities $P(f_k^+)$,
of giving the correct value for the functions
\begin{equation}
f^+_k =
a_{dit}b_{dit}\alpha_d^{a_{bit}b_{bit}+k(-1)^{a_{bit}+b_{bit}}},
\quad \textrm{ with} \quad k=0,\dots,\lfloor d/2 \rfloor -1,
\end{equation}
and $P(f_k^-)$, of giving the correct value for the functions
\begin{equation}
f^-_k = a_{dit}b_{dit}
\alpha_d^{a_{bit}b_{bit}+(k+1)(-1)^{a_{bit}+b_{bit}+1}}, \quad \textrm{ with} \quad
k=0,\dots, \lfloor d/2 \rfloor -1.
\end{equation}
That is, they aim at the maximal value of
\begin{equation}
\Delta = \sum_{k=0}^{\lfloor \frac{d}{2} \rfloor -1} \Big(1-\frac{2 k}{d-1}\Big)
\Big(P(f^+_k)-P(f^-_k)\Big). \label{delta}
\end{equation}

We will show that, if two parties use a class of classical protocols, 
the quantity
$\Delta$ introduced above, which describes a performance of the protocol,
is at most $\frac{1}{2}$, whereas if they use
two entangled qudits $\Delta$ can be larger.
Furthermore it increases with $d$.

\emph{Quantum versus classical protocol}.
Note that 
 if only one of
the independent inputs, $a_{dit}$ or $b_{dit}$, is random
the product $a_{dit}b_{dit}$ in the full functions
$f_k^{\pm}$ acquires completely random values. This is
not the case for the last factors, with inputs $a_{bit}$ and
$b_{bit}$. Thus, intuition suggests that a good classical protocol
for the two parties may be that Alice ``spends" her dit by
sending $a_{dit}$ and Bob by sending $b_{dit}$ and that they put
for the part of $f$'s dependent on the bits
 the value
most often appearing in the third column and, at the same time,
least often appearing in the fourth column of the Table \ref{T_EXP_QUDITS}. 
Moreover, because of the weight function they should
give preference to the values connected with functions for
$k=0$. The second factor of $f_0^+$ is equal to $1$ in three out of four cases,
whereas $f_0^-$ is $1$ in one of four cases. 
Thus if
each of them gives the value $a_{dit}b_{dit}$ as the
answer, $\Delta = 1 (0.75 - 0.25) = 0.5$.
\begin{table}
\begin{center}
\begin{tabular}{cccc}\hline \hline
$a_{bit}$ & $b_{bit}$ & $a_{bit}b_{bit}+k(-1)^{a_{bit}+b_{bit}}$ & $a_{bit}b_{bit}
+(k+1)(-1)^{a_{bit}+b_{bit}+1}$ \\ \hline
$0$ & $0$ & $k$ & $-(k+1)$ \\
$0$ & $1$ & $-k$ & $k+1$ \\
$1$ & $0$ & $-k$ & $k+1$ \\
$1$ & $1$ & $k+1$ & $-k$ \\ \hline \hline
\end{tabular}
\end{center}
\caption{A set of possible input values for $a_{bit}$ and $b_{bit}$ and the corresponding
values of the exponents in the functions $f_k^{\pm}$.}
\label{T_EXP_QUDITS}
\end{table}

Let us now present the class of classical protocols which
can be followed by Alice and Bob, and which contains the above
intuitive example as a special case.
Alice calculates locally any
function $a(a_{bit},\lambda_A)$ and Bob calculates locally any
function $b(b_{bit},\lambda_B)$. Here $\lambda_A$ and $\lambda_B$
are any other parameters on which their functions $a$ and $b$ may
depend. They may include random strings of numbers shared by
Alice and Bob before receiving the inputs ($\lambda$'s are independent of the inputs). 
Alice sends to Bob
$e_A = a_{dit}a$ and receives from him $e_B = b_{dit} b$. Upon
receipt of $e_A$ and $e_B$, they both give $e_Ae_B$ as their
answers (which always agree). Note, that our intuitive protocol
is reproduced by $a=1$ and $b=1$ for all inputs.

Before showing what is the maximal $\Delta$ achievable
by the classical protocols, we shall introduce its quantum
competitor. Let Alice and Bob share a pair of entangled qudits
and a suitable measuring device (see, e.g.
\cite{MULTIPORT}). This is their quantum protocol: If Alice
receives $a_{bit} = 0$, she will measure her qudit with the
apparatus which is set to measure a $d$-valued observable $A_0$.
Otherwise, i.e., for $a_{bit} = 1$, she sets her device to
measure a different $d$-valued observable $A_1$. Bob follows the same
protocol. If he receives $b_{bit} = 1$, he measures the
$d$-valued observable $B_1$ on his qudit. For $b_{bit} = 0$ he
measures a different $d$-valued observable $B_0$. We ascribe to
the outcomes of the measurements the $d$ values $1, \alpha_d,
\alpha_d^2,\dots,\alpha_d^{d-1}$. The actual value obtained by Alice
in the given measurement will be denoted again by $a$, whereas
the one of Bob's, also again, by $b$. Alice sends the dit $e_A =
a_{dit}a$ to Bob, and Bob sends dit $e_B = b_{dit}b$ to Alice.
They both broadcast $e_Ae_B$ as their answers.

The task in both the classical and quantum protocols is to maximize $\Delta$ defined by
(\ref{delta}). The probability $P(f_k^+)$ is the probability for
the product $ab$ (of the local measurement results in the quantum
case, and the local functions in the classical case) to be equal
to the part of the functions $f_k^+$ which depends only on a
$a_{bit}$ and $b_{bit}$:
\begin{eqnarray}
P(f^+_k) = \frac{1}{4} \Big[ P_{01}(ab=\gamma^{-k})+P_{11}(ab=\gamma^{k+1})
+P_{10}(ab=\gamma^{-k})+P_{00}(ab=\gamma^{k})\Big], \label{prob+}
\end{eqnarray}
where e.g. $P_{01}(ab=\gamma^{-k})$ is the probability that
$ab=\gamma^{-k}$ if she receives $a_{bit}=0$, and he receives $b_{bit}=1$.
In the quantum case the probabilities on the right-hand side of Eq. (\ref{prob+})
are probabilities for certain products of measurement results,
whereas in the classical case they are probabilities for the products of  
locally computed functions.
Recall that all four possible
combinations for $a_{bit}$ and $b_{bit}$ occur with the same
probability $\frac{1}{4}$. Similarly, the probability $P(f_k^-)$
is given by:
\begin{eqnarray}
P(f^-_k) = \frac{1}{4} \Big[ P_{01}(ab=\gamma^{k+1})+P_{11}(ab=\gamma^{-k})
+P_{10}(ab=\gamma^{k+1})+P_{00}(ab=\gamma^{-(k+1)})\Big].
\end{eqnarray}
Finally, one notices that the success measure in the task is
given by
\begin{equation}
\Delta = \frac{1}{4} I_{d}, \label{advantage}
\end{equation}
where $I_d$ is just the left-hand side of CGLMP
inequality \cite{CGLMP}. The equivalence of $I_d$ and Collins
\textsl{et al.} inequalities may not be obvious at the first
glance because in \cite{CGLMP} the authors ascribe to local
measurement results integers $0,1,\dots,d-1$ and use modulo $d$ calculus. 
However, the difference between that description and
the one used here is just in the notation. Collins \textsl{et
al.} showed that $I_d \le 2$ for all local realistic theories.

If one looks back at the family of classical protocols introduced
above, one sees that they are equivalent to a local realistic
model of the quantum protocol (the $\lambda$'s are local hidden
variables, and $a_{bit}, b_{bit}$ are local variables which
define the measurements). This implies that within the full class
of classical protocols considered here $\Delta \le \frac{1}{2}$.

Thus, the necessary and sufficient condition for the state of two
qudits to improve the success in the communication complexity
task over any classical protocol of the discussed class is that
the state violates the Bell inequality for two qudits.

It was shown in \cite{ADGL} that nonmaximally (asymmetric)
entangled states of two qudits can violate the CGLMP inequalities
stronger than the maximally entangled one. Maximal violations for
some $d$ and corresponding success measures in the CCP
are gathered in the Table \ref{TALLQ}.
\begin{table}
\begin{center}
\begin{tabular}{cccc} \hline \hline
$d$ & Maximal violation & $\Delta_Q$ & $\Delta_Q - \Delta_C$ \\ \hline
$3$ & $2.9149$ & $0.7287$ & $0.2287$ \\
$4$ & $2.9727$ & $0.7432$ & $0.2432$ \\
$5$ & $3.0157$ & $0.7539$ & $0.2539$ \\
$6$ & $3.0497$ & $0.7624$ & $0.2624$ \\
$7$ & $3.0776$ & $0.7694$ & $0.2694$ \\
$8$ & $3.1013$ & $0.7753$ & $0.2753$ \\ \hline \hline
\end{tabular}
\end{center}
\caption{Maximal violation of the CGLMP inequalities and
corresponding measures for success in the CCP. 
The $\Delta_Q$ denotes the quantum success measure
and $\Delta_C$ the classical one. 
The values of maximal violations are taken from the work of Acin \textit{et al.} \cite{ADGL}.}
\label{TALLQ}
\end{table}

In the classical protocols, even with shared random
variables, more than two dits of information exchange are
necessary to complete the task successfully with $\Delta > \frac{1}{2}$,
whereas with quantum entanglement two dits can be sufficient
for the task with the same $\Delta$. Note that the discrepancy
between the measure of success in the classical and the quantum
protocol grows with $d$.

We would like to stress that asking all $2 \lfloor d/2 \rfloor$
questions is not necessary to prove the advantage of the quantum
protocol. As showed in \cite{QCCP_QUTRIT_PRL} even one question
$f_0^+$ is sufficient for 
an advantage of quantum strategy over the classical ones, 
but asking all questions maximizes the advantage.

\chapter{Outlook and future plans}

A link between Bell inequalities and quantum communication
was described in this thesis.
Examples of quantum cryptography and communication complexity
were discussed as well as some general development in the field
of Bell inequalities themselves.

The author believes a form of Bell inequalities
will be identified in other (novel) quantum protocols and algorithms,
and they will find new applications in quantification of entanglement.
Below some 
open problems are listed which are going to be considered in the near future
(some of them are more general, others are particularly linked with this thesis):

\section{Bell's theorem}

\begin{itemize}

\item Are there entangled states which admit local realistic description?

This very general problem asks whether entanglement
and impossibility of local realistic description (sometimes called quantum nonlocality)
are the same problems.
Gisin's theorem states that all pure entangled states cannot be modeled in local realistic way.
The question remains unanswered for mixed states
and requires derivation of new series of Bell inequalities.
It was even conjectured by Peres \cite{PERES_CONJECTURE}
that entangled states which have positive partial transposes with respect
to all subsystems do have local realistic model.
No counterexample to this conjecture is known.

\item Full characterization of a polytope of local realistic models.

What is, in general, a necessary and sufficient condition
for a possibility of local realistic model?
The only experimental situation in which such a condition is known
involves arbitrary number of observers making one of two local
measurements on two-level systems.
Moreover, the condition involves correlations
between all parties.
Even the situation in which at least one of parties
is allowed not to measure is unexplored!

\end{itemize}

\section{Beyond Bell's theorem}

\begin{itemize}

\item The influence of which nonlocal parameters is essential for violation of Bell inequality?

Jarrett shows \cite{JARRETT} that locality in Bell's argument 
can be split into the conjunction of setting-dependence 
(the measurement outcome in one lab can depend on the setting in the space-like separated lab) 
and outcome-dependence (the outcome in one lab can depend on the particular outcome in the separated lab). 
It seems that for a deterministic hidden variable theory
only the setting dependence is relevant.

\end{itemize}

\section{Quantum communication complexity}

\begin{itemize}

\item Do all entanglement-based protocols have single-particle counterparts?

Can one generalize the ideas of Galv\~ao \cite{GALVAO}
to arbitrary entanglement-based quantum
communication complexity protocol?
If successful, this project will greatly reduce
experimental efforts to realize
quantum communication complexity in practice
(and will open the way to practical applications).

\item Optimality of the classical protocols for higher-dimensional systems.

The quantum communication complexity protocol 
presented here is more efficient than the broad class of classical protocols.
We conjecture that the class of classical protocols introduced includes the optimal one.
The optimality was recently proven
for the case of two-level systems  \cite{EXP_CCP},
and it is still an open problem for arbitrary dimension.

\end{itemize}

\chapter{Appendices}

\section{Appendix A: Qubits}

\subsection{Arbitrary state of qubit}

The Hilbert space of qubit states
is spanned by two orthogonal vectors, $| 0 \rangle$ and $| 1 \rangle$.
Any pure state of a qubit is given by a superposition:
\begin{equation}
| \psi \rangle = \alpha | 0 \rangle + \beta | 1 \rangle,
\label{PURE_QUBIT_STATE}
\end{equation}
with complex coefficients satisfying:
\begin{equation}
| \alpha |^2 + | \beta |^2 = 1.
\end{equation}
An arbitrary (mixed) state of a qubit
is described by a Hermitian density operator:
\begin{equation}
\rho = 
\left(
\begin{array}{cc}
p_0      & c_0 + i c_1   \\
c_0 - i c_1 & 1-p_0 
\end{array}
\right),
\label{SINGLE_QUBIT_RHO}
\end{equation}
where all three parameters $p_0,c_0,c_1$ are real.
The number $0 \le p_0 \le 1$ gives the probability
of outcome ``0'',
the numbers $c_0 \pm i c_1$
describe coherence between the basis vectors of $\rho$.
Every density operator of a qubit can be described
in terms of the Pauli matrices
\begin{equation}
\sigma_1 \equiv \sigma_x =  
\left(
\begin{array}{cc}
0      & 1   \\
1 & 0
\end{array}
\right),
\quad
\sigma_2 \equiv \sigma_y =  
\left(
\begin{array}{cc}
0      & -i   \\
i & 0
\end{array}
\right),
\quad
\sigma_3 \equiv \sigma_z =  
\left(
\begin{array}{cc}
1      & 0   \\
0 & -1
\end{array}
\right),
\end{equation}
and the identity operator
\begin{equation}
\sigma_0 = 
\left(
\begin{array}{cc}
1      & 0   \\
0 & 1
\end{array}
\right).
\end{equation}
The set of four matrices $\sigma_\mu$, with $\mu=0,1,2,3$,
forms a basis in the Hilbert-Schmidt space
with a trace scalar product. 
The density operator, decomposed in this basis, reads:
\begin{equation}
\rho = \frac{1}{2} \sum_{\mu=0}^3 m_{\mu} \sigma_{\mu},
\label{HS_DECOMP}
\end{equation}
with
\begin{equation}
m_{\mu} = {\rm Tr}(\rho \sigma_{\mu}).
\label{BLOCH_COMPONENTS}
\end{equation}
Since $\rho$ is normalized 
and the Pauli matrices are traceless
one has:
\begin{equation}
m_0 = 1.
\end{equation}
The other three 
parameters $m_k$ (with $k=1,2,3$) in (\ref{HS_DECOMP})
are linked with the decomposition (\ref{SINGLE_QUBIT_RHO})
as follows:
\begin{equation}
m_1 = 2 c_0, \quad m_2 = - 2 c_1 , \quad m_3 = 2 p_0 - 1.
\end{equation}
Note that, according to (\ref{BLOCH_COMPONENTS}), 
the numbers $m_k$ are directly experimentally accessible,
as they are given by the averages of measurements
of Pauli operators in the state $\rho$.

Thus, the formula (\ref{HS_DECOMP})
can be written as:
\begin{equation}
\rho = \frac{1}{2}[\sigma_0 + \vec m \cdot \vec \sigma],
\label{BLOCH_REPR}
\end{equation}
where $\vec m$ is a vector with components $(m_1,m_2,m_3)$
and $\vec \sigma = (\sigma_1,\sigma_2,\sigma_3)$.
The dot denotes the scalar product
\begin{equation}
\vec m \cdot \vec \sigma \equiv m_1 \sigma_1 + m_2 \sigma_2 + m_3 \sigma_3.
\end{equation}
The decomposition (\ref{BLOCH_REPR})
is a so-called Bloch representation.
An arbitrary state of a qubit is in one-to-one correspondence with a Bloch vector $\vec m$.
All Bloch vectors corresponding to physical states 
lie in a ball of unit radius.
Pure states correspond to Bloch vectors of unit length,
i.e. lie on a sphere.
Mixed states have their Bloch vectors inside the sphere,
and a maximally mixed state corresponds to the center of the ball.

The Bloch representation is of a great importance in understanding single qubits
and general measurements made upon them.
The Bloch sphere is a three dimensional object
that allows visualization of relations between the
quantum states.
As a useful example, let us derive the relation between orthogonal
states, $\langle m | m_{\perp} \rangle = 0$, and the corresponding Bloch vectors, 
$\vec m$ and $\vec m_{\perp}$.
The condition that must be satisfied by the Bloch vectors
comes from imposed orthogonality relation
between the operators $\rho = | m \rangle \langle m |$
and $\rho_{\perp} = | m_{\perp} \rangle \langle m_{\perp} |$:
\begin{equation}
{\rm Tr}(\rho \rho_{\perp}) = 0.
\end{equation}
Each of these two density operators
has its own decomposition (\ref{BLOCH_REPR}).
Since
\begin{equation}
{\rm Tr} \sigma_0 = 2, \quad {\rm Tr} \sigma_k = 0, \quad \sigma_k \sigma_k = \sigma_0,
\end{equation}
for $k=1,2,3$, and
one has the well-known spin (angular momentum) relation
\begin{equation}
\sigma_x \sigma_y = i \sigma_z
\end{equation}
and its permutations,
the orthogonality of quantum states implies
for the corresponding Bloch vectors:
\begin{equation}
\vec m \cdot \vec m_{\perp} = -1.
\end{equation}
That is, the vectors point at opposite directions.

\subsection{Arbitrary dichotomic measurement}

Consider a measurement with two possible outcomes.
The eigenvalues associated with the outcomes
can be chosen as $\pm 1$.
The operator of
this measurement has a spectral decomposition
of the form
\begin{equation}
\mathcal{M} = | m \rangle \langle m | - | m_{\perp} \rangle \langle m_{\perp} |.
\label{DICHOTOMIC_MEASUREMENT}
\end{equation}
Inserting the Bloch vectors (\ref{BLOCH_REPR})
into this equation,
keeping in mind that pure
orthogonal quantum states have opposite unit Bloch vectors,
one arrives at
\begin{equation}
\mathcal{M} = \vec m \cdot \vec \sigma, \quad {\rm with} \quad |\vec m| = 1.
\end{equation}
An arbitrary dichotomic measurement 
is parameterized by a normalized Bloch vector,
corresponding to the eigenvector
associated with one of the eigenvalues.

\subsection{Arbitrary state of many qubits}

The Hilbert space of a multiparticle states has the form
of a tensor product of spaces of individual systems.
This allows the straightforward generalization of formula (\ref{HS_DECOMP})
to the case of many qubits.
In the Hilbert-Schmidt space
of operators acting on $N$-particle pure states,
the tensor products of individual operators
\begin{equation}
\sigma_{\mu_1}^{(1)} \otimes ... \otimes \sigma_{\mu_N}^{(N)}
\quad \textrm{ with } \quad  \mu_1,...,\mu_N = 0,1,2,3
\end{equation}
form a basis with respect to the trace scalar product.
Here, $\sigma_{\mu_n}^{(n)}$ acts in the space of the $n$th qubit.
Thus, arbitrary state of $N$ qubits, decomposed in this basis,
reads
\begin{equation}
\rho = \frac{1}{2^N} \sum_{\mu_1=0}^3 ... \sum_{\mu_N=0}^3
T_{\mu_1...\mu_N} \sigma_{\mu_1}^{(1)} \otimes ... \otimes \sigma_{\mu_N}^{(N)},
\label{HS_DECOMP_MANY_QUBITS}
\end{equation}
where the real coefficients $T_{\mu_1... \mu_N}$
form the so-called correlation tensor.
According to the trace scalar product
coefficients $T_{\mu_1... \mu_N}$ are the averages of the product
of individual measurement results:
\begin{equation}
T_{\mu_1... \mu_N} = {\rm Tr}(\rho \textrm{ } \sigma_{\mu_1}^{(1)} \otimes ... \otimes \sigma_{\mu_N}^{(N)}).
\end{equation}
Since $\rho$ is normalized and the Pauli matrices are traceless
one always has
\begin{equation}
T_{0...0} = 1.
\end{equation}

A useful bound on physically allowed correlation tensors
follows from the condition
\begin{equation}
{\rm Tr}(\rho^2) \le 1,
\label{PURITY_CONDITION}
\end{equation}
which is saturated for pure states.
The square of a density operator, using decomposition (\ref{HS_DECOMP_MANY_QUBITS}),
gives
\begin{equation}
\frac{1}{2^{2N}} \sum_{\mu_1,..,\mu_N=0}^3 \sum_{\nu_1,..,\nu_N=0}^3
T_{\mu_1...\mu_N} T_{\nu_1...\nu_N} 
\sigma_{\mu_1}^{(1)} \sigma_{\nu_1}^{(1)} \otimes ... \otimes \sigma_{\mu_N}^{(N)} \sigma_{\nu_N}^{(N)}.
\end{equation}
Since the trace is a linear operation
with the general property
\begin{equation}
{\rm Tr}(\sigma_{\mu_1}^{(1)} \sigma_{\nu_1}^{(1)} \otimes ... \otimes \sigma_{\mu_N}^{(N)} \sigma_{\nu_N}^{(N)})
= {\rm Tr}(\sigma_{\mu_1}^{(1)} \sigma_{\nu_1}^{(1)}) ... {\rm Tr}(\sigma_{\mu_N}^{(N)} \sigma_{\nu_N}^{(N)}),
\end{equation}
and for every element in this product one has
\begin{equation}
{\rm Tr}(\sigma_{\mu_n}^{(n)} \sigma_{\nu_n}^{(n)}) = 2 \delta_{\mu_n \nu_n},
\end{equation}
with $\delta_{\mu_n \nu_n}$ being a Kronecker delta,
the purity condition (\ref{PURITY_CONDITION})
gives the bound
\begin{equation}
\sum_{\mu_1,...,\mu_N=0}^3 T_{\mu_1...\mu_N}^2 \le 2^N.
\label{PURITY_BOUND}
\end{equation}

\subsection{Quantum correlations}

Generally, a correlation function is defined
as the average of a product of measurement results.
As described above,
in quantum mechanics an arbitrary dichotomic measurement
is parameterized by a Bloch vector, (\ref{DICHOTOMIC_MEASUREMENT}).
Thus, the quantum correlation function of the results of
arbitrary dichotomic measurements is given by
\begin{equation}
E_{\vec m_1,...,\vec m_N}^{QM} = 
{\rm Tr}(\rho \textrm{ } \vec m_1 \cdot \vec \sigma^{(1)} \otimes ... \otimes \vec m_N \cdot \vec \sigma^{(N)}),
\end{equation}
where $\vec \sigma^{(n)}$ is a ``vector'' of local Pauli operators 
of the $n$th party:
$\vec \sigma^{(n)} = (\sigma_x^{(n)},\sigma_y^{(n)},\sigma_z^{(n)})$.
With the density matrix decomposition (\ref{HS_DECOMP_MANY_QUBITS})
one finds the relation between the quantum correlation function
and the elements of the correlation tensor of a state:
\begin{equation}
E_{\vec m_1,...,\vec m_N}^{QM} = 
\sum_{k_1=1}^3 ... \sum_{k_N=1}^3 T_{k_1...k_N} (\vec m_1)_{k_1}...(\vec m_N)_{k_N},
\label{QM_CORRELATIONS}
\end{equation}
where $(\vec m_n)_{k_n}$ is understood as the $k_n$th component ($k_n = 1,2,3$) of the Bloch vector $\vec m_n$.
The last equation can be put in the compact form
\begin{equation}
E_{\vec m_1,...,\vec m_N}^{QM} = \hat T \circ \vec m_1 \otimes ... \otimes \vec m_N,
\end{equation}
where $\hat T$ is the correlation tensor,
and $\circ$
denotes the scalar product in $\mathcal{R}^{3N}$.

Let us illustrate this formalism with an example.
We find the quantum correlation function for arbitrary
dichotomic measurements performed on two qubits in the singlet state:
\begin{equation}
| \psi^- \rangle = \frac{1}{\sqrt{2}}\Big[ | z+ \rangle_1 | z- \rangle_2 - | z- \rangle_1 | z+ \rangle_2 \Big],
\end{equation}
where $| z\pm \rangle_n$ are the eigenstates of local $\sigma_z^{(n)}$ operator.
Since the total spin of this system is zero,
individual spin measurements along the same axes 
always find the two spins to be opposite.
The average of such measurements gives $-1$.
In the language of the correlation tensor one has:
\begin{equation}
T_{11} = T_{22} = T_{33} = -1.
\end{equation}
Additionally, as for any other state, $T_{00} = 1$.
Note that the bound allowed by (\ref{PURITY_BOUND})
is already reached, which implies that all
other correlation tensor elements vanish.
Finally, there are only three terms in the sums of (\ref{QM_CORRELATIONS}),
and one easily finds that the quantum correlation function
of the singlet state reads:
\begin{equation}
E_{\vec m_1, \vec m_2}^{\psi^-} = - \vec m_1 \cdot \vec m_2.
\end{equation}

\subsection{Polarisation as qubit}

Qubits can be encoded in many physical systems.
A system represents a qubit
if any measurement made upon it
results in only one of two values,
and one can write any pure state of a system as (\ref{PURE_QUBIT_STATE}).
We show how these requirements
are satisfied by the polarization of a single photon.

Since there are only two orthogonal polarizations
one can identify
two orthogonal states of a qubit with
horizontal and vertical polarization:
\begin{equation}
|H \rangle = | 0 \rangle, \qquad |V \rangle = | 1 \rangle.
\end{equation}
Let us assume these states are the eigenstates of $\sigma_z$ operator.
Arbitrary polarization is given by a superposition
of these two with normalized coefficients and arbitrary relative phase
(in full analogy to the classical case):
\begin{equation}
| P \rangle = \alpha | H \rangle + \beta | V \rangle.
\end{equation}
The eigenbasis of the $\sigma_x$ operator
is given by polarizations rotated by $\pm 45^{\circ}$ from the horizontal one
\begin{equation}
| \pm 45 \rangle = \frac{1}{\sqrt{2}} \Big( | H \rangle \pm | V \rangle \Big),
\end{equation}
and the basis of $\sigma_y$ operator
consists of right and left circular polarizations:
\begin{equation}
| R \rangle = \frac{1}{\sqrt{2}} \Big( | H \rangle + i | V \rangle \Big), \qquad
| L \rangle = \frac{1}{\sqrt{2}} \Big( | H \rangle - i | V \rangle \Big).
\end{equation}

To measure the polarization
one needs a polarizer, quarter-wave plate, half-wave plate
and a detector.
Since a single photon gives rise to a single click (or no click)
all polarization measurements made upon it
result in only one of two values.

\section{Appendix B: Qudits}

Qudits are $d$-level quantum systems.
One way to deal with a qudit is to find a convenient physical system representing it.
A beautiful example is a photon with many accessible propagation paths \cite{MULTIPATH1,MULTIPATH2}.
Another approach is to treat many systems of lower dimensions
as a global higher-dimensional object -- a composite qudit.

\subsection{Arbitrary state of qudit}

An arbitrary physical state, a density operator,
is defined in the Hilbert-Schmidt space.
To describe a state one has to find an operator
basis in this space.
For higher-dimensional systems a possible choice of such a basis
are unitary generalizations of Pauli operators, 
$S_{kl}$ with $k,l = 0,...,d-1$.
Each of these operators can be constructed as \cite{FIVEL,PR}:
\begin{equation}
S_{kl} = S_x^k S_z^l \textrm{ with } k,l=0,...,d-1,
\end{equation}
where the action of the two operators on the right-hand side,
on the eigenvectors of $S_z$ operator, $| \kappa \rangle_z$, is defined by:
\begin{eqnarray}
S_z |\kappa \rangle_z &=& \alpha_d^{\kappa} |\kappa \rangle_z, \label{S_DEFINITION} \\
S_x |\kappa \rangle_z &=& |\kappa + 1 \rangle_z, \quad \textrm{where} \quad \kappa=0,1,...,d-1, \label{APPB_SZSX_DEF}
\end{eqnarray}
with
\begin{equation}
\alpha_d = e^{i 2\pi/d}.
\label{ALPHA}
\end{equation}
The number $\alpha_d$ is the primitive complex $d$th root of unity,
whereas the addition, here $\kappa+1$, is taken modulo $d$.
Unless explicitly stated all additions are taken modulo $d$.
For $d=2$ these operators reduce to standard Pauli operators
(which are both unitary and Hermitian).

Any quantum state can be uniquely decomposed in this basis:
\begin{equation}
\rho = \frac{1}{d} \sum_{k,l=0}^{d-1} s_{kl} S_{kl},
\label{RHO_QUDIT}
\end{equation}
where $s_{00}=1$ for normalisation
since all $S_{kl}$ operators are traceless, except the identity.
The coefficients $s_{kl}$, given by the trace formula
\begin{equation}
s_{kl} = \textrm{Tr}(S_{kl}^{\dagger} \rho),
\label{s_DEF}
\end{equation}
can be regarded as components of a generalized Bloch vector.
Contrary to the qubit case,
there is no simple relation 
which defines physically allowed generalized Bloch vectors.

One can doubt about the physical meaning of (\ref{RHO_QUDIT})
as the operators which enter the density matrix decomposition
are unitary and not Hermitian.
Do they correspond to any measurement apparatuses?
In quantum mechanics, 
different outcomes of a measurement correspond to
different orthogonal states of a system.
Due to the fact that most often measurement outcomes
are expressed in form of real numbers we are used to connect 
Hermitian operators with observables.
However, there are measurement apparatuses which \emph{do not} output a number.
Take a device which clicks if a photon is detected
or a bunch of such photo-detectors which monitor
many possible propagation paths of a photon.
The operator
associated with this apparatus has a specific spectral decomposition
(different clicks find the system in different orthogonal states).
The eigenvalues assigned to the clicks can be arbitrary,
as long as the assignment is consistent,
i.e.\ clicks of the same detector always reveal the same eigenvalue.
If one finds it useful to work with complex eigenvalues,
as it is often the case when considering higher-dimensional quantum systems,
one can use operators which are unitary,
with eigenvalues given by complex roots of unity.

With any generalized Pauli operator 
one can associate a measurement device capable to measure it.
Thus, it is possible to measure coefficients $s_{kl}$.
Any unitary operator, in particular operators the $S_{kl}$, 
has a spectral decomposition:
\begin{equation}
S_{kl} = \sum_{j=0}^{d-1} \lambda_{j} |j\rangle \langle j |,
\end{equation}
with complex eigenvalues $\lambda_{j}$.
Thus, the generalized Bloch vector components
can be written as:
\begin{equation}
s_{kl} = \textrm{Tr}(S_{kl}^{\dagger} \rho)
= \sum_{j=0}^{d-1} \lambda_{j}^* \textrm{Tr}(|j\rangle \langle j | \rho),
\label{TOMOGRAPHY}
\end{equation}
The trace on the right-hand side gives the probability, $p_{j}$, 
to obtain the $j$th outcome
in the measurement of $S_{kl}$ on 
the system prepared in the state $\rho$.

\subsection{Polarisation-path system as qudit [P2]}

Consider a qudit which is encoded in a polarized photon,
which has many possible propagation paths.\footnote{Note that only qudits of an even dimension can be realized in this way.}
First, we explicitly present devices capable to measure
all $S_{kl}$ operators in the simplest case of two paths.
Next, the setups for any number of paths are discussed.
In this way one can characterize an arbitrary state of a qudit.

Consider a polarized photon with two accessible paths.
Its state is described in a four dimensional Hilbert space,
i.e. there are fifteen different $S_{kl}$ operators to measure
(we put $s_{00} = 1$ from the very definition).
However, some of them commute (contrary to the qubit case) 
and the measurement of one of them reveals the values of the others.

We call the simplest observable,
which distinguishes what polarization has a photon
in a given path, by $S_z$.
From the definition, the eigenstates of $S_z$ are given by:
\begin{eqnarray}
|0\rangle_z = |z+\rangle_1 |z+\rangle_0, & \quad & |1\rangle_z = |z+\rangle_1 |z-\rangle_0, \nonumber \\
|2\rangle_z = |z-\rangle_1 |z+\rangle_0, & \quad & |3\rangle_z = |z-\rangle_1 |z-\rangle_0,
\label{QUQUAT}
\end{eqnarray}
where subsystem ``0'' is a polarization of a photon,
and subsystem ``1'' is a path. 
E.g. $|2 \rangle_z = |z-\rangle_1 |z+\rangle_0$
denotes a horizontally polarized photon in the path $|z-\rangle_1$.
The $z$ index inside the two-level kets 
denotes the fact that they are chosen as
the eigenstates of the individual $\sigma_z^{(n)}$ operators.
Note that polarizing beam-splitters are sufficient to perform a test of
what polarization a photon has in a certain path.
Moreover, the same device also measures the values of $S_z^2$ and $S_z^3$,
as these operators commute with $S_z$. Their eigenvalues
are powers of the $S_z$ eigenvalues.
Interestingly, the observables $S_{21}$ and $S_{23}$ can be measured in a similar way.
After expressing the eigenvectors of, say, $S_{21}$ in the $|\kappa \rangle_z$ basis,
and with definitions (\ref{QUQUAT}), one finds:
\begin{eqnarray}
|0 \rangle = |y+\rangle_1 |z-\rangle_0, & \quad & |1 \rangle = |y+\rangle_1 |z+\rangle_0, \nonumber \\
|2 \rangle = |y-\rangle_1 |z-\rangle_0, & \quad & |3 \rangle = |y-\rangle_1 |z+\rangle_0,
\end{eqnarray}
where $|y \pm \rangle_n$ is the eigenbasis of the individual $\sigma_y^{(n)}$ operator.
To measure this observable the paths meet on a beam-splitter
(which gives a phase $\pi/2$ to the reflected beam)
where different eigenstates $|y \pm \rangle_1$ are directed
into different output ports,
followed by polarizing beam-splitters.

\begin{figure}[t]
\begin{center}
\includegraphics[scale=1.3]{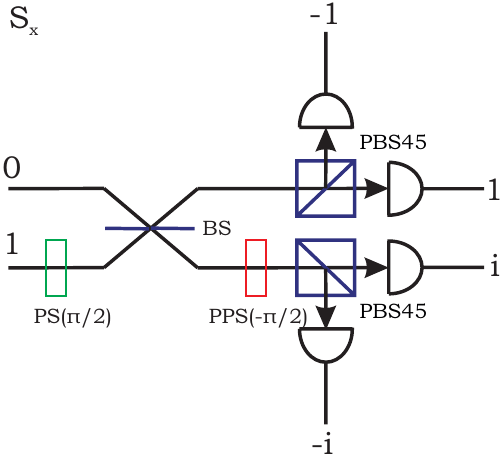}
\end{center}
\caption{
The setup which measures the operator $S_x$, for $d=4$.
The $\pi/2$ phase shift (PS($\pi/2$)) in the path 1 ($|z+\rangle_1$) and the beam-splitter (BS)
perform the path measurement, $\sigma_x^{(1)}$.
The path state $| x+ \rangle_1$ goes to the upper arm where the polarization
is measured in the $\sigma_x^{(0)}$ basis with the polarizing beam-splitter
which transmits $| x+ \rangle_0$ (denoted as PBS45).
In case of the path state $| x- \rangle_1$ the photon goes to the lower arm,
where its $| z- \rangle_0$ polarization component is phase shifted by $-\pi/2$ (PPS($-\pi/2$)).
Next, the photon enters PBS45, and is detected in one of its outputs.
The eigenvalues corresponding to clicks of each detector are also written.}
\label{SX}
\end{figure}

Consider $S_x = S_{10}$ operator.
Its eigenvectors read:
\begin{eqnarray}
|0 \rangle = |x+\rangle_1 |x+\rangle_0, & \quad & |1 \rangle = |x-\rangle_1 |y+\rangle_0, \nonumber \\
|2 \rangle = |x+\rangle_1 |x-\rangle_0, & \quad & |3 \rangle = |x-\rangle_1 |y-\rangle_0,
\label{SX_EIGEN}
\end{eqnarray}
where $|x \pm \rangle$ denotes the eigenbasis of the individual $\sigma_x^{(n)}$ operator.
Depending on the outcome of the path measurement in the $\sigma_x^{(1)}$ basis,
the polarization is measured in the $\sigma_x^{(0)}$ or $\sigma_y^{(0)}$ basis.
However,\footnote{Here comes the beauty of the approach utilizing the paths.}
this information does not have to be actively fed-forward since an
appropriate phase and a beam-splitter
drive different $\sigma_x^{(1)}$ path eigenstates
into different output ports of the beam-splitter.
It is now enough to put polarization checking devices behind the proper
outputs of the beam-splitter (see Fig. \ref{SX}).

\begin{figure}
\begin{center}
\includegraphics[scale=1.3]{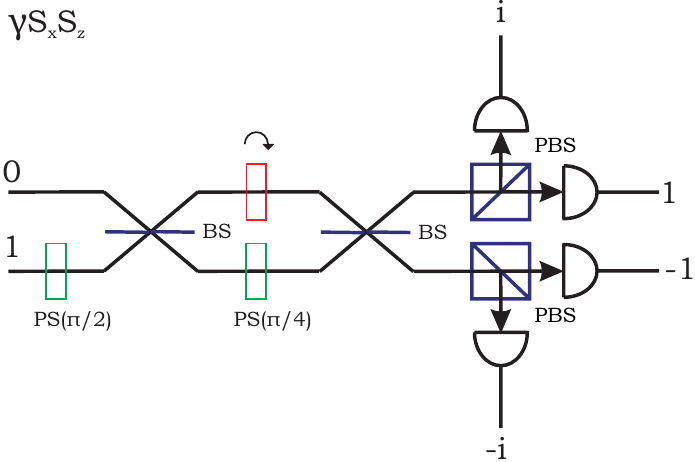}
\end{center}
\caption{Mach-Zehnder interferometer, with a polarization rotator in one arm,
followed by polarizing beam-splitters, is the most advanced device used in measurements
of $S_{kl}$, for $d=4$.
This setup, which measures the operator $\gamma S_x S_z$ (with $\gamma = e^{i \pi/4}$),
distinguishes maximally entangled states of paths and polarizations.
First, with the $\pi/2$ phase shift (PS($\pi/2$)) and the beam-splitter (BS), 
the $\sigma_x^{(1)}$ eigenstates
are converted into $\sigma_z^{(1)}$ eigenstates. 
Next, the $\pi/4$ phase (PS($\pi/4$))
is applied in the lower arm, where $| x- \rangle_1$ is directed.
In the upper arm polarization is rotated (with the plate $\curvearrowright$),
such that in both arms it is the same.
Finally, specific clicks behind the beam-splitter and polarizing beam-splitters
distinguish the states (\ref{SXSZ_EIGEN}).}
\label{SXSZ}
\end{figure}

The eigenstates of 
the last five observables are maximally entangled states of the subsystems.
Some of these observables, to keep the spectrum in the domain of fourth roots of unity,
need to be multiplied by $\gamma \equiv \alpha_4^{1/2} = e^{i \pi/4}$.
Take as an example the $S_{11}$ operator in the form $S_{11} = \gamma S_x S_z$. 
Its eigenstates are given by
\begin{equation}
\begin{array}{lcl}
|0 \rangle = \frac{1}{\sqrt{2}} \Big( |x+\rangle_1 |z-\rangle_0 - i\gamma |x-\rangle_1 |z+\rangle_0 \Big), & \quad &
|1 \rangle = \frac{1}{\sqrt{2}} \Big( |x+\rangle_1 |z+\rangle_0 - i\gamma |x-\rangle_1 |z-\rangle_0 \Big), \\
|2 \rangle = \frac{1}{\sqrt{2}} \Big( |x+\rangle_1 |z-\rangle_0 + i\gamma |x-\rangle_1 |z+\rangle_0 \Big), & &
|3 \rangle = \frac{1}{\sqrt{2}} \Big( |x+\rangle_1 |z+\rangle_0 + i\gamma |x-\rangle_1 |z-\rangle_0 \Big).
\end{array}
\label{SXSZ_EIGEN}
\end{equation}
To distinguish between these states one needs to build an interferometer 
like the one in Fig. \ref{SXSZ}.
The same setup measures $S_{22}$ and $S_{33}$,
which commute with $S_{11}$.
Finally, when different phase shifts are used, 
this setup also measures the remaining $S_{13}$ and $S_{31}$ observables.

To sum up, the most involved device,
used in the measurements of generalized Pauli operators on a composite qudit encoded in two paths
and polarization of a photon, involves a Mach-Zehnder interferometer
with a polarization rotator in one arm, followed by polarizing beam-splitters (Fig. \ref{SXSZ}).
Most of the observables are realizable with a single beam-splitter
followed by polarizing beam-splitters.

Generally, it is possible to perform an arbitrary $S_{kl}$ measurement on
polarized photons with many, $d_1$, accessible paths.
With polarizing beam-splitters in each propagation path
one transforms the initial polarization-path state $| j \rangle$
into a double-number-of-paths state $|p \rangle$, in $2d_1$ dimensional Hilbert space
(each polarizing beam-splitter generates two distinct spatial modes).
According to \cite{RECK}
one can always realize a unitary which
brings the states $|p \rangle$
to the states of well-defined propagation direction.
Thus, $2d_1$ detectors monitoring these final paths distinguish all
the eigenvectors $| j \rangle$.

\section{Appendix C: Spontaneous parametric down-conversion}

In this section we describe how a polarization entangled
state of two photons (which was used in the experimental falsification
of the class of nonlocal theories) can be generated in a nonlinear
crystal pumped with a laser field.
The description is idealized,
however it recovers all main features
of the generated photons.
This treatment was presented by the Sen family and \.Zukowski \cite{SPDC_APP}.

\subsection{Crystal-field interaction}

Consider an experiment in which a laser shines on a cubic crystal
of volume $V=L^3$.
Let us divide the volume into macroscopically small pieces $\delta  V(\vec x)$,
which however include many atoms (or molecules) of the crystal.
Since atoms are 
electrically neutral 
(so is the whole medium)
the dominant part
in the interaction Hamiltonian comes from the coupling between
electric polarization of a local volume $\delta V(\vec x)$, $\vec P(\vec x,t)$,
and a local electric field $\vec E(\vec x,t)$:
\begin{equation}
H_{int} \sim \int_{V} d \vec x \textrm{ } \vec P(\vec x,t) \cdot \vec E(\vec x,t)
= \int_{V} d \vec x \sum_{i=1}^3 P_i(\vec x,t) E_i(\vec x,t).
\label{LOCAL_INTERACTION}
\end{equation}
In strong electromagnetic fields the polarization of certain crystals
can depend on higher powers of the field:
\begin{equation}
P_i(\vec x,t) =  \sum_{j =1}^3 \chi^{(1)}_{ij} E_j(\vec x,t) + \sum_{j =1}^3 \sum_{k =1}^3 \chi^{(2)}_{ijk} E_j(\vec x,t) E_k(\vec x,t) + ....
\label{POLARIZABILITY}
\end{equation}
where one assumes the coefficients $\chi^{(m)}_{ijk...}$
are neither dependent on an actual position within the crystal
nor on time, and that the polarization in point $\vec x$
depends on the field in the same point only.
The two-photon generation process of interest 
is linked with the nonlinear
term, $\chi^{(2)}_{ijk}$, in this expansion.
Note that the electromagnetic field
should not be too strong,
as in that case even higher order emitions become non-negligible.
The nonlinear interaction Hamiltonian
can be found after inserting nonlinear dependence in (\ref{POLARIZABILITY})
into (\ref{LOCAL_INTERACTION}), and reads:
\begin{equation}
H_{int}^{(2)} \sim \int_{V} d \vec x \sum_{i=1}^3 \sum_{j =1}^3 \sum_{k =1}^3 
\chi^{(2)}_{ijk} E_i(\vec x,t) E_j(\vec x,t) E_k(\vec x,t).
\label{H2}
\end{equation}

The local field 
can be assumed to split into a classical and quantum part:
\begin{equation}
\vec E(\vec x,t) = \vec E^{cl}(\vec x,t) + \vec E^{qm}(\vec x,t),
\label{SUM_OF_FIELDS}
\end{equation}
where the classical part describes
the laser field, 
and the quantum part deals with a small number of photons.
The laser light can be taken as a monochromatic plane wave
linearly polarized along $\hat x$ direction:
\begin{equation}
\vec E^{cl}(\vec x,t) 
= E_x \cos(\vec k_0 \cdot \vec x - \omega_0 t - \varphi)
= E_x [e^{i(\vec k_0 \cdot \vec x - \omega_0 t - \varphi)} + c.c.],
\label{LASER_FIELD}
\end{equation}
where \emph{c.c.} denotes the complex conjugate,
$\vec k_0$ the wave vector and $\omega_0$ the angular frequency of the laser.

In general, one can write the quantum field 
in the interaction picture as:
\begin{eqnarray}
\vec E^{qm}(\vec x,t) & = & \sum_{p=1}^2 \int d \vec k F(\omega)
\hat \epsilon (\vec k,p) a(\vec k,p) e^{i(\vec k \cdot \vec x - \omega t)} + h.c. \nonumber \\
& \equiv & \vec E^{(+)}(\vec x,t) + \vec E^{(-)}(\vec x,t),
\label{QM_FIELD}
\end{eqnarray}
where $F(\omega) = i/\sqrt{2 \omega (2 \pi)^3}$,
the sum is taken over two orthogonal polarizations $\hat \epsilon (\vec k,p)$,
$\omega$ is the angular frequency of a photon the annihilation operator
of which is denoted by $a(\vec k,p)$, with $\vec k$ being the wave vector.
The abbreviation $h.c.$ stands for Hermitian conjugate of a previous term,
i.e. $\vec E^{(-)}(\vec x,t) = [\vec E^{(+)}(\vec x,t)]^{\dagger}$.
The principal commutation rule for 
the creation and annihilation operators is given by:
\begin{eqnarray*}
\Big[ a(\vec k , p), a^{\dagger}(\vec k', p') \Big] = \delta_{p,p'} \delta(\vec k - \vec k'), \quad
\Big[ a^{\dagger}(\vec k, p), a^{\dagger}(\vec k', p') \Big] = 0, \quad
\Big[ a(\vec{k},p), a(\vec{k}',p') \Big] =  0.
\end{eqnarray*}

Since only two-photon emitions are of interest,
after inserting the sum (\ref{SUM_OF_FIELDS}) of classical and quantum fields
into the nonlinear  Hamiltonian (\ref{H2})
and performing all the multiplications therein,
one can focus on one of the terms with two creation operators
[which come via the $\vec E^{(-)}(\vec x,t)$ part of the quantum field]:
\begin{equation}
H_{SPDC} \sim \int_{V} d \vec x \sum_{i=1}^3 \sum_{j =1}^3 \sum_{k =1}^3 
\chi^{(2)}_{ijk} E_i^{cl}(\vec x,t) E_j^{(-)}(\vec x,t) E_k^{(-)}(\vec x,t) + h.c..
\end{equation}
Writting all the fields explicitly using
formulas (\ref{LASER_FIELD}) and (\ref{QM_FIELD})
one arrives at the interaction Hamiltonian
describing the process of
spontaneous parametric down-conversion.
The terms involving two creation operators read:
\begin{eqnarray}
H_{SPDC}  \sim \sum_{j,k=1}^3 \chi^{(2)}_{1jk}
\sum_{p,p'=1}^2 \int d \vec k \int d \vec k' \mathcal{G}(\vec k, \vec k',p,p')
a^{\dagger}(\vec k,p) a^{\dagger}(\vec k',p') \nonumber \\
\times \int_V d \vec x 
\Big\{ e^{i \vec x \cdot ( \vec k_0 - \vec k - \vec k')} e^{i t ( -\omega_0 + \omega + \omega')} e^{-i \varphi} + 
e^{i \vec x \cdot ( - \vec k_0 - \vec k - \vec k')} e^{i t ( \omega_0 + \omega + \omega')} e^{i \varphi} \Big\},
\label{SPDC_HAMILTONIAN}
\end{eqnarray}
with the coupling factor 
$\mathcal{G}(\vec k, \vec k',p,p') = E_x \epsilon_j (\vec k,p) \epsilon_k (\vec k',p') F(\omega) F(\omega')$.
The variables with index zero describe the pump field,
those which are primed and unprimed describe two down-converted photons.

Let us perform the integration over the crystal volume
in the Hamiltonian (\ref{SPDC_HAMILTONIAN}):
\begin{equation}
e^{i t ( -\omega_0 + \omega + \omega')} e^{-i \varphi}
\int_V d \vec x 
e^{i \vec x \cdot ( \vec k_0 - \vec k - \vec k')} +
e^{i t ( \omega_0 + \omega + \omega')} e^{i \varphi}
\int_V d \vec x 
e^{i \vec x \cdot ( - \vec k_0 - \vec k - \vec k')}.
\end{equation}
In the limit $V \to \infty$ the two integrands approach the Dirac delta $\delta( \pm \vec k_0 - \vec k - \vec k')$.
For a finite size of the crystal one has approximate relation
$\pm \vec k_0 \approx \vec k + \vec k'$.
One can doubt about a physical meaning
of the relation with the minus sign,
in which case the generated photons propagate in the opposite
direction to the pump field.
Indeed this case is unphysical
as it will be shown when considering the
frequencies of the down-converted photons.
Practical crystals are macroscopic, with $L$ of the order of a millimeter,
and this relation (with a plus sign) holds perfectly.
It is often quoted as the momentum conservation law.

Let us describe the time evolution generated
by Hamiltonian (\ref{SPDC_HAMILTONIAN}).
All states and operators are taken in the interaction (Dirac) picture,
with the interaction Hamiltonian given by $H_{SPDC}$
(note that it explicitely depends on time, i.e. $H_{SPDC} = H_{SPDC}(t)$).

The generated two-photon state, $|\psi(t) \rangle_{12}$ (in the interaction picture), 
evolves according to the Schr\"odinger equation:
\begin{equation}
i \hbar \frac{d}{dt} |\psi(t) \rangle_{12} = H_{SPDC}(t)|\psi(t) \rangle_{12}.
\end{equation}
Therefore:
\begin{equation}
|\psi(t_f) \rangle_{12} - |\psi(t_i) \rangle_{12} = \frac{1}{i\hbar} \int_{t_i}^{t_f} H_{SPDC}(t') |\psi(t') \rangle_{12} dt',
\end{equation}
where $t_i$ ($t_f$) denotes the initial (final) interaction time.
The $H_{SPDC}$ Hamiltonian describes the interaction between the
monochromatic plane wave and the quantum field initially in the vacuum state
of no photons: $|\psi(t_i) \rangle_{12} = |\Omega \rangle$.
Since the monochromatic wave extends infinitely in time
one puts for $t_i = - \infty$
and for $t_f = \infty$.
In fact, the final time is a detection time,
but since the two-photon state is always observed outside the crystal
setting $t_f = \infty$ is a good approximation.
Using the first order of the perturbation calculus
one can replace $|\psi(t') \rangle_{12}$
on the right-hand side
with the initial vacuum state $| \Omega \rangle$.
Thus the final two-photon state $| \Psi \rangle_{12} \equiv |\psi(t_f = \infty) \rangle_{12}$ reads:
\begin{equation}
| \Psi \rangle_{12} = |\Omega \rangle + \frac{1}{i\hbar} \int_{-\infty}^{\infty} H_{SPDC}(t') dt' |\Omega \rangle.
\end{equation}
Inserting the Hamiltonian $H_{SPDC}$ and keeping in mind the momentum considerations
one notes that the time-dependent part of the two-photon state is proportional to:
\begin{equation}
e^{-i \varphi} \delta(\vec k_0 - \vec k - \vec k')
\int_{-\infty}^{\infty} dt' e^{i t' ( -\omega_0 + \omega + \omega')}
+
e^{i \varphi}\delta( - \vec k_0 - \vec k - \vec k')
\int_{-\infty}^{\infty} dt'
e^{i t ( \omega_0 + \omega + \omega')}.
\end{equation}
The two integrals are given by the Dirac delta $2 \pi \delta(\pm \omega_0+\omega+\omega')$.
Thus the allowed frequencies of the emissions satisfy the relation
$\pm \omega_0 = \omega + \omega'$.
However, the case $- \omega_0 = \omega + \omega'$
requires at least one of the frequencies $\omega$ or $\omega'$ to be negative.
This is impossible to meet.
The only physical situation left requires
\begin{eqnarray}
\vec k_0 & \approx & \vec k + \vec k', \\
\omega_0 & = & \omega + \omega'.
\end{eqnarray}
The second equation expresses energy conservation law.
These relations are known as phase matching conditions.

Additionally to phase matching conditions
one has to take into account
a dispersion relation for light
moving in a medium.
In general one has
\begin{equation}
\omega = |\vec k| c(\omega,p),
\end{equation}
where $c(\omega,p)$ is the speed of light in a given medium.
It is a function of polarization (birefringence effect)
and angular frequency (normal and anomal dispersion).
Together with the phase matching condition for frequencies
this relation leads to
\begin{equation}
|\vec k_0| c(\omega_0,p_0) \approx |\vec k| c(\omega,p) + |\vec k'| c(\omega',p').
\label{SPDC_DISPERSION}
\end{equation}
Only in specific directions one can expect correlated emissions.
From now on we consider the energy degenerate case
\begin{equation}
\omega = \omega' = \frac{\omega_0}{2},
\end{equation}
in which both down-converted photons
have the same frequency.

\subsection{Path entanglement}

\begin{figure}
\begin{center}
\includegraphics{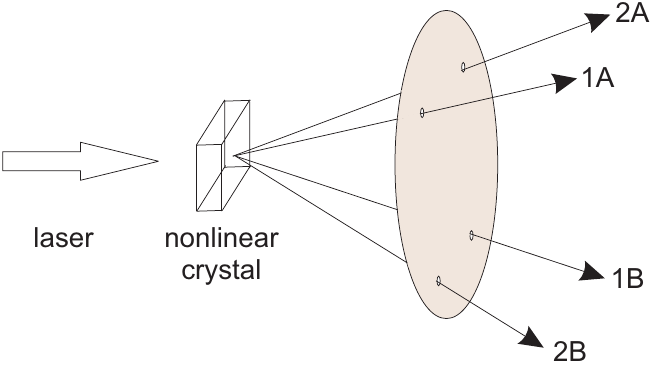}
\end{center}
\caption{Type I SPDC. The two photons have the same polarization.
They are emitted on the opposite sides of the cone centered on the laser beam.
Since the actual propagation direction  of a single photon is unknown
the emerging state is entangled.
With suitable pinholes outside the crystal one can generate
the entangled state between arbitrary number of propagation directions.
In this figure it is essentially $|1A \rangle | 1B \rangle + |2A \rangle | 2B \rangle$.}
\label{FIG_PATH_ENT}
\end{figure}

Suppose that the polarizations of the emitted photons are the same (so called Type-I SPDC).
Since we have also chosen their frequencies to be equal,
both photons propagate with the same speed, $c(\omega,p)$, inside the crystal.
In this case the dispiersion relation (\ref{SPDC_DISPERSION})
reads:
\begin{equation}
|\vec k_0| c(\omega_0,p_0) \approx 2 |\vec k| c(\omega,p).
\end{equation}
In the medium in which $c(\omega_0,p_0) > c(\omega,p)$
one has $|\vec k_0| < 2 |\vec k|$.
This, together with the wave vectors phase matching condition,
implies that directions of emitted photons
make the angle
$\alpha = \frac{|\vec k|}{|\vec k_0|/2} = \frac{c(\omega_0,p_0)}{c(\omega,p)}$
with the direction of the pump beam.
The photons are emitted on opposite
sides of the cone centered on the beam (Fig. \ref{FIG_PATH_ENT}).
Their joint state outside the crystal
can be found from the interaction Hamiltonian $H_{SPDC}$ to read:
\begin{equation}
| \Psi \rangle_{12} \sim \int d \vec k \mathcal{G}(\vec k, \vec k_0 - \vec k,p,p)
a^{\dagger}(\vec k, p) a^{\dagger}(\vec k_0 - \vec k, p) | \Omega \rangle,
\end{equation}
i.e. it is a \emph{coherent} superposition of emitions
into opposide directions of the cone,
in which the actual direction of a single photon is not fixed.
This is a ``path'' entangled state.

\subsection{Polarisation entanglement}

All the crystals in which parametric
down-conversion takes place are birefringent.
If the molecules of the medium are centro-symmetric
the coefficients $\chi^{(2)}_{ijk}$ vanish,
and one cannot observe the process.

\begin{figure}
\begin{center}
\includegraphics[scale=0.5]{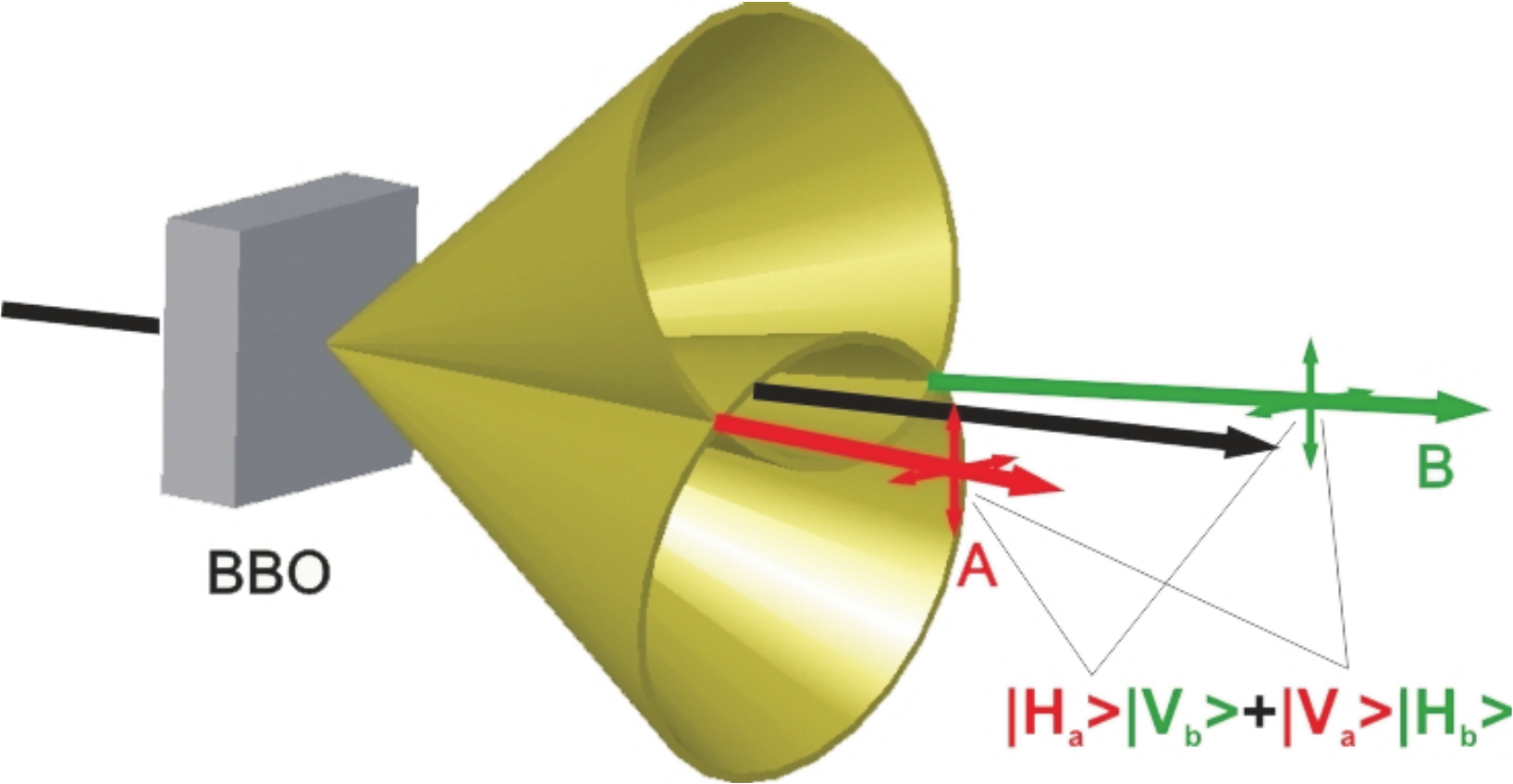}
\end{center}
\caption{Type II SPDC. The two photons have orthogonal polarizations.
They are emitted on two different cones.
Entangled state emerges from the intersection of the cones.
Because of the birefringence of the crystal
to see quantum interference one needs to compensate
different time of flight of $H$ and $V$ polarized photons (not shown in this Figure).}
\label{FIG_TYPE_II_PDC}
\end{figure}

For a suitable angle between
the laser beam and the optical axis of the crystal
the down-converted photons have orthogonal polarizations (Type II SPDC).
One of them has polarization
of the ordinary beam,
the other - of the extraordinary beam.
The
photons with orthogonal polarizations
appear on \emph{two} different cones (Fig. \ref{FIG_TYPE_II_PDC}).

Let us focus on the light generated in the intersection points of the cones.
There, one of the photons has an orthogonal polarization
to the other, but the polarization of a single photon
is not defined.
However, in principle one can learn the polarization of a single photon
with the time-of-flight through the crystal measurement
(since differently polarized photons propagate with different
velocities in the birefringent media).
Thus, to observe the polarization entanglement
one has to erase the time-of-flight information.
This can be achieved with a compensation outside the crystal.
One simply rotates the polarization
and lets the photons pass through a half-width crystal.
The final state essentially reads \cite{TYPE-II-SPDC}:
\begin{equation}
|\psi \rangle_{12} = \frac{1}{\sqrt{2}} \Big[ | H \rangle_1 | V \rangle_2 + e^{i \phi} | V \rangle_1 | H \rangle_2 \Big],
\end{equation}
where $H$ and $V$ denote horizontal and vertical polarization, respectively.
The relative phase, $\phi$, can be arbitrarily engineered
e.g. by using an additional phase shifter.
This source was used to disprove the class of nonlocal theories
of the main text.

\end{document}